\documentclass[a4paper,11pt]{article}
\pdfoutput=1 
\usepackage{jcappub}
\usepackage[T1]{fontenc} 
\usepackage[usenames,dvipsnames]{xcolor}
\usepackage{amsmath,amssymb}
\usepackage{caption}
\usepackage{subcaption}
\usepackage{multirow}
\usepackage{multicol}
\usepackage{booktabs}
\usepackage{rotating} 
\usepackage[commentmarkup=footnote]{changes}
\usepackage{pifont}
\usepackage{graphicx}
\usepackage{hyperref}
\usepackage[capitalize]{cleveref}
\usepackage{xspace}
\preprint{CERN-TH-2024-173}
\definechangesauthor[name=RB, color=teal]{rb}

\title{
Stellar-mass black-hole binaries in LISA: characteristics and complementarity with current-generation interferometers
}

\author[a,b,1]{R.~Buscicchio,}
\author[c,d,1]{J.~Torrado,}
\note{Corresponding authors}
\author[e,f]{C.~Caprini,}
\author[g]{G.~Nardini,}
\author[h]{N.~Karnesis,}
\author[f]{M.~Pieroni,}
\author[a,b]{A.~Sesana}

\newcommand{\Bic}{Dipartimento di Fisica “G. Occhialini”, Università degli Studi di Milano-Bicocca, Piazza della Scienza 3, 20126 Milano, Italy}
\newcommand{\Infn}{INFN, Sezione di Milano-Bicocca, Piazza della Scienza 3, 20126 Milano, Italy}

\affiliation[a]{\Bic}
\affiliation[b]{\Infn}
\affiliation[c]{Dipartimento di Fisica e Astronomia “G. Galilei”, Università degli Studi di Padova, via Marzolo 8, I--35131 Padova, Italy}
\affiliation[d]{INFN, Sezione di Padova, via Marzolo 8, I--35131 Padova, Italy}
\affiliation[e]{Université de Genève, Département de Physique Théorique and Centre for Astroparticle Physics, 24 quai Ernest-Ansermet, CH-1211 Genéve 4, Switzerland}
\affiliation[f]{CERN, Theoretical Physics Department, 1 Esplanade des Particules, CH-1211 Genéve 23,
Switzerland}
\affiliation[g]{Department of Mathematics and Physics, University of Stavanger, NO-4036 Stavanger, Norway}
\affiliation[h]{Department of Physics, Aristotle University of Thessaloniki, Thessaloniki 54124, Greece}
\emailAdd{riccardo.buscicchio@unimib.it}
\emailAdd{jesus.torrado@pd.infn.it}

\newcommand{\diff}{\mathrm{d}}
\newcommand{\sun}{\odot}
\newcommand{\msun}{M_\sun}
\newcommand{\yr}{\mathrm{yr}}

\newcommand{\PP}{\textsc{Power Law\,+\,Peak}\xspace}
\newcommand{\cat}{\textsc{GWTC-3}\xspace}
\newcommand{\srcname}{sBHB}
\newcommand{\srcnames}{sBHBs}
\abstract{
Stellar-mass black-hole binaries are the most numerous gravitational-wave sources observed to date. 
Their properties make them suitable for observation both by ground- and space-based detectors.
Starting from synthetic catalogues constructed based on
observational constraints from ground-based detectors, we
explore the detection rates and the characteristic parameters of the stellar-mass black-hole binaries observable by LISA during their inspiral, using signal-to-noise ratio thresholds as a detection criterion.
We find that only a handful of these sources will be detectable 
with signal-to-noise ratio larger than 8: 
about 5 sources on average in 4 years of mission duration, among which only one or two are multiband ones (i.e. merging in less than 15 years). 
We find that detectable sources have chirp mass $10\,\msun\lesssim \mathcal{M}_c\lesssim 100\, \msun$, residual time-to-coalescence $4\mathrm{yr}\lesssim \tau_c\lesssim 100 \mathrm{yr}$, and redshift $z\lesssim 0.1$, much closer than those observed up to now by ground-based detectors. 
We also explore correlations between the number of LISA detectable sources and the parameters of the population, suggesting that a joint measurement with the stochastic signal might be informative of the population characteristics.
By performing parameter estimation on a subset of sources from the catalogues, we conclude that, even if LISA measurements will not be directly informative on the population due to the low number of resolvable sources, it will characterise a few, low-redshift candidates with great precision. 
Furthermore, we construct for the first time the LISA waterfall plot for low chirp-mass systems, as a function of their time to coalescence and inclination.
We demonstrate that LISA will also be able to
discriminate and characterize, through very precise parameter estimation, a population of binaries with higher masses, $\mathcal{M}_c\sim \mathcal{O}(10^3)\msun$, at the boundary of ground-based detectors sensitivity.
}

\begin{document}
\maketitle
\flushbottom

\section{Introduction}
\label{sec:intro}

During their lifetime from formation to coalescence, stellar-mass black-hole binaries (\srcnames{}) emit gravitational waves (GWs) over a broad frequency range, crossing both LISA and ground-based interferometer sensitivities~\cite{Sesana:2016ljz}. 
However, unlike ground-based interferometers probing \srcnames{} at merger~\cite{KAGRA:2021duu}, LISA will detect \srcnames{} during the inspiralling phase, with GW signals behaving differently depending on the source chirp mass and orbital period at the start of LISA observations.
Within the mission lifetime, most \srcnames{} will produce quasi-monochromatic GW signals, while a minority will show a slow, positive frequency drift, either crossing a significant portion of the LISA band or eventually moving out of it.
These latter are the sources with higher signal-to-noise ratio (SNR), and can transition into the ground-based detector band over a timescale short enough to allow for multiband detection~\cite{Sesana:2016ljz,2019PhRvD..99j3004G,Buscicchio:2021dph, Toubiana:2022vpp}.

In this paper, we
investigate
what LISA can reveal about the
\srcname{} population.
The vast majority of \srcnames{} have too low SNR to be individually resolvable by LISA, and contribute to a stochastic GW background, which we characterised in~\cite{Babak:2023lro}.
Here we work on the same 
assumptions and simulation framework
of Ref.~\cite{Babak:2023lro}, but instead of focusing on the stochastic background component, we focus on the individual \srcnames{} detected by LISA: we 
estimate their number, describe their properties, compare them with those of the \srcnames{} detected so far by LIGO/Virgo/KAGRA (LVK), and perform a few example parameter estimations to characterise LISA capabilities.

In \cref{subsec:gwtc3-post}, we summarise the population simulation framework that we developed
in~\cite{Babak:2023lro} to construct the \srcname{} catalogues.
In particular, we adopt the \PP LVK model of the \srcname{} population~\cite{KAGRA:2021duu} based on the analysis of the \cat catalogue~\cite{KAGRA:2021vkt}, further supplemented with an extension of the merger-rate redshift going beyond a single power law, based on the star formation rate of~\cite{Madau:2016jbv}.
Besides the merger rate, we disregard any possible redshift dependence of other population
parameters, and adopt a uniform distribution for the residual time-to-coalescence evaluated at the start of the LISA mission and in the detector frame.
The criterion we adopt to select, out of the synthetic catalogues, the \srcnames{} detectable by LISA is solely based on SNR, which we detail in \cref{subsec:rapid-pop-sim}. To compute it, we further assume that all binaries are circular. In \cref{subsec:source-distr}, we present the number density distributions of the \srcnames{} detectable by LISA for three values of detection threshold, as a function of the residual time-to-coalescence, chirp mass, redshift, and mass ratio.
We predict a median value of 5 resolvable sources with $\mathrm{SNR}\ge 8$, increasing to about 40 for $\mathrm{SNR}\ge 4$, a threshold value compatible with archival searches~\cite{Toubiana:2022vpp,Wong:2018uwb}.
These numbers are in broad agreement with previous analyses \cite{Seto:2022xmh,Lehoucq:2023zlt,Ruiz-Rocha:2024xjt} (see \cref{subsec:source-distr}). According to our results, most resolvable sources have residual time-to-coalescence $4 \,\mathrm{yr}\lesssim \tau_c\lesssim 100\,\mathrm{yr}$, with a median of only one or two suitable for multiband detection, i.e. with~$\tau_c\leq 15$\,yr.
Their chirp mass distribution peaks between $10\,\msun\lesssim \mathcal{M}_c\lesssim 50\, \msun$ with a tail up to $\mathcal{M}_c\sim 100\, \msun$, and they are characterised by low redshift $z\lesssim 0.1$, a requisite to yield sufficiently high SNR in LISA.
We further find that the catalogues providing the highest number of events contain sources that are chirping and appear at relatively low frequency in the LISA band: 
regardless of the actual population model, these would be the two features that need to be fulfilled in the real Universe realisation to maximise the number of resolvable sources with LISA. 
  
In \cref{sec:pop-model}, we analyse how the number of resolved \srcnames{} depends on the population parameters. We find that the strongest correlation is with the index $\kappa$ of the merger rate power-law dependence with redshift, while weaker correlations appear with the index $\alpha$ of the power-law distribution of the primary component mass, and with the mixture parameter $\lambda_{\mathrm{peak}}$ of its high-mass peak.
We further analyse the role of a possible time delay $t_d$ between the binary formation and its merger, finding that dependence of the number of resolved LISA sources on the time delay is negligible, since it is dominated by the population posterior variance.

In \cref{sec:senscompar} we perform a comparison between the sources in LVK \cat and those detectable by LISA \emph{if the \srcname{} population obeys the LVK-inferred model} \PP.
We find that statistically LISA is only sensitive to very close-by \srcnames{} with $z\lesssim 0.1$, in the low-redshift tail of the sources' distribution, where LVK has observed no events, yet.
While seemingly contradictory, we argue this is to be expected (see discussion in \cref{subsec:source-distr}).
Regarding their chirp mass and mass ratio, the sample of \srcnames{} detectable by LISA is distributed similarly to the LVK events. 
This indicates that LISA parameter reach does not extend beyond that of LVK if the \PP population model is confirmed.
\footnote{We remind that even if LVK and LISA probe similar statistical samples of the \srcname{} population, observing them at both the inspiral and merger stages would constrain important features of the \srcname{} evolution (e.g.~eccentricity) \cite{LISA:2022yao} as well as cosmological and general relativity observables~\cite{LISACosmologyWorkingGroup:2022jok,LISA:2022kgy}.}
On the other hand, if the characteristic chirp masses extend beyond $10^2\,\msun$ (see e.g.\ the population of beyond-gap binaries of \cite{Mangiagli:2019sxg}), a fraction of the population gradually exits the LVK detection region while remaining within the LISA one.
We demonstrate this by producing, for the first time, the LISA waterfall plot in the chirp-mass range $1\, \msun \lesssim \mathcal{M}_c \lesssim 5\times10^3\,\msun$.
We do so exploring the dependence on $\tau_c$ and inclination $\iota$ (cfr.~\cite{2020NatAs...4..260J} for optimally oriented sources).
Due to the peculiar shape of the LISA sensitivity curve combined with the presence of the Galactic foreground~\cite{LISA:2017pwj, Colpi:2024xhw}, the waterfall plot peaks in redshift at around
$100\,\msun$, decreases, and then grows again for chirp masses higher than $10^3\,\msun$. 
We show that this is a feature specific to sources with merger times larger than the LISA observation time. 

In \cref{subsec:param-estim} we perform parameter estimation on a subset of 12 sources extracted from the catalogues with the restriction that they have an SNR larger than 8 and will merge in less than $15\,\mathrm{yr}$ from the start of LISA observations. Both of these restrictions place such sources at $z\leq0.1$.
We find that LISA can characterise these sources very well, with $\sim 10^{-4}$ relative uncertainty on the chirp mass and on the residual time-to-coalescence, $20-30$\% relative uncertainty on the redshift, and $\sim 10^{-6}$ relative uncertainty on the initial frequency of entrance in the LISA band. 
In addition, we perform parameter estimation on three sources with $\mathcal{O}(10^3)\,\msun$. 
Even though they are well beyond the higher mass cutoff imposed within the \PP model, we nevertheless consider them to confirm what is hinted by the waterfall plot, i.e.~that LISA will be able to detect and characterise such a population of high mass sources, if it exists. 
The parameters of these sources are also very well measured, featuring $\sim 10^{-6}$ relative uncertainty on the chirp mass and $10\%$ on the redshift, for the highest mass source with redshifted $\mathcal{M}_c=1800\,\msun$.
Finally, in \cref{sec:conclusions} we present our conclusions.

\section{Expected sources}
\label{sec:exp-sources}

In this section we detail the \srcname{} astrophysical population model of our catalogues, and revise our procedure to extract the \srcnames{} detectable by LISA.
We then analyze the statistical properties of the resolvable-source catalogues, provide the sources number density distribution, and describe their distribution in redshift, chirp mass, and residual time-to-coalescence.
We conclude by exploring how these properties depend on the choice of population model and its posterior.

\subsection{\srcname{} population model}
\label{subsec:gwtc3-post}

Our analysis is based on the \srcname{} population synthesis approach presented in Ref.~\cite{Babak:2023lro}, which assumes the fiducial \srcname{} population model inferred by the LVK Collaboration~\cite{KAGRA:2021duu}.
Even though this model might be subject to changes as the number and accuracy of LVK detections increase, it currently yields the largest evidence, hence we adopt it as the fiducial one.

The procedure detailed in~\cite{Babak:2023lro} to produce synthetic \srcname{} catalogues involves several assumptions, which we summarise in what follows.
Previous studies~\cite{KAGRA:2021kbb,Mapelli:2019bnp,KAGRA:2021duu} have found no evidence of redshift dependence in the \srcname{} population parameters (though other works mildly hint towards such an effect~\cite{Rinaldi:2023bbd,Fishbach:2018edt,Karathanasis:2022rtr}).
Including redshift dependency in our analysis would come at much higher computational cost,
since we could not take advantage of the factorization of parts of the number density distribution to generate the corresponding source parameters independently.
Therefore, here we neglect any potential dependence on redshift in the distribution of the population parameters.\footnote{Our analysis of the \srcname{} confusion noise in~\cite{Babak:2023lro} requires this assumption to hold up to $z\approx5$, while in the present analysis, as we will prove, it is enough that it holds up to $z\approx0.5$, since the likelihood of LISA resolving a source at $z>0.1$ is very low.}
Moreover, we assume that binary formation and evolution are in a steady state for the range of residual time-to-coalescence relevant to the LISA band ($\lesssim 10^5\,\mathrm{yr}$), meaning the rate of binaries merging or emitting at any allowed frequency is statistically constant; equivalently, $\tau_c$, the residual time-to-coalescence in the source frame is uniformly distributed across the \srcname{} population. 
Note that this steady-state assumption would hold even in the case of a redshift dependence in the distribution of the population properties, since that dependency would not have a significant effect across such a short time span.

Additionally, we work under the assumption of \srcname{} with negligible eccentricity.
It is well-established that LISA has significantly greater potential to infer sources' eccentricities than LVK-like detectors. 
The reason is two-fold: circularization naturally reduces the eccentricity when a system approaches its merger~\cite{Peters:1963ux}, and the large number of waveform cycles observable by LISA makes it highly sensitive to the source eccentricity in its early inspiral~\cite{Klein:2022rbf}.
However, by assuming no eccentricity, we reduce the computational cost and focus on other sets of population parameters that LISA can characterize complementarily to LVK. Furthermore, since eccentricity is not constrained by current observations, results would be heavily dependent on our prior model assumptions.
Note, however, that substantial evidence for small eccentricity on systems observed by LVK has been provided in literature~\cite{Romero-Shaw:2020thy,Gayathri:2020coq,Bhaumik:2024cec}.

Within the aforementioned assumptions the differential number of \srcnames{} 
with parameters $\xi$ at redshift $z$ merging after a time $\tau_c$ (in the source frame) reads~\cite{Babak:2023lro}
\begin{equation}
  \label{eq:d3N}
  \frac{\diff^3N(z, \tau_c, \xi,\Lambda)}{\diff\xi\, \diff z\, \diff\tau_c} = R(z, \tau_c)\left[\frac{\diff V_c}{\diff z}(z) \right] p(\xi | \Lambda)\, .
\end{equation}
Here, the dependencies on the population model are enclosed in the \srcname{} merger rate $R(z, \tau_c)$, and in the probability density function $p(\xi | \Lambda)$ of the source parameters $\xi$, both intrinsic (masses, spins, polarisations, orbital phase and frequency) and extrinsic (sky position, inclination, distance), as a function of the parameters $\Lambda$ describing the population model.
The cosmology is instead accounted for in the Universe comoving volume dependence on redshift, $\textrm{d} V_c/\textrm{d} z$. 
For consistency with LVK and~\cite{Babak:2023lro}, we use the best-fit $\Lambda$CDM cosmological model inferred from the ``Planck 2015 + external'' data combination of~\cite{Planck:2015fie}. 
Note that in \cref{eq:d3N} we can use the merger rate density in the form that LVK is constraining~\cite{KAGRA:2021duu} because within our assumptions, all values of $\tau_c$ are equally likely at any $z$, and thus the time interval $ \diff \tau_c$ is equivalent to a generic time interval $\diff t$ (see discussion in~\cite{Babak:2023lro}).

The choice of the functional forms on the right-hand side of \cref{eq:d3N} sets our population model. 
As in~\cite{Babak:2023lro}, for the distribution $p(\xi | \Lambda)$ of source parameters we follow the nowadays standard~\cite{KAGRA:2021duu} combination of a \PP model for the mass distributions, a tapered inverse power-law for the mass ratio, and independent Beta distributions for the spin amplitudes and isotropic plus truncated Gaussian mixtures for the tilts. The extrinsic source characteristics are drawn from uniform and isotropic distributions.

Finally, our $R(z,\tau_c)$ follows the functional form of the Madau-Fragos star-formation rate (SFR) $R_{\rm SFR}$~\cite{Madau:2016jbv}, with the modifications described in App.\ A of~\cite{Babak:2023lro}. 
At redshifts $z\lesssim 1.5$, this is equivalent to the power-law-like merger rate adopted by LVK~\cite{KAGRA:2021duu}. 
Since, as we will confirm later, LISA does not individually resolve \srcnames{} at $z\gg 0.5$, for all practical purposes 
we use in our analyses
a rate $R(z) = R_0 (1 + z)^{\kappa}$, with parameters $R_0$ and $\kappa$ 
matching
those inferred by LVK. The importance of this choice will be discussed in \Cref{sec:pop-model}.

\subsection{Rapid population simulation}
\label{subsec:rapid-pop-sim}

Our goal is to investigate within the population model described above the number and properties of the \srcnames{} that LISA is expected to detect individually in its nominal 4-year mission duration~\cite{LISA:2017pwj, Colpi:2024xhw}.
We 
study the distribution of the properties of these individually resolvable compact binaries across population parameters compatible with the posterior inferred from \cat~\cite{KAGRA:2021duu}.
To do so, we would have to generate a large number of synthetic catalogues of \srcnames{} emitting in the LISA frequency band, varying the population parameters.
These catalogues contain tens to hundreds of millions of events. 
Though we make use of the fast \textsc{extrapops} simulation\footnote{Here we use the term \textit{simulation} to refer to the generation of synthetic populations from a probability distribution for the source parameters, as opposed to the term \textit{synthesis}, which in the literature usually refers to the creation of synthetic populations from astrophysical simulation codes, involving e.g.~stellar formation and evolution.}
code~\cite{2023ascl.soft05003T},\footnote{Available at \url{https://github.com/JesusTorrado/extrapops}} which can produce millions of events per second and CPU core, the cost in terms of CPU time and storage of this direct approach would be 
prohibitive.
We therefore aim to restrict the simulation to events above a minimum SNR.
Parameter inference analyses simulating the LISA data analysis pipeline~\cite{Buscicchio:2021dph,Toubiana:2022vpp} have established that an SNR of 8 is a viable detection threshold.
In order to extend our study to sources susceptible to be identified via archival searches in LISA data (i.e.\ using the detection of the post-LISA signal at ground-based interferometers), we consider also SNRs down to 6 and 4~\cite{Wong:2018uwb, Toubiana:2022vpp}, paying special attention to sources with a residual time-to-coalescence smaller than 15 years after the start of LISA observation, which could be suitable for multiband detection.

Only a very small fraction of the sources of each synthetic population, consisting of all \srcname{} emitting in the LISA band, survive these SNR cuts.
Therefore, imposing them when generating synthetic populations greatly reduces computational costs.
However, calculations of realistic SNRs are expensive.
To impose this cut, we therefore first use an analytic, approximate SNR calculation, based on the leading-order Stationary Phase Approximation waveform for inspiralling compact binaries.
In addition, for this calculation, we assume sources to be optimally-oriented (\textit{face-on}, i.e.\ 
cosine-
inclination $\cos\iota = 1$), in order to minimize the chances that we miss an event for which the approximate SNR would be 
underestimated.
Following~\cite{Babak:2021mhe}, averaging over polarizarion and sky position, the approximate SNR squared reads:
\begin{equation}
\label{eq:AnalyticalSNR}
\left< \mathrm{SNR}^2 \right>_{\cos\iota = 1} =
2 \,\frac{16}{c^3} \left(\frac{ \sqrt{\frac{5}{24}}\, G^{5/6}\mathcal{M}_c^{5/6}}{\pi^{2/3} d_L} \right)^2 \int^{f_\mathrm{max}}_{f_\mathrm{min}} \frac{f^{-7/3}}{S_{h,X} (f)}\, \diff f,
\end{equation}
with $\mathcal{M}_c$ being the chirp mass of the source, $d_L$ its luminosity distance, and $S_{h,X}$ the LISA detector sensitivity in the long-wavelength limit for a single TDI 1.5 channel (Eq.\ (56) in~\cite{Babak:2021mhe}), with the factor 2 in front of \cref{eq:AnalyticalSNR} accounting for the two effective TDI channels of LISA in this approximation.
At this stage, we do not include the effect of the confusion noise coming from unresolved population members, because it is at sub-percent level with respect to the noise sensitivity~\cite{Babak:2023lro}.
The integration is performed over 
the intersection between the LISA frequency band and the span of frequencies over which the source has drifted during the 4 years of observation time.

Since in our catalogues we want to capture all sources with realistic SNR above $4$, for robustness the cut we impose is rather $\left(\left< \mathrm{SNR}^2 \right>_{\cos\iota = 1}\right)^{1/2} = 2$ from \cref{eq:AnalyticalSNR}.
Furthermore, we achieve even faster simulation by imposing that the sources in the catalogues have redshift smaller than $1$ and residual times to coalescence shorter than $2000\,\yr$.
Outside these limits it is extremely unlikely to find sources with realistic SNR above 4.

Within these restrictions, using \textsc{extrapops}, we generate one population realisation for each of the approximately 11 thousand equal-weight posterior samples from the publicly available \cat catalogue~\cite{ligo_scientific_collaboration_and_virgo_2021_5655785} for the fiducial population model described in \cref{subsec:gwtc3-post}~\cite{KAGRA:2021duu}.
For each realisation, we re-compute the SNR of each source using the simulated instrument data analysis pipeline presented in \cite{Karnesis:2021tsh}, which provides a more realistic evaluation of the SNR in LISA than \cref{eq:AnalyticalSNR}.
This pipeline is based on an iterative process, where the stochastic signal plus instrumental noise is directly computed from the power spectral density (PSD) of the data.
Then, any source with SNR higher than a given threshold is classified as resolvable, and thus is subtracted from the data. The noise levels are estimated again and the loud sources are subtracted in the same manner as before. The procedure stops when either convergence is reached at the noise PSD,
or there are no more sources left to subtract.
In this evaluation, we add to the LISA instrumental noise the confusion noise due to unresolved members from the corresponding population, computed as described in ~\cite{Babak:2023lro} -- despite this only having a small effect on the final SNR. 
We finally 
group
the sources taking 4, 6 and 8 as SNR thresholds, according to the realistic SNR computation, and discard the rest.
The full set of synthetic populations, together with a Python notebook demonstrating parts of the present analysis, is available at \url{https://zenodo.org/records/13974091} \cite{torrado_2024_13974091}.

\subsection{Sources distribution}
\label{subsec:source-distr}

As explained in the previous section, we have simulated at least one population per equally weighted sample from the \cat posterior, and computed the number of detectable sources $N_\mathrm{det}$ for each of the SNR threshold considered: $\rho_0=4,\,6$ and $8$. With $\Lambda$ being the population parameters referred to in \cref{eq:d3N} (here also including the ones defining the merger rate), the resulting set of tuples $(N_\mathrm{det}, \Lambda)$ represents a fair sample from the joint distribution $p(N_\mathrm{det},\Lambda \mid \rho_0,\mathrm{LVK})$, where the condition ``$\mathrm{LVK}$'' denotes the \cat population posterior. 
We construct the expected distribution for the number of \srcnames{} resolvable by LISA and compatible with the \cat observations by marginalizing $p(N_\mathrm{det},\Lambda\mid\rho_0,\mathrm{LVK})$ over the population parameter space $\Omega_\Lambda$, or, equivalently, taking the expected value
of the number of observable sources conditioned to the population parameters
over the posterior $p\left(\Lambda \mid \mathrm{LVK}\right)$ from \cat:
\begin{equation}\label{eq:Ndist}
p\left(N_\mathrm{det} \mid \rho_0, \mathrm{LVK}\right) = \int_{\Omega_\Lambda} p\left(N_\mathrm{det} \mid \Lambda,\rho_0\right) p\left(\Lambda \mid \mathrm{LVK}\right)\, \diff\Lambda\;.
\end{equation}
Numerically, this operation amounts to simply dropping the population parameters from the $(N_\mathrm{det}, \Lambda)$ tuples constructed as described above.

\begin{figure}[ht]
    \centering
    \includegraphics[width=0.7\textwidth]{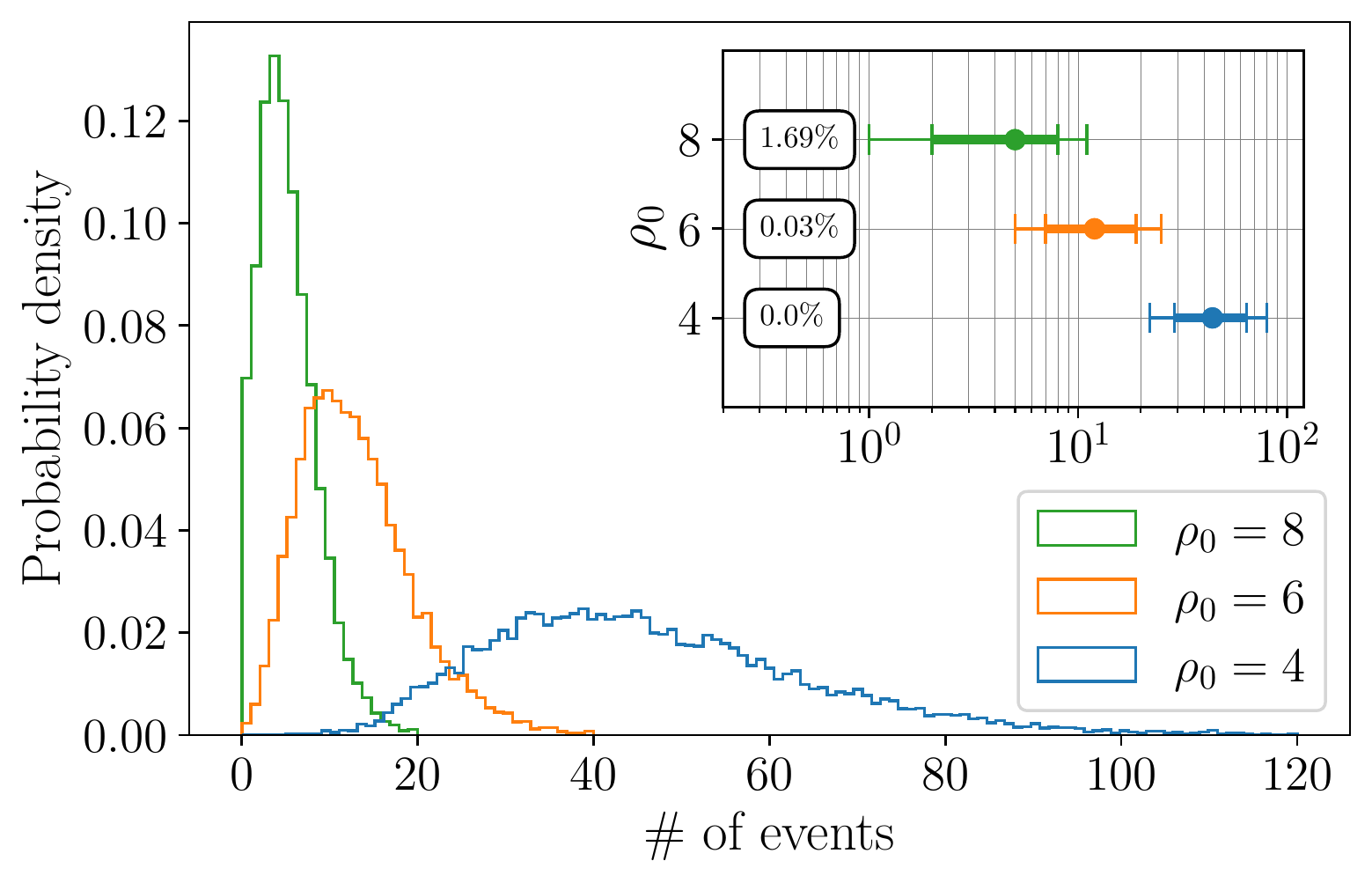}
    \caption{
    Histograms of the predicted distribution of number of sources resolvable by LISA,
    inferred from \cat observations, using synthetic catalogues generated as explained in \cref{subsec:rapid-pop-sim}.
    The blue (orange, green) probability density refers to an SNR threshold $\rho_0= 4\,(6,8)$.
    (\textit{Inset plot}) Point estimates of the number of detectable sources: bullet markers correspond to the median of each distribution in the main plot, while thick (thin) lines denote the $68\% (90\%)$ confidence interval.
    Percentages in each row denote the probability of zero detectable sources for each given $\rho_0$ threshold.}
    \label{fig:nsources}
\end{figure}

The resulting distribution for $N_{\rm det}$
is shown in \cref{fig:nsources}.
The number of expected resolvable sources with SNR threshold $\rho_0=8$, has a median value around $5$, having a $\sim 2\%$ probability for no sources being resolved, and a $10\%$ probability for more than 10 sources being resolved.
We compare our findings with previous results in the literature: using a similar merger rate and an ad-hoc chirp mass distribution based on the \srcname{} masses in GWTC-3, Ref.~\cite{Seto:2022xmh} finds an analytical point estimate of approximately $4$ detectable sources;
Using the population model that we analyze here, Ref.~\cite{Lehoucq:2023zlt} obtains $6^{+3}_{-2}$ detectable \srcnames{} taking into account the $90\%$ confidence interval of the merger rate at $z=0.2$ under \cat, and fixing the rest of the parameters, capturing roughly half of the full posterior uncertainty of $5^{+6}_{-4}$ shown in our study. When the detection threshold is lowered to values suitable for archival searches, we expect a number of sources of the order of $10$ and $50$ for SNR threshold of $6$ and $4$,respectively.

\begin{figure}[ht!]
     \centering
     \begin{subfigure}[c]{0.48\textwidth}
         \centering
         \includegraphics[width=\textwidth]{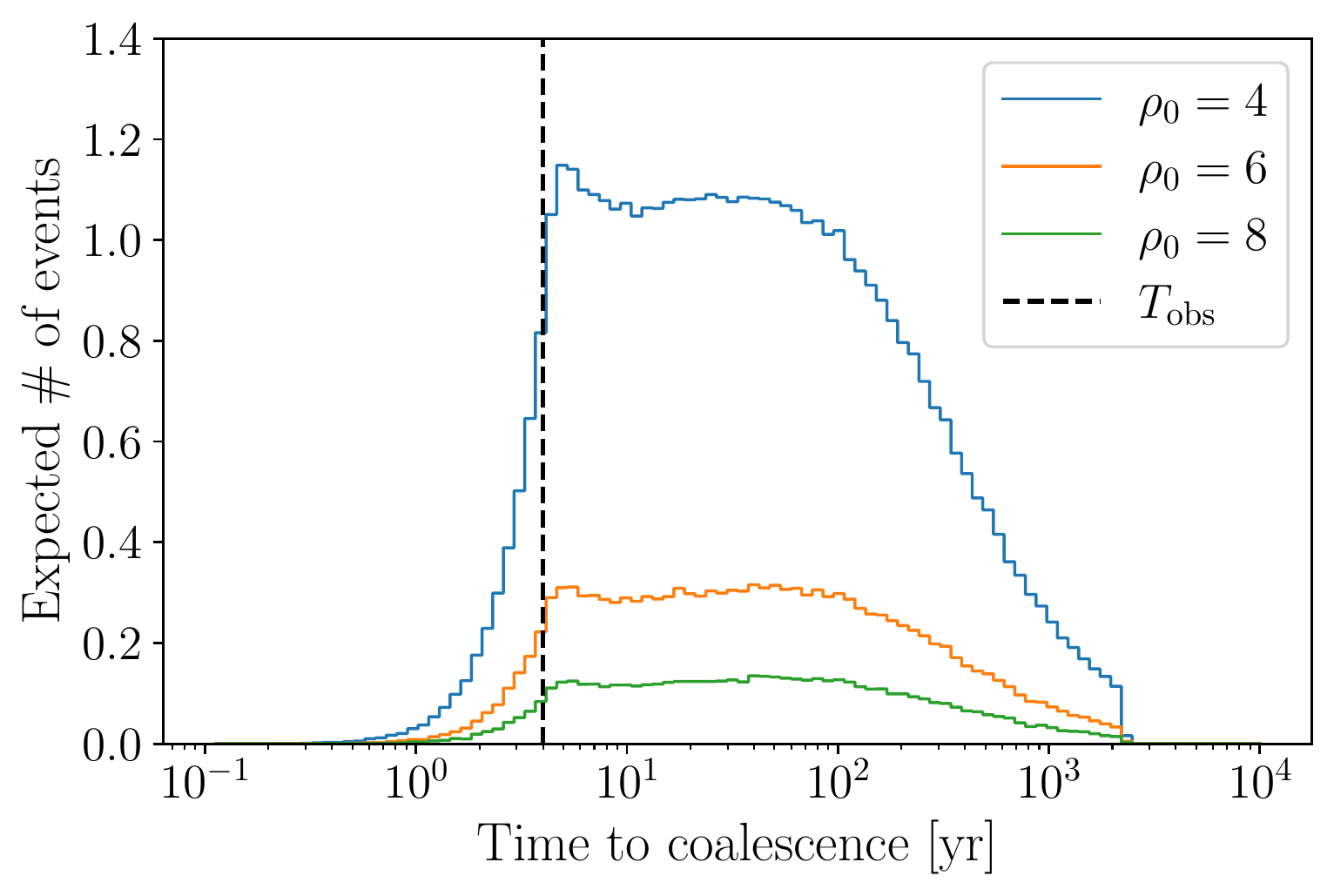}
         \caption{}
         \label{fig:tchist}
     \end{subfigure}
     \hfill
     \begin{subfigure}[c]{0.48\textwidth}
         \centering
         \includegraphics[width=\textwidth]{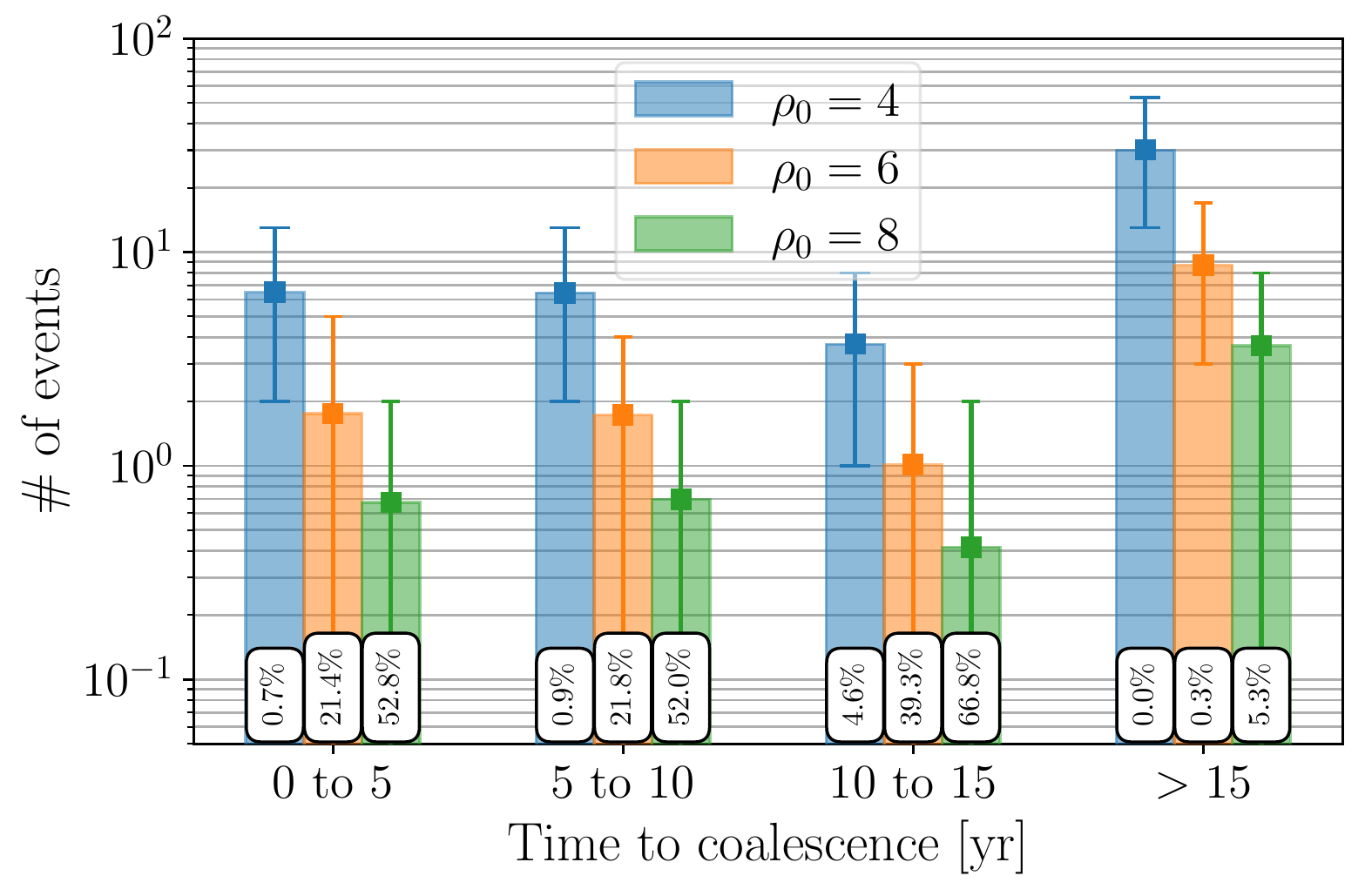}
         \caption{}
         \label{fig:tchist_numsources}
     \end{subfigure}
     \caption{
      (a) Distribution over 
      residual time-to-coalescence in detector frame of sources number counts across all catalogues for SNR thresholds $\rho_0=4\,\text{(blue)},\,6\,\text{(orange)},\,8\,\text{(green)}$. The dashed black line denotes the nominal mission duration $T_{\rm obs}=4\,\mathrm{yr}$.
       (b) Number of detectable events aggregated by residual time-to-coalescence intervals, including median (bar top edge) and $90\%$ confidence intervals (thin capped lines), for SNR thresholds $\rho_0=4,6,8$ (same colours).
       Percentages at the bottom of each bar denote the fraction of catalogues, for each residual time-to-coalescence range and $\rho_0$ threshold, yielding no detectable sources.
       Notably, for $\rho_0= 4$, all catalogues contain at least one source with $\tau_c>15$\,yr.
       Percentages do not necessarily add up to 100\%, as some catalogues yield no detectable sources in multiple $\tau_c$ and $\rho_0$ intervals.}
       \label{fig:tcdist}
\end{figure}

\cref{fig:tchist} shows the number density distributions as a function of the residual (log-)time-to-coalescence $\tau_c$.
The SNR of inspirals in LISA strongly depends on $\tau_c$. 
On one hand, their emission amplitude increases with decreasing $\tau_c$.
On the other hand, their emission frequency also increases with decreasing $\tau_c$, as does, in turn, the GW-driven frequency drift; this has the opposite effect: as sources spend less time in the LISA band, those with smaller $\tau_c$ have smaller accumulated SNR.
As shown in~\cref{fig:tchist}, the switch between these two regimes occurs at $\tau_c$ comparable to the mission duration, which we have set to 4\,yr~\cite{Tamanini:2019usx}
.
Therefore, given the large range of $\tau_c$ resulting in detectable sources for all SNR thresholds, the counts are dominated by sources that merge long after the LISA mission,  with $\tau_c$ of the order of 10 -- 100\,yr.\footnote{The sharp cut at $\tau_c=2000\,\mathrm{yr}$ in \cref{fig:tchist} corresponds to the limit imposed on $\tau_c$ for the generation of the catalogues, and does not affect the sources of interest.}  Most of the resolvable \srcnames{} are then detected as nearly monochromatic signals by LISA~\cite{Seto:2022xmh}.
On the other hand, sources merging before the end of the mission, characterised by a large frequency drift, make up roughly $10\%$ of the total, for all SNR thresholds. 

We compare our findings on the number of sources merging within the mission with results in literature. In Ref.~\cite{Ruiz-Rocha:2024xjt}, the authors populated with \srcnames{} the star formation regions resulting from the Illustris hydrodynamic cosmological simulation \cite{Vogelsberger:2014dza}, using the binary star evolution code MOBSE \cite{Giacobbo:2017qhh} with a fiducial model approximately matching the mean merger rate recovered by \cat. For sources merging before the end of the LISA mission ($4\,\yr$), they find point estimates of $2$, $5$ and $18$ for the SNR thresholds $\rho_0=8, 6$ and $4$. 
That is at least a factor of $4$ greater than our result, likely due to modeling differences (e.g. greater likelihood of massive sources, a merger rate consistently greater than the mean \cat one for $z<2$).

If LISA detects sources with a small enough residual time-to-coalescence,
in our convention $\tau_c \leq 15\,\mathrm{yr}$,
future ground-based GW experiments should be able to observe them as they merge.
Let us now focus on these \emph{multiband} sources, either those individually resolved by LISA ($\rho_0=8$), or potentially found in archival searches in the LISA data ($\rho_0= 4,6$).
These sources account for roughly $1/3$ of the total for all SNR thresholds.
\cref{fig:tchist_numsources} shows the number of resolvable sources aggregated by residual time-to-coalescence intervals: for SNR detection threshold $\rho_0= 4$ we find $16^{+14}_{-11}$ multiband (i.e.~$\tau_c \leq 15\,\mathrm{yr}$) sources at $90\%$ confidence interval, for $\rho_0= 6$ the number decreases to $4^{+6}_{-3}$. 
For $\rho_0=8$, we find ${2}^{+3}_{-2}$ multiband sources. 
Even though the posterior lower bound is compatible with zero, $79\%$ of the catalogues yield at least one source with $\tau_c \leq 15$ and SNR above $\rho_0=8$. 
The percentages shown in~\cref{fig:tchist_numsources} denote the fraction of catalogues yielding no detectable sources aggregated by $\tau_c$ ranges.

\begin{figure}[t]
    \centering
    \includegraphics[width=0.99\textwidth]{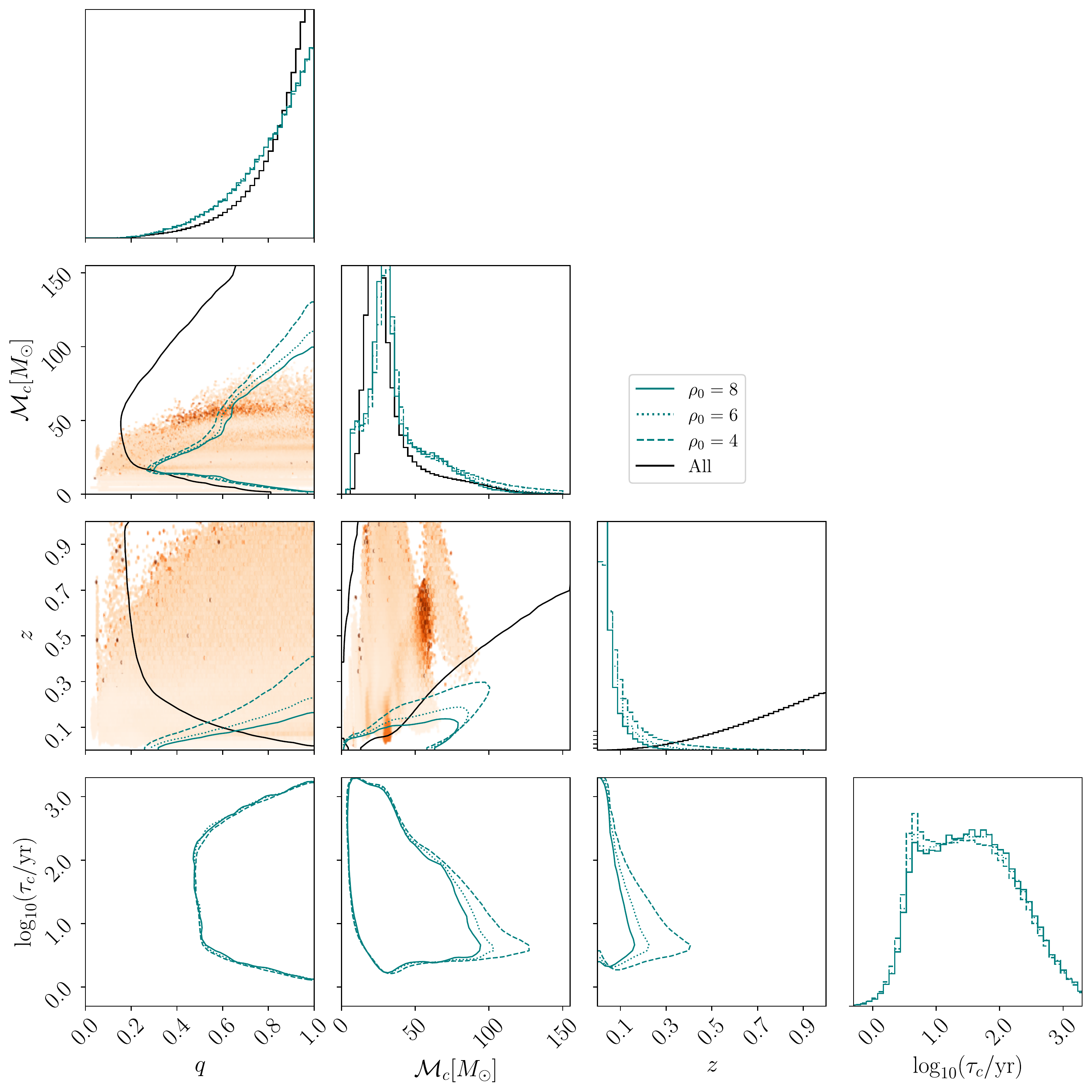}
    \caption{Predicted distribution of LISA detectable sources. Teal solid (dotted, dashed) lines enclose $90\%$ of sources with SNR above 8 (6,4) across all catalogues. Solid black lines are the $90\%$ contours for a full population (including unresolved sources) simulated with the median population parameters from \cat.
    Orange shaded areas correspond to the density of posterior samples from individual events in \cat.
    }
  \label{fig:LISAcorner}
\end{figure}

\Cref{fig:LISAcorner} further illustrates 
the complementarity between space- and ground-based detectors. It displays the source parameter distributions across our catalogues. 
The $90\%$ probability contours 
for the most relevant source parameters 
of the expected resolved events (i.e.~chirp mass ${\cal M}_c$, redshift $z$, mass ratio $q$ and residual time-to-coalescence $\tau_c$) are shown as a teal solid (dotted, dashed) line
for $\rho_0=8$ (6,4).
To gauge whether LISA is complementary to LVK in
probing these parameters,
we overplot in orange the
individual events posteriors from the \cat catalogue.
This allows for a straightforward
comparison with LISA detection potential
for sources belonging to the same population~\cite{KAGRA:2021duu}.
It is apparent that
the LISA parameter reach is roughly within the LVK one except for the high-mass tail of the chirp mass distribution. 
In fact, the bulk of the LISA sources have chirp mass $10\,\msun\lesssim \mathcal{M}_c\lesssim 50\, \msun$, similarly to LVK, but the tail of the mass distribution extends to up to $\mathcal{M}_c\sim 100\, \msun$.
The reach of LISA is strongly
limited in redshift: it is unlikely to resolve events with redshift larger than 0.1, a much smaller horizon than LVK.
LISA will also be less sensitive to smaller mass ratios.

The seeming inconsistency between the observed events in \cat and the support for low redshift and masses of the population observed by LISA, illustrated by the little-to-no overlap in the $(\mathcal{M}_c, z)$ subplot in \cref{fig:LISAcorner}, is readily explained.
While the LVK network has collected data in observing mode for 393 (171) days with two (three) detectors, the detectability of sources by LISA is based solely on their emission frequency, i.e.~including events merging in up to thousands of years. 
In other words, among the hundreds of millions of sources within that range of residual time-to-coalescence, only $\mathcal{O}(10)$ might be
resolved by LISA. Sampling the same portion of the parameter space with ground-based experiments would require a number of the order of 1000 hypothetical non-simultaneous experimental runs.
In \cref{fig:LISAcorner} we also show the population of all predicted sources (solid black line), for a representative catalogue with the median population parameters from \cat, but without any selection due to detectability by LISA. 
The black line encompasses the sources detected in GWTC-3, but not the potentially detectable ones by LISA, because they fall at the far tail of the distribution (c.f.~again the $({\cal M}_c,z)$ subplot).

\begin{figure}[t]
    \centering\includegraphics[width=0.6\textwidth]{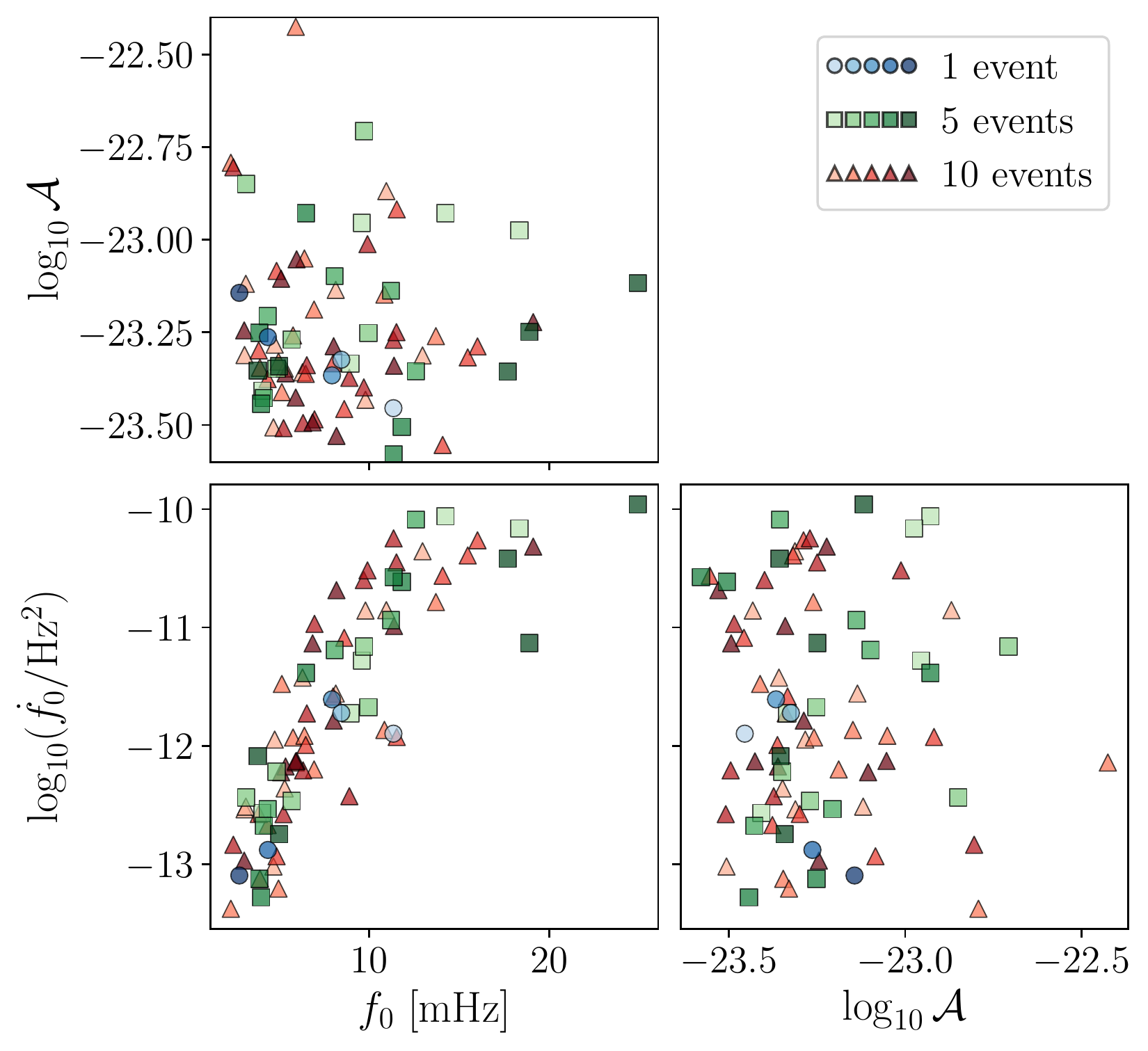}
    \caption{Scatter plots of individual events across catalogues, in terms of phenomenological binary parameters: the
    signal
    frequency at the start of LISA observation, 
    its 
    amplitude, and the frequency derivative. Circle (squares, triangles) denote parameters from 5 random catalogues with $N_{\rm det}=1 (5,10)$, each catalogue shown with a different shade of blue (green, red). 
}\label{fig:scatter-cats}
\end{figure}

In order to understand the features of LISA \srcname{} detectability, it is also useful to analyse phenomenological source parameters such as the waveform amplitude, the initial GW frequency $f_0$ and and its derivative $\dot f$ (both evaluated at the start of the LISA observation in the detector frame).
In \cref{fig:scatter-cats} we show their values
for three sets of 5 catalogues, randomly sampled from those yielding 1, 5 and 10 resolvable sources, respectively.
Clearly, as $f_0$ increases, the frequency derivative $\dot f$ increases as well, as predicted by the leading-order inspiral relation
\begin{equation}
    \dot{f}=\frac{96}{5}\pi^{8/3}\left(\frac{G\,\mathcal{M}_c}{c^3}\right)^{5/3}f^{11/3}\,.
    \label{eq:fdot}
\end{equation}
Furthermore, the sources pertaining to resolvable-source-poor catalogues tend to have smaller $\dot{f}$, consistently with what is expected for the whole population, which is composed mainly of slowly-evolving binaries. 
On the other hand, the correlation between the plus (cross) polarization amplitude ${\cal A}_+$ (${\cal A}_\times$) and frequency, given by
\begin{equation}
\mathcal{A}_{\left\{+,\times\right\}}
=\frac{4}{d_L}
\left(\frac{G\,\mathcal{M}_c}
{c^2}\right)^{5/3}
\left(\frac{\pi\, f(t)}
{c}\right)^{2/3}\left\{\frac{1+
\cos^2\iota}{2},\cos\iota
\right\}\,,  
\end{equation}
is less clear in~\cref{fig:scatter-cats}, probably because other physical parameters, e.g.\ the luminosity distance and the inclination, play a relevant role.
It is anyway appreciable from the figure that the sources with low frequency $f_0$ tend to cluster at low amplitude, because they stay in band longer and therefore accumulate more SNR. \cref{fig:scatter-cats} also highlights a large variability of source properties across catalogues, so that even the catalogues with the smallest number of resolvable \srcnames{} appear to reflect the characteristics of the whole population.

\subsection{Dependency on the population parameters 
and time delay
} \label{sec:pop-model}

In this section, we first investigate the correlation between the number of LISA-detectable sources and the population parameters as inferred from \cat.
We then analyse the impact of a possible time delay between the formation of the binary of stars and their evolution into \srcname{} systems.

\begin{figure}[ht]
    \centering
    \includegraphics[width=1.0\textwidth]{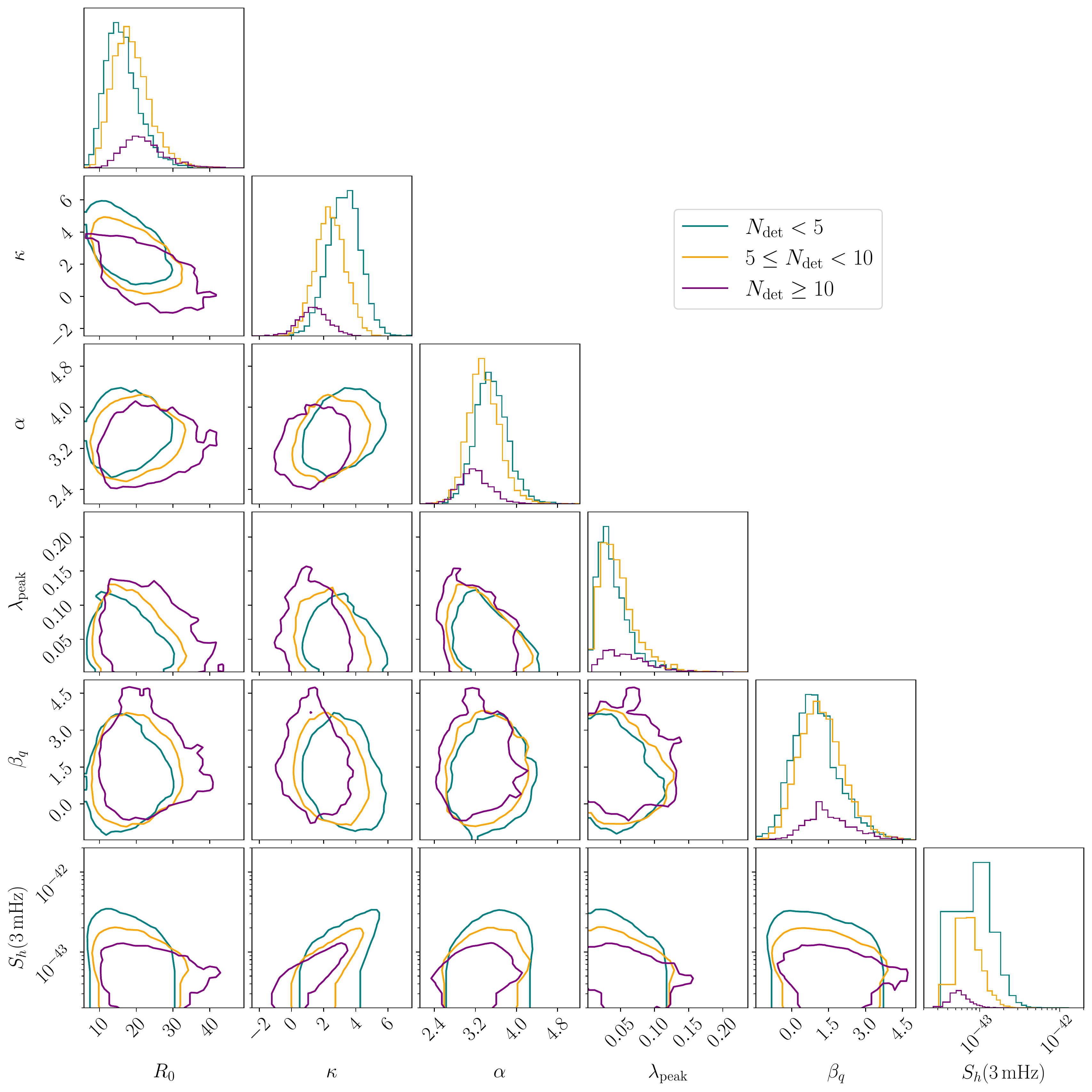}
    \caption{
    Distribution of some population parameters and expected SGWB power with respect to the number of detectable sources. Teal (orange, purple) contour lines denote $90\%$ confidence intervals for population parameters of catalogues with $N_{\rm det}<5$  ($5\leq N_{\rm det}<10$, $N_{\rm det}\geq10$) expected detectable sources by LISA. The one dimensional histograms along the diagonal are normalized to the number of catalogues for each chosen $N_{\rm det}$, accordingly.
    (\textit{Bottom row}): The expected SGWB power spectral density at a reference frequency of $3\ {\rm mHz}$. We observe systematic anti-correlation between the number of detectable sources and the residual SGWB amplitude, and a positive correlation between the power-law index $\kappa$ and the background amplitude. 
    }\label{fig:CorrCorner}
\end{figure}

\begin{figure}
     \centering
     \begin{subfigure}[b]{0.495\textwidth}
         \centering
         \includegraphics[width=\textwidth]{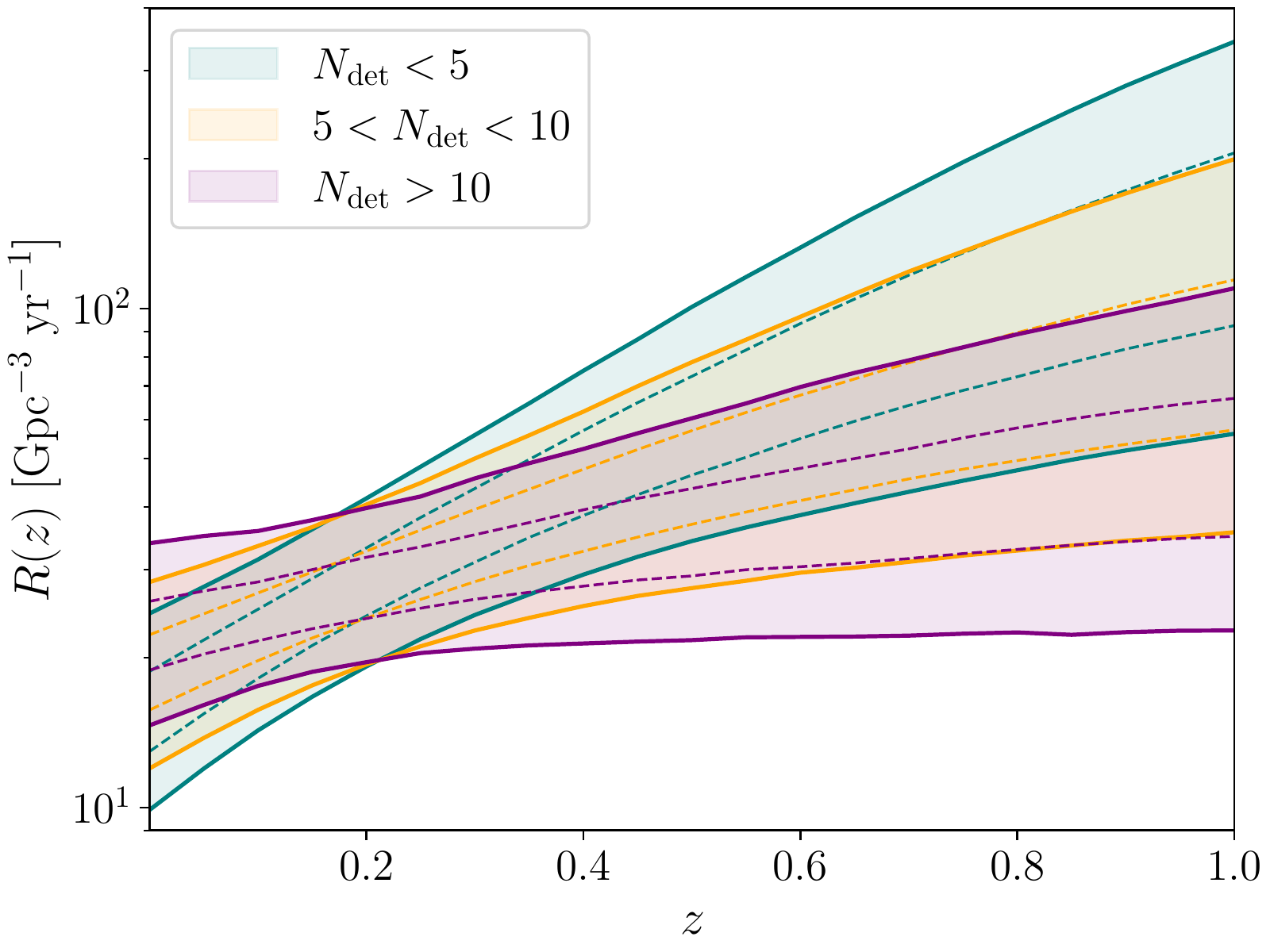}
         \caption{}
         \label{fig:corr_rate}
     \end{subfigure}
     \hfill
     \begin{subfigure}[b]{0.495\textwidth}
         \centering
         \includegraphics[width=\textwidth]{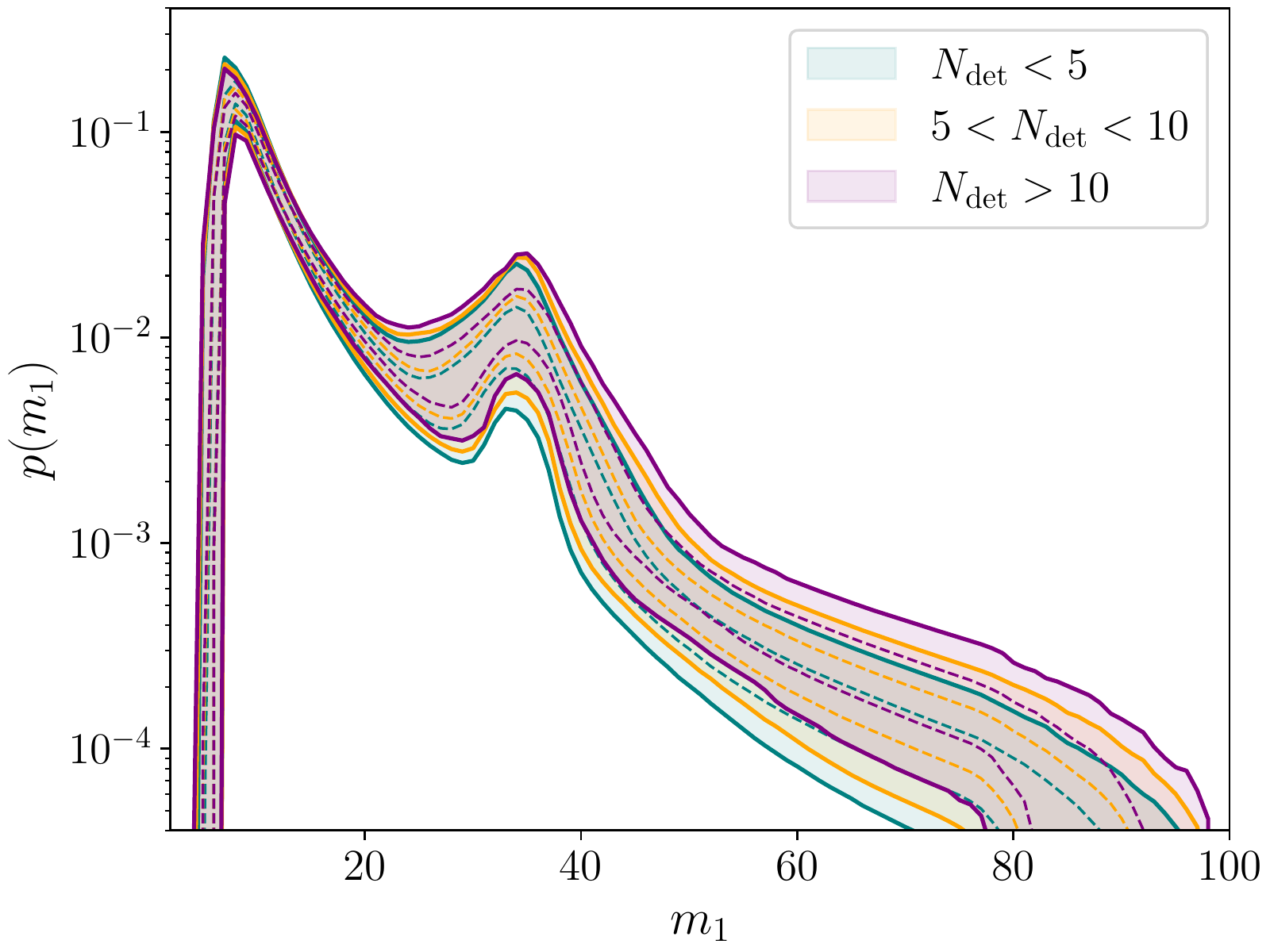}
         \caption{}
         \label{fig:corr_mass1}
     \end{subfigure}
    \caption{
    (a) Effect of the negative correlation between the merger rate power-law index $\kappa$ and the number of loud ($\rho_0= 8$) sources in LISA: $50\%$ and $90\%$ confidence intervals of the merger rate versus redshift, $R(z)$, for catalogues with $N_{\rm det}<5$, $5\leq N_{\rm det}<10$, $N_{\rm det}\geq 10$. Larger number of detectable sources are associated with larger rates at redshift below $z=0.2$, which correspond to lower values of the power-law index $\kappa$: indeed, LVK constrains best the merger rate at $z=0.2$, which produces the anticorrelation.
    (b) 
    The probability distribution of the binary primary component mass, for catalogues yielding $N_{\rm det}<5$, $5<N_{\rm det}<10$, $N_{\rm det}>10$. 
    Larger number of loud events are associated with less negative indices for the power-law component of the distribution, $\alpha$, and in a smaller measure positively correlated with the relative amplitude of the Gaussian peak (given by larger values of the mixture parameter $\lambda_{\mathrm{peak}}$) and its mean mass.
  }
  \label{fig:corr_params}
\end{figure}

\Cref{fig:CorrCorner} shows the $90\%$ confidence intervals on the population parameters for the synthetic catalogues, grouped by the number of detectable sources in each, with $N_{\rm det}<5$, $5\leq N_{\rm det}<10$, $N_{\rm det}\geq 10$, and assuming $\rho_0=8$.
We find a significant negative correlation between the expected number of sources above threshold and the merger rate power-law index $\kappa$ in the low-redshift regime: $R(z)\propto (1 + z)^{\kappa}$; this effect is even greater at lower detection thresholds.
This correlation is caused by the combination of (i) the merger rate being best constrained by LVK at $z=0.2$, so that lower values of $\kappa$ yield an excess of sources below that redshift, and a defect above it; and (ii) the fact that LISA is more sensitive to sources at redshift $z\lesssim0.2$.
\Cref{fig:corr_rate} illustrates it, showing the $50\%$ and $90\%$ confidence intervals of the merger rate versus redshift, $R(z)$, for catalogues with $N_{\rm det}<5$, $5\leq N_{\rm det}<10$, $N_{\rm det}\geq10$. The role of the pivot at $z=0.2$ is apparent.

We further focus on the dependence on the mass distribution parameters.
As stated in \cref{subsec:gwtc3-post}, in our analysis we assume that the mass $m_1$ of the primary component follows the same fiducial \PP model as in the \cat population study~\cite{KAGRA:2021duu}: a mixture of a large power-law component ($m_1^{-\alpha}$), and a small, narrow Gaussian component centered at $m_1\approx30\,$--$40\,M_\odot$, with some overall tapering at low masses.
The higher sensitivity of LISA to large-mass binaries with respect to ground-based detectors produces two significant correlations: 
catalogues with higher number of sources tend to have a heavier-tailed power law component (smaller $\alpha$) and larger a contribution from the Gaussian component (larger mixture parameter $\lambda_\mathrm{peak}$), though the latter correlation is less pronounced.
\cref{fig:corr_mass1} illustrates these correlations.
We point out that a small positive correlation is observed with the power-law index $\beta_q$ controlling the mass ratio $q=m_2/m_1$ distribution, indicating a preference for near equal-mass binaries, as shown in~\cref{fig:LISAcorner}.

In summary, the  sensitivity of LISA, favouring the detection of low-redshift, massive sources, produces correlations between the parameters describing the population and the number of detected sources.
However, the expected number counts are too small to discern the effect of such correlations over Poisson noise, and therefore the detection of individual sources has limited statistical power to constrain the population model. 
On the other hand, the amplitude of the stochastic gravitational wave background (SGWB)  due to the unresolved \srcnames{} presents similar correlations, but with opposite signs. 
This can be appreciated from the bottom row of \cref{fig:CorrCorner} showing the amplitude of the SGWB from \srcnames{} at a reference frequency of 3\,mHz,  computed following~\cite{Babak:2023lro}.
Therefore, a potential synergy between the observation of the \srcname{} background and resolvable sources can be exploited to gain insight on the population model.
For example, detecting a high background amplitude together with a high number of detected sources might indicate that a revision of the merger rate amplitude and power-law assumption is needed.
These considerations, however, are subject to the caveat that recent confusion background estimates due to extra-galactic white dwarfs may hinder a precise determination of that from unresolved \srcnames{}~\cite{Staelens:2023xjn}.

\begin{figure}
     \centering
     \begin{subfigure}[b]{0.42\textwidth}
         \centering
         \includegraphics[width=\textwidth]{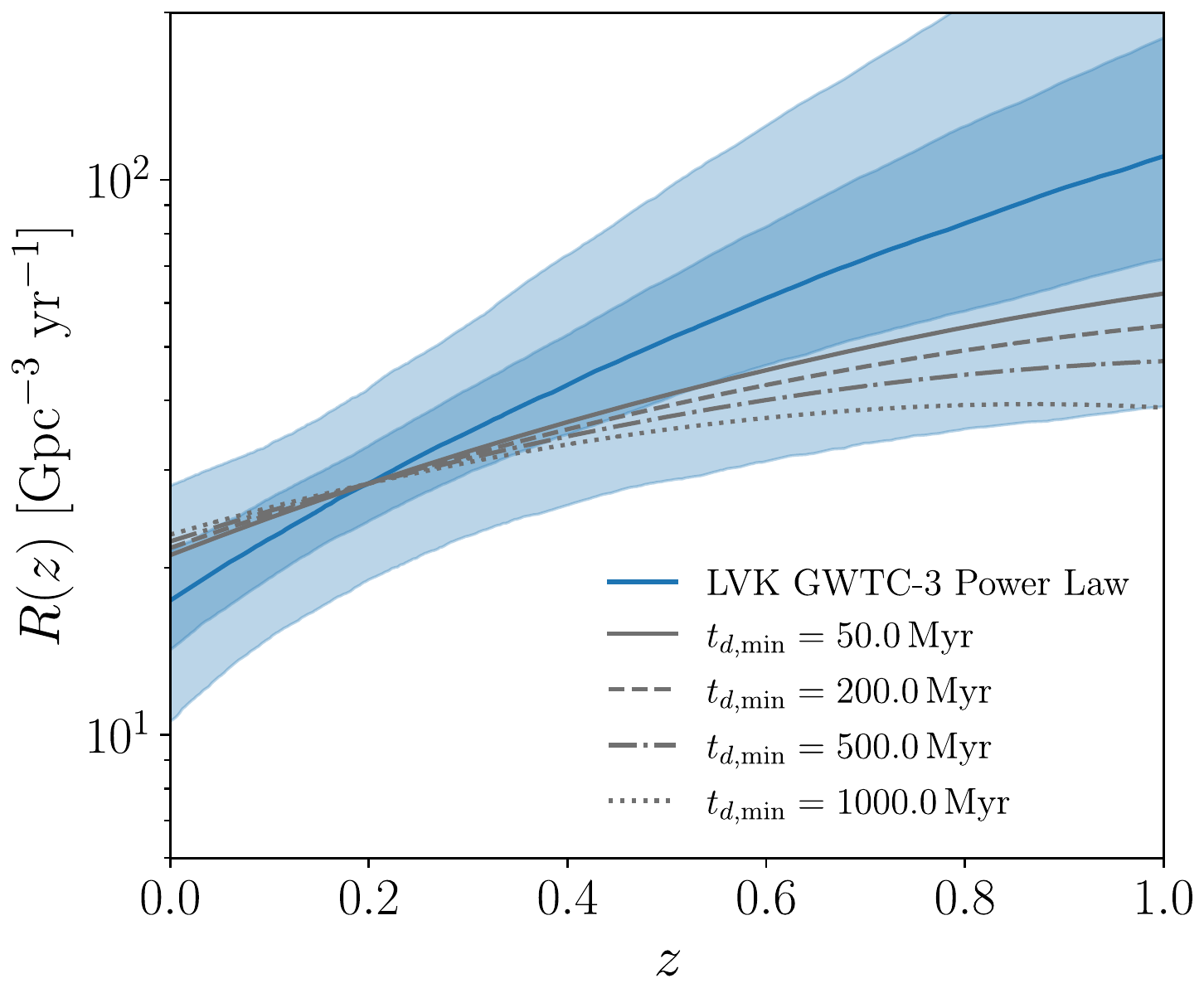}
         \caption{\label{fig:comp_delay_rate}}
     \end{subfigure}
     \hfill
     \begin{subfigure}[b]{0.525\textwidth}
         \centering
         \includegraphics[width=\textwidth]{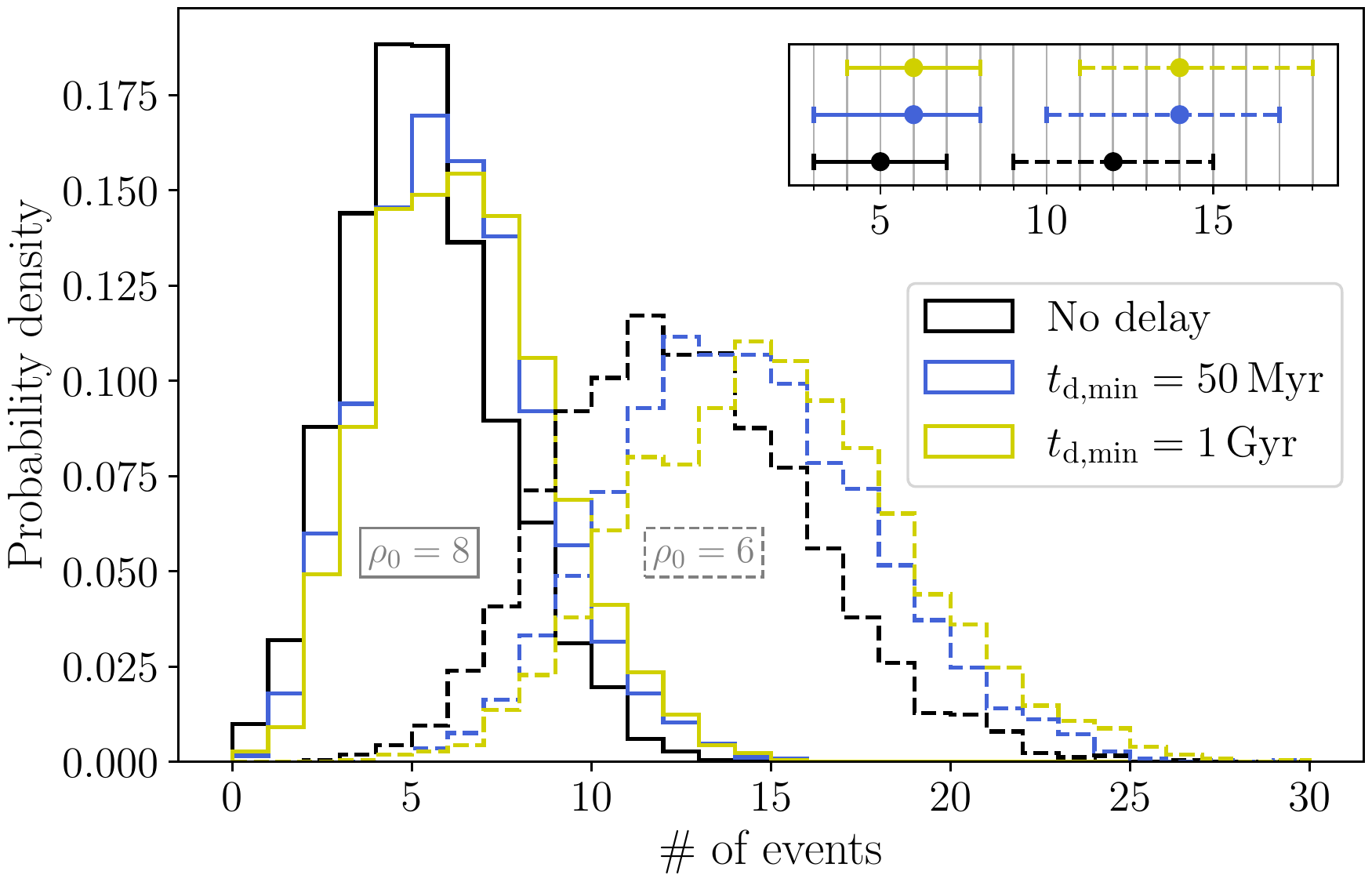}
         \caption{\label{fig:comp_delay_counts}}
     \end{subfigure}
    \caption{(a) 
    \cat posterior for the merger rate (blue-shaded) compared to the median \cat merger rate including different minimum time-delays with the inverse-time model. (b) Dependence of the number of sources with SNR above threshold $\rho_0= 8$ and $6$ on the minimum time delay for the median of the population parameters from the \cat posterior; the effect is smaller than the posterior variance (cf.\ \cref{fig:nsources}); the inset shows the $68\%$ CI.
    }
        \label{fig:comp_delay}
\end{figure}

The presence of a time delay between the formation of the stellar binary and their evolution to \srcnames{} can also modify the merger rate, with an effect similar to a shallower power-law index $\kappa$.
As found above, the value of $\kappa$ strongly influences the number of resolvable sources. 
We therefore investigate whether time delays can alter this number within our model.
To do that, we augment the merger rate used in our study (following the functional form of the star formation rate in Madau-Fragos~\cite{Madau:2016jbv}, see \cref{subsec:gwtc3-post}) with a time-delay log-uniform distribution, 
leading to~\cite{Dvorkin:2016wac, Fishbach:2021mhp}
\begin{equation}
R(z) = R_{\rm ref} \int_{t_{d,\mathrm{min}}}^{t_{d,\mathrm{max}}} R_\mathrm{SFR}(t(z)+t_d) \frac{1}{t_d}\,\mathrm{d} t_d \, ,
\end{equation}
where $R_\mathrm{SFR}(z)$ is the star formation rate, and 
the integration is performed between a chosen minimum and a large maximum time-delay, $t_{\rm d,min}$ and $t_{\rm d,max}$, respectively.
We fix the normalisation of the merger rate to the median rate of the \cat posterior at $z=0.2$, where it is best constrained.
We then explore physically motivated time-delay models in the literature by considering $t_{\rm d,min} = 0, 50, 200, 500, 1000 {\rm \,Myr}$ (due to the $1/t_d$ dependence, we take $t_{\rm d,max}$ large enough such that it does not influence the integral).
The resulting merger rates are presented in~\cref{fig:comp_delay_rate}: they are 
well within the blue band corresponding to the LVK posterior, and also compatible with the SGWB upper bound by LVK~\cite{KAGRA:2021duu}.

We assess the effect of a time-delay on the expected number of LISA detectable sources by fixing the population parameters to the median values obtained from population analyses in \cat for the fiducial model presented in \cref{subsec:gwtc3-post}, while replacing the merger rate with the ``delayed'' models, for the discrete set of $t_{\rm d,min}$ listed above.
We follow the procedure described in \cref{subsec:rapid-pop-sim} to evaluate the number of detectable sources with SNRs above $\rho_0=8$ and $6$, performing 2500 simulations for each delayed model.
In this context, given the large number of expensive realistic LISA SNR evaluations needed, we adopt the following procedure to speed up the source selection.
Starting for the sources in our main (non-delayed) set of catalogues, for which we have \emph{already} computed both the realistic SNR and the inclination-averaged one of \cref{eq:AnalyticalSNR}, we fit a multivariate polynomial function to the ratio of both SNRs,\footnote{The specific fitting function consists of the product of three polynomials accounting separately for the effect of the parameters most correlated with the difference between realistic and approximate SNR: one of degree 6 for the inclination, one of degree 6 for the ecliptic latitude, and one of degree 4 for the residual log-time-to-coalescence.}
reaching an accuracy better than a few percent.
Then, for the newly-generated sources predicted by the delayed model, we compute approximate realistic LISA SNRs from the product of the fast inclination-averaged SNRs and the fitting function.
Results are shown in \cref{fig:comp_delay_counts}.
Comparing with \cref{fig:nsources}, we can see that the observed variability of the number of detections is too weak to be directly informative on the time-delay distribution, confirming what was already hinted in~\cite{Lehoucq:2023zlt} using a simpler analysis.

\section{Comparison of LISA and LVK sensitivity to \srcnames{}}
\label{sec:senscompar}

We turn our attention to 
the characterization of the complementarity 
between ground- and space-based detectors in detecting \srcname{} signals, without making any assumption on the underlying population.
For this purpose, we compare the sensitivity of LISA and LVK, evaluated in terms of SNR, in the $(\mathcal{M}_c,z)$ parameter space by constructing so-called ``waterfall plots''.
We assume LISA operating at nominal sensitivity for $4\, {\rm yr}$~\cite{LISA:2017pwj, Colpi:2024xhw}.
While Einstein Telescope~\cite{2010CQGra..27s4002P} and Cosmic Explorer~\cite{2023arXiv230613745E} readiness in the late '30s represents a compelling scenario thanks to their improved low-frequency sensitivity, their observation window remains uncertain. 
Therefore, we opt to conservatively consider the current LVK network operating at design sensitivity~\cite{KAGRA:2013rdx}

\begin{figure}[th]
    \centering
    \includegraphics[width=0.7\textwidth]{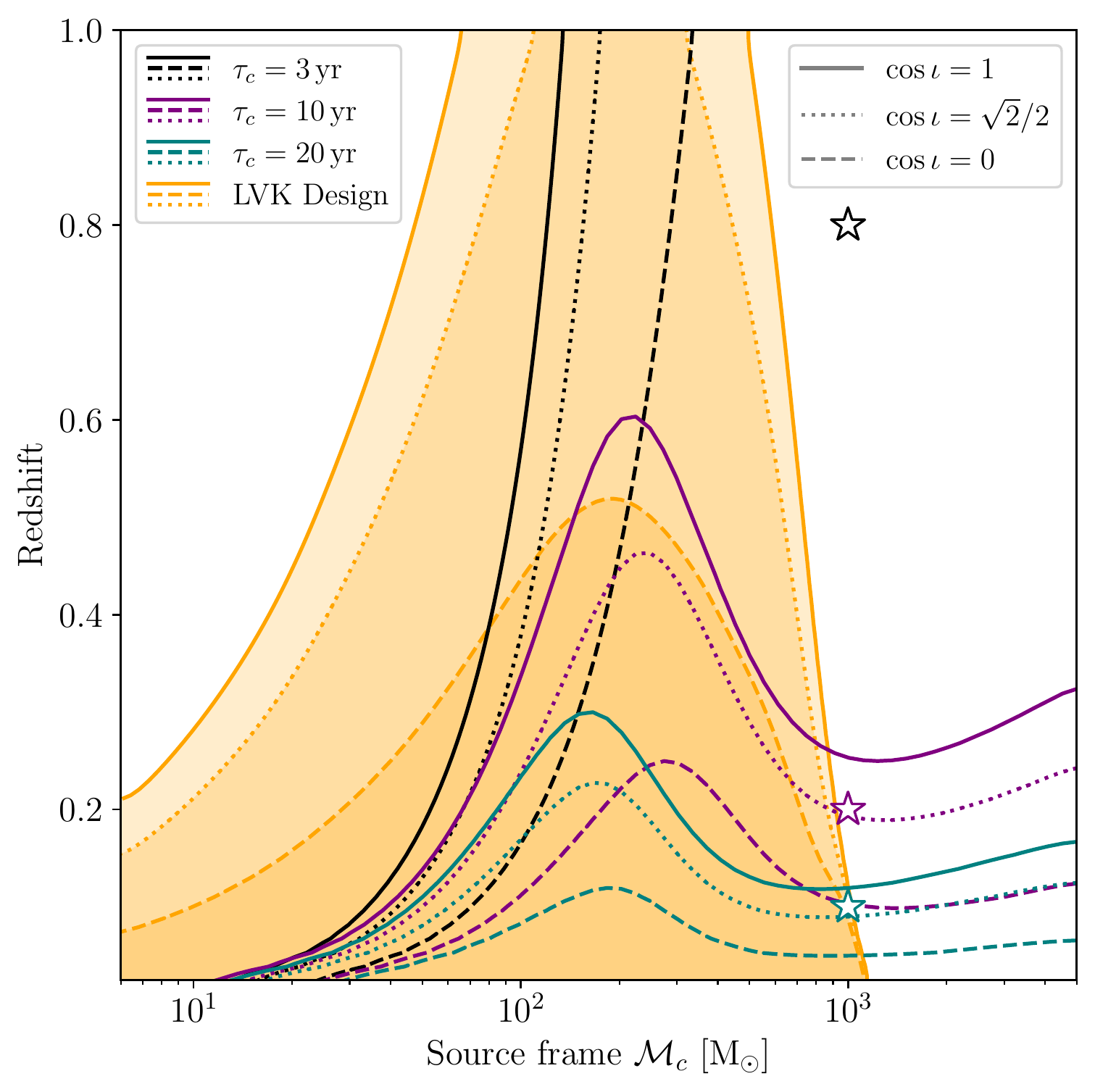}
    \caption{
    Comparison of LISA and LVK sensitivities 
    in the $(\mathcal{M}_c,z)$ plane. Black (purple, teal) lines denote the detectability horizon of LISA  with a threshold SNR of 8
    for sources with fixed residual time-to-merger of $\tau_c=3 {\,\rm yr}$ ($\tau_c=10 {\,\rm yr}$, $\tau_c=20 {\,\rm yr}$). Solid (dotted and dashed) lines denote sources whose orbital plane is inclined at $0 \deg$, $45 \deg$, and $90\deg$, respectively. 
    The orange shaded regions denote the detectability horizon of the LVK detector network at design sensitivity, with an SNR threshold of 14, corresponding approximately to three single SNRs\,=\,8. These orange regions are computed by omitting sources above 1000\,$\msun$ as their SNRs
    heavily depend on the LVK detectors' stability at very low frequencies below $10\,\mathrm{Hz}$; the Gaussian smoothing introduces the small tail
    visible in the figure.
    Black (purple, teal) stars denote three massive sources selected for parameter estimation at the boundaries of the LVK detectors design sensitivity, with source-frame chirp mass ${\cal M}_c = 1000\,\msun$, redshift $z=0.1$ ($0.2, 0.8$) and $\tau_c=20\, \mathrm{yr}$ ($10\, \mathrm{yr}, 3\, \mathrm{yr}$).
    }
    \label{fig:LISALIGOwater}
\end{figure}

\begin{figure}[th]
    \centering
    \includegraphics[width=0.9\textwidth]{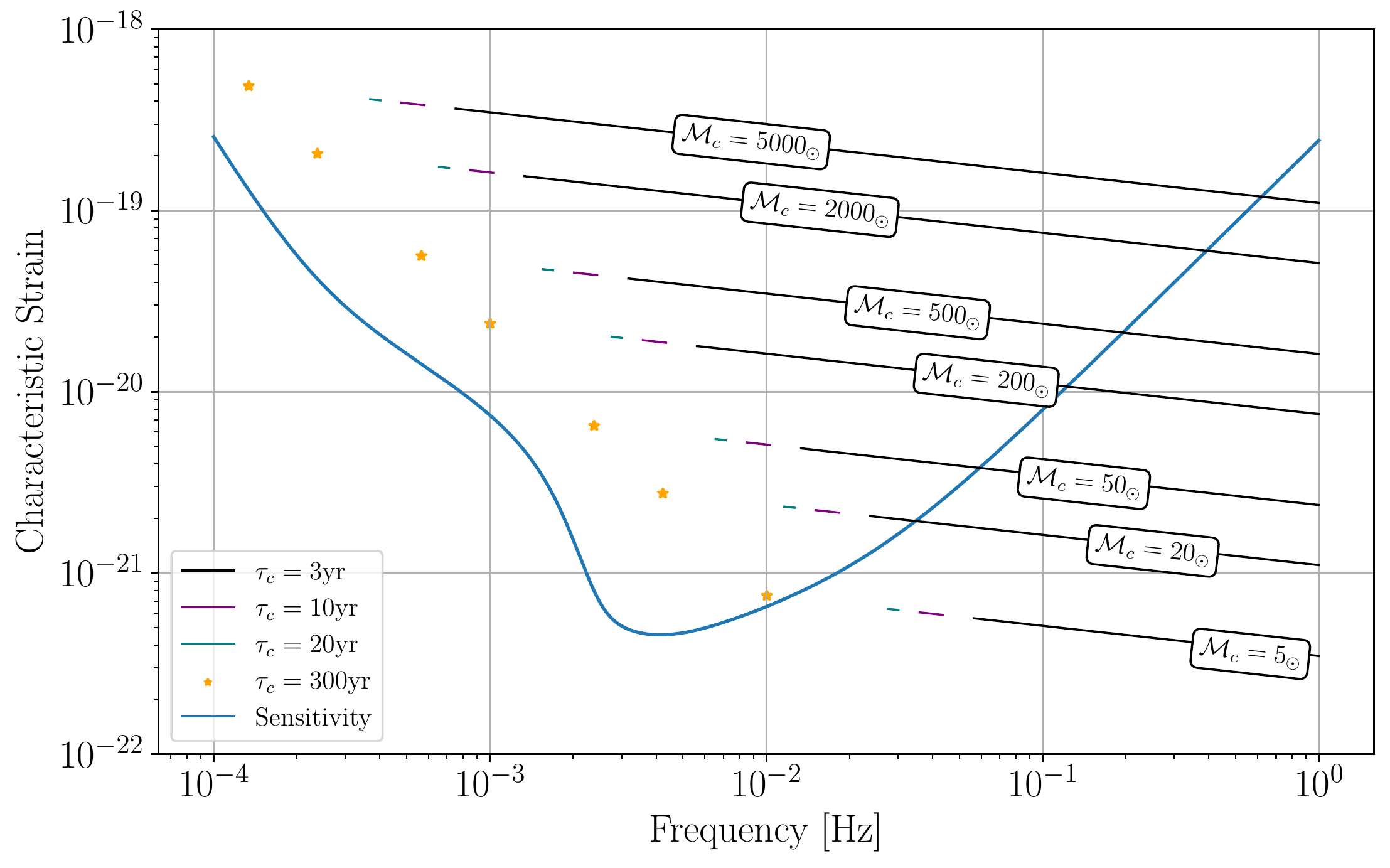}
    \caption{
    Characteristic strain 
    in LISA
    of sources with different residual time-to-coalescence 
    and chirp mass
    . 
    The blue line is the sum of the detector noise and the confusion noise from the Galaxy.
    Segments and points denote the inclination-averaged characteristic strain of 7 sources, all at {redshift 0.3},  
    with different detector frame chirp mass ${\cal M}_c$, according to the labels.
    Black (purple, teal) lines correspond to sources injected at a fixed time to merger of $\tau_c=3 {\,\rm yr}$ ($\tau_c=10 {\,\rm yr}$, $\tau_c=20 {\,\rm yr}$); orange dots denote quasi-monochromatic sources very early in their inspiral phase, with $\tau_c=300 {\,\rm yr}$. 
    The dip around 1000 solar masses in \cref{fig:LISALIGOwater} corresponds to the 
    amplitude of the Galactic
    confusion noise as a function of frequency, 
    which dominates the instrumental noise
    at frequencies between $0.2 {\, \rm mHz}$ and $3 {\, \rm mHz}$. The black lines extend out of the LISA band and reach their merger within the mission duration.
    For simplicity, no
    LISA response is applied 
    when plotting the characteristic strain
    as the lines would become quite cluttered (see e.g. Fig.1 in~\cite{Buscicchio:2021dph}).
    }
    \label{fig:LISAtracks}
\end{figure}
We 
characterize the detectability of individual GWs as a function of source-frame chirp mass ${\cal M}_c$ and redshift, see~\cref{fig:LISALIGOwater}.
As in the previous section, a GW signal is deemed detectable by LISA if its SNR is greater than 8.
Note that, in constructing the waterfall plot, we do not rely on the approximate expression in~\cref{eq:AnalyticalSNR}: we instead generate each GW signal and couple it to LISA with an accurate frequency-domain response, coherently with the parameter estimations presented in~\cref{subsec:param-estim}.
Similarly, we flag an event as detectable by LVK if its network SNR is greater than 14, roughly equivalent to a three-detector network threshold ${\rm SNR}=8$.

The parameters most affecting the SNR, apart from $\mathcal{M}_c$ and $z$, are:
(i) the source cosine inclination $\cos\iota$ with respect to the line-of-sight, which we bracket with three fiducial values: $0,1,$ and $\sqrt{2}/2$;
(ii) in the case of LISA, the binary's time-to-coalescence, for which we consider the values  $\tau_c=3 {\,\rm yr}\,,~\tau_c=10 {\,\rm yr}$, and $\tau_c=20 {\,\rm yr}$.
Concerning (i), it can be observed in \cref{fig:LISALIGOwater} that the SNR loss as the binary orientation moves away from face-on or face-off, decreases the maximal redshift at which sources can be detected.
Concerning (ii), as also
shown in \cref{fig:LISALIGOwater}, sources close to $3\,\mathrm{yr}$ from the merger are detectable up to large distances if ${\cal M}_c \ge 100 M_\odot$.
However, the horizon reduces to below $z\sim 0.6 (0.25)$ if the same sources are observed at $10 (20) {\rm yr}$ from the merger. 
Such reduction originates from the signal frequency evolution over the mission duration for systems with masses below $\sim1000 M_\odot$, exemplified by the colorful lines in \cref{fig:LISAtracks} (see description below). 
The slow frequency evolution shortens the frequency support and lowers the resulting integrated SNR. 

Furthermore, across all inclinations, we observe a peak at redshift $z\lesssim 0.6$ ($\lesssim 0.3$) for residual time-to-coalescence equal to 10\,(20)\,yr. \Cref{fig:LISAtracks} helps clarify the origin of this peculiar structure.
Therein, we pick a reference redshift equal to 0.3 and show the characteristic strain for sources with chirp mass between $5$ and $5000 \,M_\odot$ and 4 initial times-to-coalescence. 
For systems that are far from coalescence, the presence of a minimum (corresponding to maximal sensitivity) in the LISA noise curve at around $3{\rm mHz}$ induces the peak structure in the horizon distance plot in \cref{fig:LISALIGOwater}.
While the signal amplitude (and thus its SNR) increases with $\mathcal{M}_c$, systems with masses between a few hundred and a few thousand solar masses, emit at frequencies between $0.5$ and $3\,{\rm mHz}$, where the confusion noise from Galactic binaries dominates over the instrumental noise. 
This additional noise source sharply increases the overall noise curve, strongly suppressing the SNR for these systems. Binaries with larger masses emit at even lower frequencies, where the confusion noise is negligible and the instrumental noise dominates the total noise budget, yielding an ${\rm SNR}$ that monotonically increases with the mass of the system.
A similar peak is absent for $\tau_c = 3\,$yr because the SNR is dominated by the part of the characteristic strain signal extending to high frequency. 

Note that
the LVK network horizon does not feature the same peak structure. Indeed, no equivalent of the Galactic foreground exists for LVK and, furthermore, every  system entering the LVK frequency band 
merges in less than a few seconds (at the relevant masses) leading to a negligible signal duration compared to the total observing time.
Finally, we highlight that the detectability of binary systems with hundreds up to thousands of solar masses is highly 
dependent on the ground-based detectors' stability at and below $5\,{\rm Hz}$. 
Therefore, the high-mass end of the LVK waterfall plot should be treated with caution.  
In \cref{subsec:param-estim}, we expand on the parameter estimation accuracy for such sources by LISA.

\section{Parameter estimation on representative sources}
\label{subsec:param-estim}

In order to assess how well LISA will be able to characterise individual \srcnames{}, we have also performed Bayesian parameter estimation on a sample of representative sources.
We simulate and infer on the GW signals produced by each source using \textsc{Balrog}, following closely the procedure described in~\cite{Buscicchio:2021dph, Klein:2022rbf}, which we summarize here for completeness.
We simulate the GW emission in the frequency domain and perform Bayesian inference on the three 
time-delay-interferometry (TDI) data streams $d=\{d_k; k=A,E,T\}$, 
in terms of which the noise covariance matrix is diagonal for an equal-arms constellation with equal noise levels in the six inter-satellite links 
~\cite{Prince:2002hp,Muratore:2020mdf,Muratore:2022nbh,Hartwig:2023pft}. The likelihood is parameterized by the 11 parameters describing individual binaries, each defined in the Solar System Barycenter frame:
the chirp mass ${\cal M}_c=(m_1m_2)^{3/5}/(m_1 + m_2)^{1/5}$, where $m_{1,2}$ are the binary component masses;
the dimensionless mass difference $\delta\mu=(m_1-m_2)/(m_1+m_2)$ and the component spins $\chi_{1,2}$;
the initial orbital frequency $f^0_{\rm orb}$.
The intrinsic parameters are further completed by the extrinsic ones:
the luminosity distance $d_L$, 
the cosine inclination $\iota$ of the binary orbital plane with respect to the line-of-sight, described by its (sine)ecliptic latitude $\sin b$ and longitude $l$.
Finally, the GW signal is completely specified by choosing a polarization angle $\psi$ and an initial orbital phase $\phi_{\rm orb}$.
The likelihood therefore reads
\begin{equation}
\label{eq:likelihood}
\ln{\cal L}(d\mid \theta ) =
-\sum_k \frac{(d_k-s_k(\theta)\mid d_k-s_k(\theta))_k}{2} + {\rm const} ,
\end{equation}
with $s_k(\theta)$ the TDI output associated to a GW signal $h(f;\theta)$.
The inner product in \cref{eq:likelihood} is defined as
\begin{equation}\label{eq:inner-product}
    (a\mid b)_k = 2\int_{f_{\rm min}(\theta)}^{f_{\rm max}(\theta)} \frac{a_k(f)b^*_k(f) + a^*_k(f)b_k(f)}{S_k(f)}\,,
\end{equation}
where the integration boundaries $f_{\rm min}(\theta), f_{\rm max}(\theta)$ denote the lower and upper end of the injected signal with parameters $\theta$, spanned over the nominal LISA mission duration of $4 {\rm yr}$ (as in~\cref{eq:AnalyticalSNR}).
Our choice of parameterization is
aimed to (partially) reduce correlations between parameters, hence speeding up the likelihood surface exploration.
We set broad uniform priors on mildly constrained parameters (i.e.~$\delta \mu, \chi_1, \chi_2, \cos \iota, \phi_{\rm orb}, \psi$).
For the remaining parameters (i.e.~${\cal M}_c, d_L, \sin b, l$ and $f^0_{\rm orb})$ we choose priors small enough to keep the computational cost at a minimum, and sufficiently large to explore the posterior in the region dominated by the likelihood.
This is motivated by the assumption that, in a realistic data analysis
setup, a preliminary search for such signals will be available and provide rough initial guesses on the source parameters (see  e.g.~\cite{Bandopadhyay:2023gkb,Bandopadhyay:2024lwv}).
Posterior samples for each source considered are obtained through nested sampling as implemented in \textsc{Nessai}~\cite{Williams:2021qyt}, and are distributed according to
\begin{equation}\label{eq:posterior}
    p(\theta \mid d) = \frac{{\cal L}(d\mid \theta)\pi(\theta)}{{\cal Z}(d)}\,,
\end{equation}
where ${\cal Z}(d)$ is a normalization constant dependent only on the fixed data.

We select the sources on which to run parameter estimation as follows.
We first filter catalogue events with SNRs between 8 and 12, to warrant detectability,
and source-frame chirp masses between $20~\msun$ and $40~\msun$, to explore the chirp-mass range where the bulk of the \srcname{} population probability is (see~\cref{fig:LISAcorner}).
We further group the selected sources by three $\tau_c$ ranges,  $\tau_c \leq 5~{\rm yr}$, $5~{\rm yr} < \tau_c \leq 10~{\rm yr}$, and $10~{\rm yr} < \tau_c \leq 15~{\rm yr}$, respectively.
Though these are not the most representative values for $\tau_c$, as we expect most \srcnames{} to have higher ones (see~\cref{fig:tcdist}), we select \srcnames{} closer to merger because they have the highest SNR and better chances to be multiband.
For each $\tau_c$ class, we select two representative sources close to face-on or face-off ($|\cos \iota|\simeq 1$), and two close to edge-on ($|\cos \iota|\simeq 0$).
Among the sources available in our catalogues, we pick those that are low on the Ecliptic plane, with $\sin b<1/2$.

Overall, we select a set of 12 sources, whose injection parameters are listed in \cref{tab:injections}.
The SNR cut selects, among the catalogues, sources at very low redshift overall, 
with larger chirp masses allowing for larger redshifts at comparable SNRs.
The initial orbital frequency $f^0_{\rm orb}$ tends to increase for the low chirp mass sources, to maintain them drifting (see \cref{eq:fdot}) and thereby place them in the higher SNR region. 
On the other hand, the SNR cut clearly plays a minor role in constraining the source inclination (and all the remaining parameters, for which no sizable correlation with SNR is observed).

\begin{figure}[ht]
    \centering
    \includegraphics[width=\textwidth]{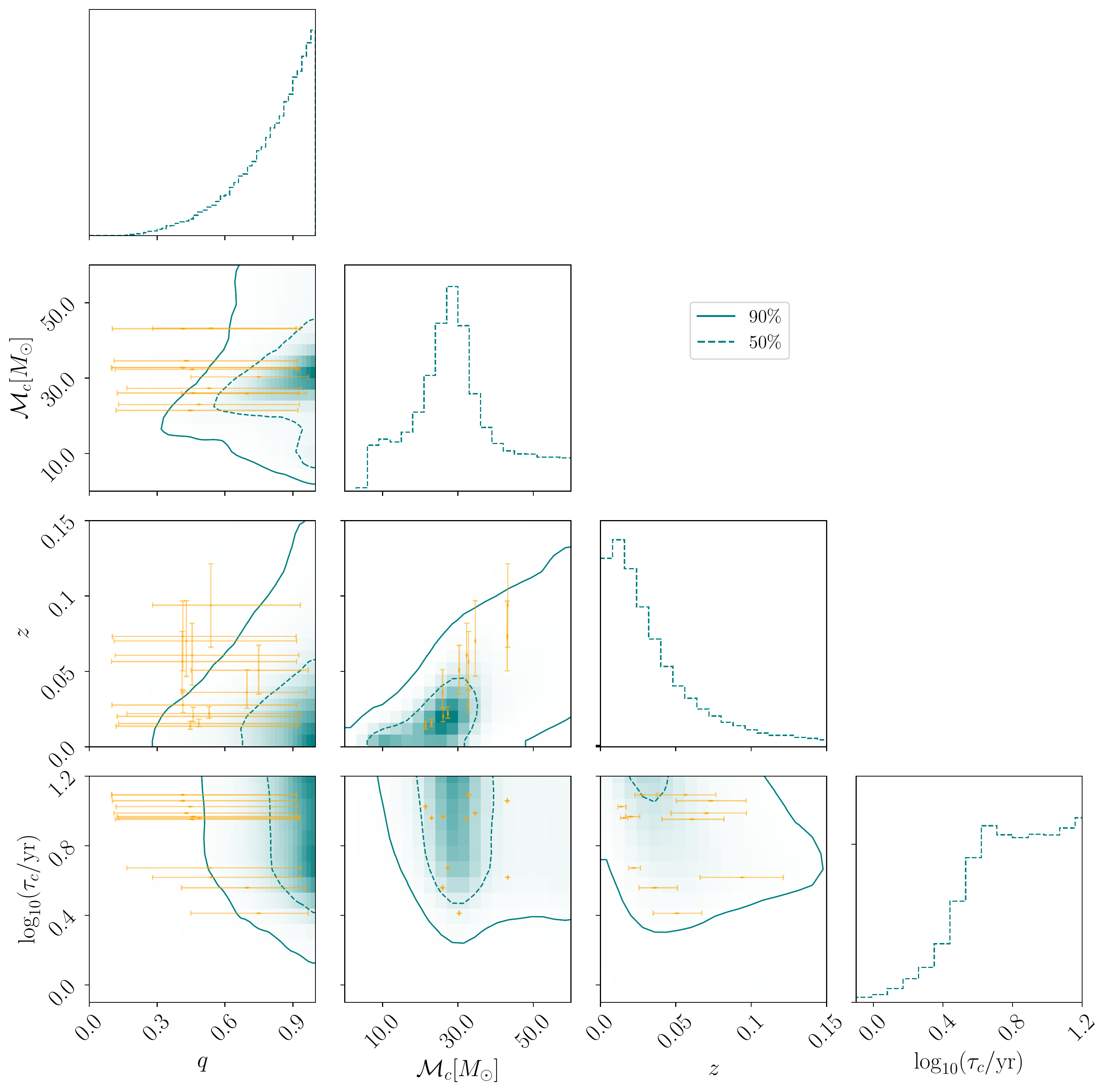}
    \caption{
    Results of the Bayesian parameter estimation on the
    set of 12 representative 
    sources, selected from the catalogues as
    presented in \cref{subsec:param-estim}. Yellow dots (error bars) denote the posterior median ($90\%$ confidence intervals) on each individual parameter. We focus on the best measured parameters, i.e.~the chirp mass, residual time-to-coalescence, redshift, and mass ratio 
    (see \cref{tab:pe-intrinsic} and \cref{tab:pe-extrinsic} for the individual parameters posterior median and uncertainties on the entire parameter set). The solid (dashed) blue lines represent the 90\% (50\%) contours enclosing 
    the expected distribution of detectable sources, as in \cref{fig:LISAcorner}. 
    The selected sources have smaller $q$ as compared the rest of the population, due to the SNR selection and chirp-mass cuts described in~\cref{subsec:param-estim}. The chirp mass and residual time-to-coalescence are measured very precisely, while the redshift posterior uncertainty increases with the median redshift. 
    }\label{fig:LISAcorner2}
\end{figure}

Results of individual parameter estimations are presented in~\cref{fig:LISAcorner2},~\cref{tab:pe-intrinsic} and~\cref{tab:pe-extrinsic}. 
As expected (see~\cite{Colpi:2024xhw}), LISA yields individual chirp-mass (initial orbital frequencies, residual time-to-coalescence) measurements with relative precision greater than 
$2\times 10^{-4}$ ($9\times 10^{-7}$, $4\times 10^{-4}$)
across all sources 
predicted in the catalogues,
and up to 
$2\times 10^{-5}$ ($3\times 10^{-7}$, $3\times10^{-5}$)
for those at less than $5\,{\rm yr}$ from merger.
Indeed, the intrinsic parameters of the latter are overall better determined, because of the larger drifting within the LISA band.
For sources with high and intermediate $\tau_c$, the relative errors on $\mathcal{M}_c$, $f^0_{\rm orb}$ and $\tau_c$ decrease with decreasing $\mathcal{M}_c$, due to the fact that they chirp more ($f^0_{\rm orb}$ increases as shown in \cref{tab:injections}). At low $\tau_c$ the tendency is inverted and errors decrease with increasing chirp mass (indeed, the trend in $f^0_{\rm orb}$ observed for high and intermediate $\tau_c$ sources, is no longer apparent).

On the other hand, the relative redshift errors vary between 20\% and 50\% across all sources, both with low and high $\tau_c$.
The inclination also does not influence the parameter estimation.
\Cref{fig:LISAcorner2} shows the selected source parameters with respect to the 90\% (and 50\%) of all the detectable \srcnames{} in the catalogues. 
Their chirp mass, redshift, and residual time-to-coalescence are broadly capturing the whole population distribution.
However, as a consequence of SNR selection and the chirp mass cuts described previously, they have lower mass ratios as compared to the rest of the population.
From this figure, one can appreciate the exquisite precision with which both the chirp mass and residual time-to-coalescence are determined, while the redshift is measured much better for close-by sources.

To explore the high-mass region of LISA sensitivity, we additionally
consider three equal-mass binary systems with source-frame chirp masses ${\cal M}_c =1000\, M_\odot$ and redshifts 0.1, 0.2, 0.8, respectively.
The chirp mass is explicitly chosen at the high-end of the LVK design sensitivity, just above the highest chirp-masses observable by LVK at design sensitivity, where LISA has the best chances to complement LVK detections (see~\cref{fig:LISALIGOwater}).
We choose arbitrarily to set these sources as face-on, non-spinning systems and place them at the Ecliptic north pole.
The three high-mass sources appear in the LISA band at much lower frequencies than the stellar-mass ones.
Being more massive, they have high SNR even though they are about a factor of 10 farther away in redshift (all 12 sources considered previously have redshifts lower than $z=0.1$).
The relative measurement precision of their chirp-mass (initial orbital frequencies, residual time-to-coalescence) is $7\times 10^{-4}$ ($5\times 10^{-6}$, $2\times 10^{-3}$) for the two closest sources, with detector-frame chirp mass of $1100\, \msun$ and $1200\,\msun$, and decreases to $5\times 10^{-6}$ ($9\times 10^{-7}$, $6\times 10^{-6}$) for the most distant source, with detector-frame ${\cal M}_c$ of 1800 solar masses. 
The increase in mass plays therefore a dominant role in the precision of LISA parameter estimation.
The relative error on the redshift also decreases from about 25\% -- a precision similar to that of some catalogue sources -- down to 7\% for the farthest one.

In summary, we find that
LISA grants a much better precision on the chirp-mass measurement than LVK (see e.g.~\cite{KAGRA:2021vkt}),
on the \srcnames{} that we expect it will detect.
This is true within the assumptions justifying the priors we have adopted
~\cite{Buscicchio:2021dph,Toubiana:2022vpp}.
The redshift, on the other hand, is determined with similar precision, but sources detectable by LVK extend up to higher redshifts, $z\lesssim 1$. 
The sub-per mille precision with which LISA can measure the residual time-to-coalescence of \srcnames{} with $\tau_c\leq 15$ yr 
opens up the possibility of multiband searches. 
While the extremely high precision in the chirp mass measurement is not expected to revolutionise our understanding of the bulk of the \srcname{} population observed so far, it has a great potential for signals at the low-frequency end of the LVK design sensitivity, with 
$\mathcal{M}_c \sim  10^3\, \msun$.
Indeed, if such a population exists, LISA will be able to observationally characterise it and distinguish it from \srcnames{} with lighter masses, i.e.
$\mathcal{M}_c\lesssim \mathcal{O}(10^2)\, M_\odot$.

\newcommand{\RunIMBHBlackSNR}{\ensuremath{90.12}}
\newcommand{\RunIMBHBlackBHBOneDimensionlessSpinOneZeroAbsWidth}{\ensuremath{1.6}}
\newcommand{\RunIMBHBlackBHBOneDimensionlessSpinTwoZeroAbsWidth}{\ensuremath{1.8}}
\newcommand{\RunIMBHBlackBHBOneEclipticLatitudeZeroAbsWidth}{\ensuremath{0.05}}
\newcommand{\RunIMBHBlackBHBOneEclipticLongitudeZeroAbsWidth}{\ensuremath{0.01}}
\newcommand{\RunIMBHBlackBHBOneInclinationZeroAbsWidth}{\ensuremath{0.5}}
\newcommand{\RunIMBHBlackBHBOneInitialOrbitalPhaseZeroAbsWidth}{\ensuremath{4.0}}
\newcommand{\RunIMBHBlackBHBOneLuminosityDistanceZeroAbsWidth}{\ensuremath{1012.8606}}
\newcommand{\RunIMBHBlackBHBOneMassRatioZeroAbsWidth}{\ensuremath{0.09}}
\newcommand{\RunIMBHBlackBHBOneMergerTimeOrInitialOrbitalFrequencyZeroAbsWidth}{\ensuremath{6\!\times\!10^{-7}}}
\newcommand{\RunIMBHBlackBHBOneMergerTimeZeroAbsWidth}{\ensuremath{2\!\times\!10^{-5}}}
\newcommand{\RunIMBHBlackBHBOnePolarizationZeroAbsWidth}{\ensuremath{4.0}}
\newcommand{\RunIMBHBlackBHBOneRedshiftZeroAbsWidth}{\ensuremath{0.1}}
\newcommand{\RunIMBHBlackBHBOneRedshiftedChirpMassZeroAbsWidth}{\ensuremath{0.009}}
\newcommand{\RunIMBHBlackBHBOneRedshiftedMassOneZeroAbsWidth}{\ensuremath{205.246}}
\newcommand{\RunIMBHBlackBHBOneRedshiftedMassTwoZeroAbsWidth}{\ensuremath{181.331}}
\newcommand{\RunIMBHBlackBHBOneReducedMassDifferenceZeroAbsWidth}{\ensuremath{0.09}}
\newcommand{\RunIMBHBlackBHBOnecosInclinationZeroAbsWidth}{\ensuremath{0.2}}
\newcommand{\RunIMBHBlackBHBOnesinEclipticLatitudeZeroAbsWidth}{\ensuremath{0.05}}
\newcommand{\RunIMBHBlackBHBOneDimensionlessSpinOneZeroInj}{\ensuremath{0.0}}
\newcommand{\RunIMBHBlackBHBOneDimensionlessSpinTwoZeroInj}{\ensuremath{0.0}}
\newcommand{\RunIMBHBlackBHBOneEclipticLatitudeZeroInj}{\ensuremath{0.0}}
\newcommand{\RunIMBHBlackBHBOneEclipticLongitudeZeroInj}{\ensuremath{3.1}}
\newcommand{\RunIMBHBlackBHBOneInclinationZeroInj}{\ensuremath{1.6}}
\newcommand{\RunIMBHBlackBHBOneInitialOrbitalPhaseZeroInj}{\ensuremath{1}}
\newcommand{\RunIMBHBlackBHBOneLuminosityDistanceZeroInj}{\ensuremath{5162.1658}}
\newcommand{\RunIMBHBlackBHBOneMassRatioZeroInj}{\ensuremath{1.0}}
\newcommand{\RunIMBHBlackBHBOneMergerTimeOrInitialOrbitalFrequencyZeroInj}{\ensuremath{0.7}}
\newcommand{\RunIMBHBlackBHBOneMergerTimeZeroInj}{\ensuremath{3.0}}
\newcommand{\RunIMBHBlackBHBOnePolarizationZeroInj}{\ensuremath{0.0}}
\newcommand{\RunIMBHBlackBHBOneRedshiftZeroInj}{\ensuremath{0.8}}
\newcommand{\RunIMBHBlackBHBOneRedshiftedChirpMassZeroInj}{\ensuremath{1800.0}}
\newcommand{\RunIMBHBlackBHBOneRedshiftedMassOneZeroInj}{\ensuremath{2067.8638}}
\newcommand{\RunIMBHBlackBHBOneRedshiftedMassTwoZeroInj}{\ensuremath{2067.4503}}
\newcommand{\RunIMBHBlackBHBOneReducedMassDifferenceZeroInj}{\ensuremath{1\!\times\!10^{-4}}}
\newcommand{\RunIMBHBlackBHBOnecosInclinationZeroInj}{\ensuremath{1}}
\newcommand{\RunIMBHBlackBHBOnesinEclipticLatitudeZeroInj}{\ensuremath{0.0}}
\newcommand{\RunIMBHBlackBHBOneDimensionlessSpinOneZeroPost}{\ensuremath{0.05^{+0.75}_{-0.84}}}
\newcommand{\RunIMBHBlackBHBOneDimensionlessSpinTwoZeroPost}{\ensuremath{-0.05^{+0.93}_{-0.85}}}
\newcommand{\RunIMBHBlackBHBOneEclipticLatitudeZeroPost}{\ensuremath{-0.0009^{+0.0248}_{-0.0254}}}
\newcommand{\RunIMBHBlackBHBOneEclipticLongitudeZeroPost}{\ensuremath{3.139^{+0.006}_{-0.006}}}
\newcommand{\RunIMBHBlackBHBOneInclinationZeroPost}{\ensuremath{0.4^{+0.2}_{-0.3}}}
\newcommand{\RunIMBHBlackBHBOneInitialOrbitalPhaseZeroPost}{\ensuremath{2.9^{+2.0}_{-2.0}}}
\newcommand{\RunIMBHBlackBHBOneLuminosityDistanceZeroPost}{\ensuremath{4684^{+428}_{-584}}}
\newcommand{\RunIMBHBlackBHBOneMassRatioZeroPost}{\ensuremath{0.4^{+0.6}_{-0.4}}}
\newcommand{\RunIMBHBlackBHBOneMergerTimeOrInitialOrbitalFrequencyZeroPost}{\ensuremath{0.7119232^{+0.0000003}_{-0.0000003}}}
\newcommand{\RunIMBHBlackBHBOneMergerTimeZeroPost}{\ensuremath{2.999996^{+0.000004}_{-0.000015}}}
\newcommand{\RunIMBHBlackBHBOnePolarizationZeroPost}{\ensuremath{-0.3^{+2.0}_{-2.0}}}
\newcommand{\RunIMBHBlackBHBOneRedshiftZeroPost}{\ensuremath{0.74^{+0.05}_{-0.08}}}
\newcommand{\RunIMBHBlackBHBOneRedshiftedChirpMassZeroPost}{\ensuremath{1800.0018^{+0.0064}_{-0.0027}}}
\newcommand{\RunIMBHBlackBHBOneRedshiftedMassOneZeroPost}{\ensuremath{2157^{+126}_{-79}}}
\newcommand{\RunIMBHBlackBHBOneRedshiftedMassTwoZeroPost}{\ensuremath{1982^{+75}_{-110}}}
\newcommand{\RunIMBHBlackBHBOneReducedMassDifferenceZeroPost}{\ensuremath{0.04^{+0.06}_{-0.04}}}
\newcommand{\RunIMBHBlackBHBOnecosInclinationZeroPost}{\ensuremath{0.90^{+0.08}_{-0.12}}}
\newcommand{\RunIMBHBlackBHBOnesinEclipticLatitudeZeroPost}{\ensuremath{-0.0009^{+0.0248}_{-0.0254}}}
\newcommand{\RunIMBHBlackBHBOneEclipticLongitudeZeroWidth}{\ensuremath{0.004}}
\newcommand{\RunIMBHBlackBHBOneInclinationZeroWidth}{\ensuremath{0.3}}
\newcommand{\RunIMBHBlackBHBOneInitialOrbitalPhaseZeroWidth}{\ensuremath{4.0}}
\newcommand{\RunIMBHBlackBHBOneLuminosityDistanceZeroWidth}{\ensuremath{0.2}}
\newcommand{\RunIMBHBlackBHBOneMassRatioZeroWidth}{\ensuremath{0.09}}
\newcommand{\RunIMBHBlackBHBOneMergerTimeOrInitialOrbitalFrequencyZeroWidth}{\ensuremath{9\!\times\!10^{-7}}}
\newcommand{\RunIMBHBlackBHBOneMergerTimeZeroWidth}{\ensuremath{6\!\times\!10^{-6}}}
\newcommand{\RunIMBHBlackBHBOneRedshiftZeroWidth}{\ensuremath{0.2}}
\newcommand{\RunIMBHBlackBHBOneRedshiftedChirpMassZeroWidth}{\ensuremath{5\!\times\!10^{-6}}}
\newcommand{\RunIMBHBlackBHBOneRedshiftedMassOneZeroWidth}{\ensuremath{0.10}}
\newcommand{\RunIMBHBlackBHBOneRedshiftedMassTwoZeroWidth}{\ensuremath{0.09}}
\newcommand{\RunIMBHBlackBHBOneReducedMassDifferenceZeroWidth}{\ensuremath{929.147}}
\newcommand{\RunIMBHBlackBHBOnecosInclinationZeroWidth}{\ensuremath{0.2}}
\newcommand{\RunHiTauBadIncOneSNR}{\ensuremath{10.04}}
\newcommand{\RunHiTauBadIncOneBHBOneDimensionlessSpinOneZeroInj}{\ensuremath{0.5}}
\newcommand{\RunHiTauBadIncOneBHBOneDimensionlessSpinTwoZeroInj}{\ensuremath{0.2}}
\newcommand{\RunHiTauBadIncOneBHBOneEclipticLatitudeZeroInj}{\ensuremath{0.01}}
\newcommand{\RunHiTauBadIncOneBHBOneEclipticLongitudeZeroInj}{\ensuremath{4.8}}
\newcommand{\RunHiTauBadIncOneBHBOneInclinationZeroInj}{\ensuremath{1.4}}
\newcommand{\RunHiTauBadIncOneBHBOneInitialOrbitalPhaseZeroInj}{\ensuremath{0.8}}
\newcommand{\RunHiTauBadIncOneBHBOneLuminosityDistanceZeroInj}{\ensuremath{113.970}}
\newcommand{\RunHiTauBadIncOneBHBOneMassRatioZeroInj}{\ensuremath{0.8}}
\newcommand{\RunHiTauBadIncOneBHBOneMergerTimeOrInitialOrbitalFrequencyZeroInj}{\ensuremath{5.1}}
\newcommand{\RunHiTauBadIncOneBHBOneMergerTimeZeroInj}{\ensuremath{12.37}}
\newcommand{\RunHiTauBadIncOneBHBOnePolarizationZeroInj}{\ensuremath{3.7}}
\newcommand{\RunHiTauBadIncOneBHBOneRedshiftZeroInj}{\ensuremath{0.03}}
\newcommand{\RunHiTauBadIncOneBHBOneRedshiftedChirpMassZeroInj}{\ensuremath{32.71}}
\newcommand{\RunHiTauBadIncOneBHBOneRedshiftedMassOneZeroInj}{\ensuremath{41.52}}
\newcommand{\RunHiTauBadIncOneBHBOneRedshiftedMassTwoZeroInj}{\ensuremath{34.08}}
\newcommand{\RunHiTauBadIncOneBHBOneReducedMassDifferenceZeroInj}{\ensuremath{0.10}}
\newcommand{\RunHiTauBadIncOneBHBOnecosInclinationZeroInj}{\ensuremath{0.2}}
\newcommand{\RunHiTauBadIncOneBHBOnesinEclipticLatitudeZeroInj}{\ensuremath{0.01}}
\newcommand{\RunHiTauBadIncOneBHBOneDimensionlessSpinOneZeroPost}{\ensuremath{-0.05^{+0.93}_{-0.84}}}
\newcommand{\RunHiTauBadIncOneBHBOneDimensionlessSpinTwoZeroPost}{\ensuremath{-0.004^{+0.90}_{-0.89}}}
\newcommand{\RunHiTauBadIncOneBHBOneEclipticLatitudeZeroPost}{\ensuremath{0.004^{+0.089}_{-0.093}}}
\newcommand{\RunHiTauBadIncOneBHBOneEclipticLongitudeZeroPost}{\ensuremath{4.850^{+0.009}_{-0.009}}}
\newcommand{\RunHiTauBadIncOneBHBOneInclinationZeroPost}{\ensuremath{1.4^{+0.1}_{-0.2}}}
\newcommand{\RunHiTauBadIncOneBHBOneInitialOrbitalPhaseZeroPost}{\ensuremath{2.4^{+1.8}_{-1.7}}}
\newcommand{\RunHiTauBadIncOneBHBOneLuminosityDistanceZeroPost}{\ensuremath{125^{+44}_{-22}}}
\newcommand{\RunHiTauBadIncOneBHBOneMassRatioZeroPost}{\ensuremath{0.4^{+0.5}_{-0.3}}}
\newcommand{\RunHiTauBadIncOneBHBOneMergerTimeOrInitialOrbitalFrequencyZeroPost}{\ensuremath{5.123816^{+0.000002}_{-0.000002}}}
\newcommand{\RunHiTauBadIncOneBHBOneMergerTimeZeroPost}{\ensuremath{12.366^{+0.001}_{-0.003}}}
\newcommand{\RunHiTauBadIncOneBHBOnePolarizationZeroPost}{\ensuremath{0.5^{+1.6}_{-1.7}}}
\newcommand{\RunHiTauBadIncOneBHBOneRedshiftZeroPost}{\ensuremath{0.028^{+0.009}_{-0.005}}}
\newcommand{\RunHiTauBadIncOneBHBOneRedshiftedChirpMassZeroPost}{\ensuremath{32.715^{+0.005}_{-0.002}}}
\newcommand{\RunHiTauBadIncOneBHBOneRedshiftedMassOneZeroPost}{\ensuremath{60^{+73}_{-20}}}
\newcommand{\RunHiTauBadIncOneBHBOneRedshiftedMassTwoZeroPost}{\ensuremath{25^{+11}_{-11}}}
\newcommand{\RunHiTauBadIncOneBHBOneReducedMassDifferenceZeroPost}{\ensuremath{0.4^{+0.4}_{-0.4}}}
\newcommand{\RunHiTauBadIncOneBHBOnecosInclinationZeroPost}{\ensuremath{0.2^{+0.2}_{-0.1}}}
\newcommand{\RunHiTauBadIncOneBHBOnesinEclipticLatitudeZeroPost}{\ensuremath{0.004^{+0.089}_{-0.093}}}
\newcommand{\RunHiTauBadIncOneBHBOneDimensionlessSpinOneZeroWidth}{\ensuremath{3.4}}
\newcommand{\RunHiTauBadIncOneBHBOneDimensionlessSpinTwoZeroWidth}{\ensuremath{8.1}}
\newcommand{\RunHiTauBadIncOneBHBOneEclipticLatitudeZeroWidth}{\ensuremath{12.38}}
\newcommand{\RunHiTauBadIncOneBHBOneEclipticLongitudeZeroWidth}{\ensuremath{0.004}}
\newcommand{\RunHiTauBadIncOneBHBOneInclinationZeroWidth}{\ensuremath{0.2}}
\newcommand{\RunHiTauBadIncOneBHBOneInitialOrbitalPhaseZeroWidth}{\ensuremath{4.5}}
\newcommand{\RunHiTauBadIncOneBHBOneLuminosityDistanceZeroWidth}{\ensuremath{0.6}}
\newcommand{\RunHiTauBadIncOneBHBOneMassRatioZeroWidth}{\ensuremath{1.0}}
\newcommand{\RunHiTauBadIncOneBHBOneMergerTimeOrInitialOrbitalFrequencyZeroWidth}{\ensuremath{7\!\times\!10^{-7}}}
\newcommand{\RunHiTauBadIncOneBHBOneMergerTimeZeroWidth}{\ensuremath{4\!\times\!10^{-4}}}
\newcommand{\RunHiTauBadIncOneBHBOnePolarizationZeroWidth}{\ensuremath{0.9}}
\newcommand{\RunHiTauBadIncOneBHBOneRedshiftZeroWidth}{\ensuremath{0.6}}
\newcommand{\RunHiTauBadIncOneBHBOneRedshiftedChirpMassZeroWidth}{\ensuremath{2\!\times\!10^{-4}}}
\newcommand{\RunHiTauBadIncOneBHBOneRedshiftedMassOneZeroWidth}{\ensuremath{2.3}}
\newcommand{\RunHiTauBadIncOneBHBOneRedshiftedMassTwoZeroWidth}{\ensuremath{0.7}}
\newcommand{\RunHiTauBadIncOneBHBOneReducedMassDifferenceZeroWidth}{\ensuremath{7.9}}
\newcommand{\RunHiTauBadIncOneBHBOnecosInclinationZeroWidth}{\ensuremath{1.6}}
\newcommand{\RunHiTauBadIncOneBHBOnesinEclipticLatitudeZeroWidth}{\ensuremath{12.36}}
\newcommand{\RunHiTauBadIncTwoSNR}{\ensuremath{12.08}}
\newcommand{\RunHiTauBadIncTwoBHBOneDimensionlessSpinOneZeroInj}{\ensuremath{0.4}}
\newcommand{\RunHiTauBadIncTwoBHBOneDimensionlessSpinTwoZeroInj}{\ensuremath{0.3}}
\newcommand{\RunHiTauBadIncTwoBHBOneEclipticLatitudeZeroInj}{\ensuremath{0.06}}
\newcommand{\RunHiTauBadIncTwoBHBOneEclipticLongitudeZeroInj}{\ensuremath{2.9}}
\newcommand{\RunHiTauBadIncTwoBHBOneInclinationZeroInj}{\ensuremath{1.4}}
\newcommand{\RunHiTauBadIncTwoBHBOneInitialOrbitalPhaseZeroInj}{\ensuremath{3.6}}
\newcommand{\RunHiTauBadIncTwoBHBOneLuminosityDistanceZeroInj}{\ensuremath{57.04}}
\newcommand{\RunHiTauBadIncTwoBHBOneMassRatioZeroInj}{\ensuremath{0.6}}
\newcommand{\RunHiTauBadIncTwoBHBOneMergerTimeOrInitialOrbitalFrequencyZeroInj}{\ensuremath{7.1}}
\newcommand{\RunHiTauBadIncTwoBHBOneMergerTimeZeroInj}{\ensuremath{10.59}}
\newcommand{\RunHiTauBadIncTwoBHBOnePolarizationZeroInj}{\ensuremath{0.4}}
\newcommand{\RunHiTauBadIncTwoBHBOneRedshiftZeroInj}{\ensuremath{0.01}}
\newcommand{\RunHiTauBadIncTwoBHBOneRedshiftedChirpMassZeroInj}{\ensuremath{21.42}}
\newcommand{\RunHiTauBadIncTwoBHBOneRedshiftedMassOneZeroInj}{\ensuremath{31.62}}
\newcommand{\RunHiTauBadIncTwoBHBOneRedshiftedMassTwoZeroInj}{\ensuremath{19.37}}
\newcommand{\RunHiTauBadIncTwoBHBOneReducedMassDifferenceZeroInj}{\ensuremath{0.2}}
\newcommand{\RunHiTauBadIncTwoBHBOnecosInclinationZeroInj}{\ensuremath{0.2}}
\newcommand{\RunHiTauBadIncTwoBHBOnesinEclipticLatitudeZeroInj}{\ensuremath{0.06}}
\newcommand{\RunHiTauBadIncTwoBHBOneDimensionlessSpinOneZeroPost}{\ensuremath{-0.02^{+0.89}_{-0.84}}}
\newcommand{\RunHiTauBadIncTwoBHBOneDimensionlessSpinTwoZeroPost}{\ensuremath{0.05^{+0.86}_{-0.93}}}
\newcommand{\RunHiTauBadIncTwoBHBOneEclipticLatitudeZeroPost}{\ensuremath{0.04^{+0.06}_{-0.11}}}
\newcommand{\RunHiTauBadIncTwoBHBOneEclipticLongitudeZeroPost}{\ensuremath{2.879^{+0.005}_{-0.005}}}
\newcommand{\RunHiTauBadIncTwoBHBOneInclinationZeroPost}{\ensuremath{1.34^{+0.09}_{-0.13}}}
\newcommand{\RunHiTauBadIncTwoBHBOneInitialOrbitalPhaseZeroPost}{\ensuremath{2.2^{+2.8}_{-1.7}}}
\newcommand{\RunHiTauBadIncTwoBHBOneLuminosityDistanceZeroPost}{\ensuremath{61^{+15}_{-9}}}
\newcommand{\RunHiTauBadIncTwoBHBOneMassRatioZeroPost}{\ensuremath{0.4^{+0.5}_{-0.3}}}
\newcommand{\RunHiTauBadIncTwoBHBOneMergerTimeOrInitialOrbitalFrequencyZeroPost}{\ensuremath{7.077528^{+0.000002}_{-0.000002}}}
\newcommand{\RunHiTauBadIncTwoBHBOneMergerTimeZeroPost}{\ensuremath{10.5881^{+0.0008}_{-0.0017}}}
\newcommand{\RunHiTauBadIncTwoBHBOnePolarizationZeroPost}{\ensuremath{0.3^{+1.6}_{-3.0}}}
\newcommand{\RunHiTauBadIncTwoBHBOneRedshiftZeroPost}{\ensuremath{0.014^{+0.003}_{-0.002}}}
\newcommand{\RunHiTauBadIncTwoBHBOneRedshiftedChirpMassZeroPost}{\ensuremath{21.416^{+0.002}_{-0.001}}}
\newcommand{\RunHiTauBadIncTwoBHBOneRedshiftedMassOneZeroPost}{\ensuremath{37^{+41}_{-12}}}
\newcommand{\RunHiTauBadIncTwoBHBOneRedshiftedMassTwoZeroPost}{\ensuremath{17^{+7}_{-7}}}
\newcommand{\RunHiTauBadIncTwoBHBOneReducedMassDifferenceZeroPost}{\ensuremath{0.4^{+0.4}_{-0.3}}}
\newcommand{\RunHiTauBadIncTwoBHBOnecosInclinationZeroPost}{\ensuremath{0.23^{+0.13}_{-0.09}}}
\newcommand{\RunHiTauBadIncTwoBHBOnesinEclipticLatitudeZeroPost}{\ensuremath{0.04^{+0.06}_{-0.11}}}
\newcommand{\RunHiTauBadIncTwoBHBOneDimensionlessSpinOneZeroWidth}{\ensuremath{4.7}}
\newcommand{\RunHiTauBadIncTwoBHBOneDimensionlessSpinTwoZeroWidth}{\ensuremath{5.3}}
\newcommand{\RunHiTauBadIncTwoBHBOneEclipticLatitudeZeroWidth}{\ensuremath{2.6}}
\newcommand{\RunHiTauBadIncTwoBHBOneEclipticLongitudeZeroWidth}{\ensuremath{0.004}}
\newcommand{\RunHiTauBadIncTwoBHBOneInclinationZeroWidth}{\ensuremath{0.2}}
\newcommand{\RunHiTauBadIncTwoBHBOneInitialOrbitalPhaseZeroWidth}{\ensuremath{1.3}}
\newcommand{\RunHiTauBadIncTwoBHBOneLuminosityDistanceZeroWidth}{\ensuremath{0.4}}
\newcommand{\RunHiTauBadIncTwoBHBOneMassRatioZeroWidth}{\ensuremath{1.3}}
\newcommand{\RunHiTauBadIncTwoBHBOneMergerTimeOrInitialOrbitalFrequencyZeroWidth}{\ensuremath{5\!\times\!10^{-7}}}
\newcommand{\RunHiTauBadIncTwoBHBOneMergerTimeZeroWidth}{\ensuremath{2\!\times\!10^{-4}}}
\newcommand{\RunHiTauBadIncTwoBHBOnePolarizationZeroWidth}{\ensuremath{12.49}}
\newcommand{\RunHiTauBadIncTwoBHBOneRedshiftZeroWidth}{\ensuremath{0.4}}
\newcommand{\RunHiTauBadIncTwoBHBOneRedshiftedChirpMassZeroWidth}{\ensuremath{1\!\times\!10^{-4}}}
\newcommand{\RunHiTauBadIncTwoBHBOneRedshiftedMassOneZeroWidth}{\ensuremath{1.7}}
\newcommand{\RunHiTauBadIncTwoBHBOneRedshiftedMassTwoZeroWidth}{\ensuremath{0.7}}
\newcommand{\RunHiTauBadIncTwoBHBOneReducedMassDifferenceZeroWidth}{\ensuremath{3.1}}
\newcommand{\RunHiTauBadIncTwoBHBOnecosInclinationZeroWidth}{\ensuremath{1.0}}
\newcommand{\RunHiTauBadIncTwoBHBOnesinEclipticLatitudeZeroWidth}{\ensuremath{2.6}}
\newcommand{\RunHiTauGoodIncOneSNR}{\ensuremath{10.01}}
\newcommand{\RunHiTauGoodIncOneBHBOneDimensionlessSpinOneZeroInj}{\ensuremath{0.4}}
\newcommand{\RunHiTauGoodIncOneBHBOneDimensionlessSpinTwoZeroInj}{\ensuremath{0.3}}
\newcommand{\RunHiTauGoodIncOneBHBOneEclipticLatitudeZeroInj}{\ensuremath{0.4}}
\newcommand{\RunHiTauGoodIncOneBHBOneEclipticLongitudeZeroInj}{\ensuremath{4.1}}
\newcommand{\RunHiTauGoodIncOneBHBOneInclinationZeroInj}{\ensuremath{0.3}}
\newcommand{\RunHiTauGoodIncOneBHBOneInitialOrbitalPhaseZeroInj}{\ensuremath{0.4}}
\newcommand{\RunHiTauGoodIncOneBHBOneLuminosityDistanceZeroInj}{\ensuremath{392.599}}
\newcommand{\RunHiTauGoodIncOneBHBOneMassRatioZeroInj}{\ensuremath{0.9}}
\newcommand{\RunHiTauGoodIncOneBHBOneMergerTimeOrInitialOrbitalFrequencyZeroInj}{\ensuremath{4.4}}
\newcommand{\RunHiTauGoodIncOneBHBOneMergerTimeZeroInj}{\ensuremath{11.44}}
\newcommand{\RunHiTauGoodIncOneBHBOnePolarizationZeroInj}{\ensuremath{3.5}}
\newcommand{\RunHiTauGoodIncOneBHBOneRedshiftZeroInj}{\ensuremath{0.08}}
\newcommand{\RunHiTauGoodIncOneBHBOneRedshiftedChirpMassZeroInj}{\ensuremath{43.10}}
\newcommand{\RunHiTauGoodIncOneBHBOneRedshiftedMassOneZeroInj}{\ensuremath{53.48}}
\newcommand{\RunHiTauGoodIncOneBHBOneRedshiftedMassTwoZeroInj}{\ensuremath{45.87}}
\newcommand{\RunHiTauGoodIncOneBHBOneReducedMassDifferenceZeroInj}{\ensuremath{0.08}}
\newcommand{\RunHiTauGoodIncOneBHBOnecosInclinationZeroInj}{\ensuremath{1.0}}
\newcommand{\RunHiTauGoodIncOneBHBOnesinEclipticLatitudeZeroInj}{\ensuremath{0.4}}
\newcommand{\RunHiTauGoodIncOneBHBOneDimensionlessSpinOneZeroPost}{\ensuremath{-0.03^{+0.91}_{-0.85}}}
\newcommand{\RunHiTauGoodIncOneBHBOneDimensionlessSpinTwoZeroPost}{\ensuremath{0.007^{+0.89}_{-0.90}}}
\newcommand{\RunHiTauGoodIncOneBHBOneEclipticLatitudeZeroPost}{\ensuremath{0.39^{+0.02}_{-0.02}}}
\newcommand{\RunHiTauGoodIncOneBHBOneEclipticLongitudeZeroPost}{\ensuremath{4.15^{+0.01}_{-0.01}}}
\newcommand{\RunHiTauGoodIncOneBHBOneInclinationZeroPost}{\ensuremath{0.7^{+0.4}_{-0.5}}}
\newcommand{\RunHiTauGoodIncOneBHBOneInitialOrbitalPhaseZeroPost}{\ensuremath{2.7^{+2.0}_{-2.0}}}
\newcommand{\RunHiTauGoodIncOneBHBOneLuminosityDistanceZeroPost}{\ensuremath{341^{+116}_{-110}}}
\newcommand{\RunHiTauGoodIncOneBHBOneMassRatioZeroPost}{\ensuremath{0.4^{+0.5}_{-0.3}}}
\newcommand{\RunHiTauGoodIncOneBHBOneMergerTimeOrInitialOrbitalFrequencyZeroPost}{\ensuremath{4.441422^{+0.000002}_{-0.000002}}}
\newcommand{\RunHiTauGoodIncOneBHBOneMergerTimeZeroPost}{\ensuremath{11.436^{+0.001}_{-0.003}}}
\newcommand{\RunHiTauGoodIncOneBHBOnePolarizationZeroPost}{\ensuremath{-0.4^{+2.0}_{-2.0}}}
\newcommand{\RunHiTauGoodIncOneBHBOneRedshiftZeroPost}{\ensuremath{0.07^{+0.02}_{-0.02}}}
\newcommand{\RunHiTauGoodIncOneBHBOneRedshiftedChirpMassZeroPost}{\ensuremath{43.095^{+0.007}_{-0.003}}}
\newcommand{\RunHiTauGoodIncOneBHBOneRedshiftedMassOneZeroPost}{\ensuremath{78^{+94}_{-27}}}
\newcommand{\RunHiTauGoodIncOneBHBOneRedshiftedMassTwoZeroPost}{\ensuremath{32^{+15}_{-15}}}
\newcommand{\RunHiTauGoodIncOneBHBOneReducedMassDifferenceZeroPost}{\ensuremath{0.4^{+0.4}_{-0.4}}}
\newcommand{\RunHiTauGoodIncOneBHBOnecosInclinationZeroPost}{\ensuremath{0.8^{+0.2}_{-0.3}}}
\newcommand{\RunHiTauGoodIncOneBHBOnesinEclipticLatitudeZeroPost}{\ensuremath{0.38^{+0.02}_{-0.02}}}
\newcommand{\RunHiTauGoodIncOneBHBOneDimensionlessSpinOneZeroWidth}{\ensuremath{4.5}}
\newcommand{\RunHiTauGoodIncOneBHBOneDimensionlessSpinTwoZeroWidth}{\ensuremath{6.4}}
\newcommand{\RunHiTauGoodIncOneBHBOneEclipticLatitudeZeroWidth}{\ensuremath{0.1}}
\newcommand{\RunHiTauGoodIncOneBHBOneEclipticLongitudeZeroWidth}{\ensuremath{0.005}}
\newcommand{\RunHiTauGoodIncOneBHBOneInclinationZeroWidth}{\ensuremath{2.9}}
\newcommand{\RunHiTauGoodIncOneBHBOneInitialOrbitalPhaseZeroWidth}{\ensuremath{10.38}}
\newcommand{\RunHiTauGoodIncOneBHBOneLuminosityDistanceZeroWidth}{\ensuremath{0.6}}
\newcommand{\RunHiTauGoodIncOneBHBOneMassRatioZeroWidth}{\ensuremath{0.9}}
\newcommand{\RunHiTauGoodIncOneBHBOneMergerTimeOrInitialOrbitalFrequencyZeroWidth}{\ensuremath{7\!\times\!10^{-7}}}
\newcommand{\RunHiTauGoodIncOneBHBOneMergerTimeZeroWidth}{\ensuremath{4\!\times\!10^{-4}}}
\newcommand{\RunHiTauGoodIncOneBHBOnePolarizationZeroWidth}{\ensuremath{1.2}}
\newcommand{\RunHiTauGoodIncOneBHBOneRedshiftZeroWidth}{\ensuremath{0.6}}
\newcommand{\RunHiTauGoodIncOneBHBOneRedshiftedChirpMassZeroWidth}{\ensuremath{2\!\times\!10^{-4}}}
\newcommand{\RunHiTauGoodIncOneBHBOneRedshiftedMassOneZeroWidth}{\ensuremath{2.3}}
\newcommand{\RunHiTauGoodIncOneBHBOneRedshiftedMassTwoZeroWidth}{\ensuremath{0.6}}
\newcommand{\RunHiTauGoodIncOneBHBOneReducedMassDifferenceZeroWidth}{\ensuremath{10.06}}
\newcommand{\RunHiTauGoodIncOneBHBOnecosInclinationZeroWidth}{\ensuremath{0.5}}
\newcommand{\RunHiTauGoodIncOneBHBOnesinEclipticLatitudeZeroWidth}{\ensuremath{0.10}}
\newcommand{\RunHiTauGoodIncTwoSNR}{\ensuremath{9.18}}
\newcommand{\RunHiTauGoodIncTwoBHBOneDimensionlessSpinOneZeroInj}{\ensuremath{0.4}}
\newcommand{\RunHiTauGoodIncTwoBHBOneDimensionlessSpinTwoZeroInj}{\ensuremath{0.1}}
\newcommand{\RunHiTauGoodIncTwoBHBOneEclipticLatitudeZeroInj}{\ensuremath{0.2}}
\newcommand{\RunHiTauGoodIncTwoBHBOneEclipticLongitudeZeroInj}{\ensuremath{3.3}}
\newcommand{\RunHiTauGoodIncTwoBHBOneInclinationZeroInj}{\ensuremath{2.8}}
\newcommand{\RunHiTauGoodIncTwoBHBOneInitialOrbitalPhaseZeroInj}{\ensuremath{0.7}}
\newcommand{\RunHiTauGoodIncTwoBHBOneLuminosityDistanceZeroInj}{\ensuremath{294.343}}
\newcommand{\RunHiTauGoodIncTwoBHBOneMassRatioZeroInj}{\ensuremath{0.8}}
\newcommand{\RunHiTauGoodIncTwoBHBOneMergerTimeOrInitialOrbitalFrequencyZeroInj}{\ensuremath{5.1}}
\newcommand{\RunHiTauGoodIncTwoBHBOneMergerTimeZeroInj}{\ensuremath{12.38}}
\newcommand{\RunHiTauGoodIncTwoBHBOnePolarizationZeroInj}{\ensuremath{2.7}}
\newcommand{\RunHiTauGoodIncTwoBHBOneRedshiftZeroInj}{\ensuremath{0.06}}
\newcommand{\RunHiTauGoodIncTwoBHBOneRedshiftedChirpMassZeroInj}{\ensuremath{32.81}}
\newcommand{\RunHiTauGoodIncTwoBHBOneRedshiftedMassOneZeroInj}{\ensuremath{42.14}}
\newcommand{\RunHiTauGoodIncTwoBHBOneRedshiftedMassTwoZeroInj}{\ensuremath{33.78}}
\newcommand{\RunHiTauGoodIncTwoBHBOneReducedMassDifferenceZeroInj}{\ensuremath{0.1}}
\newcommand{\RunHiTauGoodIncTwoBHBOnecosInclinationZeroInj}{\ensuremath{-0.9}}
\newcommand{\RunHiTauGoodIncTwoBHBOnesinEclipticLatitudeZeroInj}{\ensuremath{0.2}}
\newcommand{\RunHiTauGoodIncTwoBHBOneDimensionlessSpinOneZeroPost}{\ensuremath{-0.04^{+0.92}_{-0.85}}}
\newcommand{\RunHiTauGoodIncTwoBHBOneDimensionlessSpinTwoZeroPost}{\ensuremath{0.004^{+0.89}_{-0.90}}}
\newcommand{\RunHiTauGoodIncTwoBHBOneEclipticLatitudeZeroPost}{\ensuremath{0.23^{+0.03}_{-0.03}}}
\newcommand{\RunHiTauGoodIncTwoBHBOneEclipticLongitudeZeroPost}{\ensuremath{3.30^{+0.01}_{-0.01}}}
\newcommand{\RunHiTauGoodIncTwoBHBOneInclinationZeroPost}{\ensuremath{2.5^{+0.5}_{-0.4}}}
\newcommand{\RunHiTauGoodIncTwoBHBOneInitialOrbitalPhaseZeroPost}{\ensuremath{2.9^{+2.0}_{-2.0}}}
\newcommand{\RunHiTauGoodIncTwoBHBOneLuminosityDistanceZeroPost}{\ensuremath{261^{+97}_{-87}}}
\newcommand{\RunHiTauGoodIncTwoBHBOneMassRatioZeroPost}{\ensuremath{0.4^{+0.5}_{-0.3}}}
\newcommand{\RunHiTauGoodIncTwoBHBOneMergerTimeOrInitialOrbitalFrequencyZeroPost}{\ensuremath{5.112832^{+0.000002}_{-0.000002}}}
\newcommand{\RunHiTauGoodIncTwoBHBOneMergerTimeZeroPost}{\ensuremath{12.379^{+0.001}_{-0.003}}}
\newcommand{\RunHiTauGoodIncTwoBHBOnePolarizationZeroPost}{\ensuremath{0.2^{+2.0}_{-2.0}}}
\newcommand{\RunHiTauGoodIncTwoBHBOneRedshiftZeroPost}{\ensuremath{0.06^{+0.02}_{-0.02}}}
\newcommand{\RunHiTauGoodIncTwoBHBOneRedshiftedChirpMassZeroPost}{\ensuremath{32.807^{+0.005}_{-0.002}}}
\newcommand{\RunHiTauGoodIncTwoBHBOneRedshiftedMassOneZeroPost}{\ensuremath{60^{+75}_{-21}}}
\newcommand{\RunHiTauGoodIncTwoBHBOneRedshiftedMassTwoZeroPost}{\ensuremath{25^{+11}_{-11}}}
\newcommand{\RunHiTauGoodIncTwoBHBOneReducedMassDifferenceZeroPost}{\ensuremath{0.4^{+0.4}_{-0.4}}}
\newcommand{\RunHiTauGoodIncTwoBHBOnecosInclinationZeroPost}{\ensuremath{-0.8^{+0.3}_{-0.2}}}
\newcommand{\RunHiTauGoodIncTwoBHBOnesinEclipticLatitudeZeroPost}{\ensuremath{0.23^{+0.03}_{-0.03}}}
\newcommand{\RunHiTauGoodIncTwoBHBOneDimensionlessSpinOneZeroWidth}{\ensuremath{4.0}}
\newcommand{\RunHiTauGoodIncTwoBHBOneDimensionlessSpinTwoZeroWidth}{\ensuremath{15.81}}
\newcommand{\RunHiTauGoodIncTwoBHBOneEclipticLatitudeZeroWidth}{\ensuremath{0.3}}
\newcommand{\RunHiTauGoodIncTwoBHBOneEclipticLongitudeZeroWidth}{\ensuremath{0.006}}
\newcommand{\RunHiTauGoodIncTwoBHBOneInclinationZeroWidth}{\ensuremath{0.3}}
\newcommand{\RunHiTauGoodIncTwoBHBOneInitialOrbitalPhaseZeroWidth}{\ensuremath{5.5}}
\newcommand{\RunHiTauGoodIncTwoBHBOneLuminosityDistanceZeroWidth}{\ensuremath{0.6}}
\newcommand{\RunHiTauGoodIncTwoBHBOneMassRatioZeroWidth}{\ensuremath{1.0}}
\newcommand{\RunHiTauGoodIncTwoBHBOneMergerTimeOrInitialOrbitalFrequencyZeroWidth}{\ensuremath{7\!\times\!10^{-7}}}
\newcommand{\RunHiTauGoodIncTwoBHBOneMergerTimeZeroWidth}{\ensuremath{4\!\times\!10^{-4}}}
\newcommand{\RunHiTauGoodIncTwoBHBOnePolarizationZeroWidth}{\ensuremath{1.5}}
\newcommand{\RunHiTauGoodIncTwoBHBOneRedshiftZeroWidth}{\ensuremath{0.6}}
\newcommand{\RunHiTauGoodIncTwoBHBOneRedshiftedChirpMassZeroWidth}{\ensuremath{2\!\times\!10^{-4}}}
\newcommand{\RunHiTauGoodIncTwoBHBOneRedshiftedMassOneZeroWidth}{\ensuremath{2.3}}
\newcommand{\RunHiTauGoodIncTwoBHBOneRedshiftedMassTwoZeroWidth}{\ensuremath{0.7}}
\newcommand{\RunHiTauGoodIncTwoBHBOneReducedMassDifferenceZeroWidth}{\ensuremath{7.1}}
\newcommand{\RunHiTauGoodIncTwoBHBOnecosInclinationZeroWidth}{\ensuremath{-0.6}}
\newcommand{\RunHiTauGoodIncTwoBHBOnesinEclipticLatitudeZeroWidth}{\ensuremath{0.3}}
\newcommand{\RunLowTauBadIncOneSNR}{\ensuremath{10.59}}
\newcommand{\RunLowTauBadIncOneBHBOneDimensionlessSpinOneZeroInj}{\ensuremath{0.3}}
\newcommand{\RunLowTauBadIncOneBHBOneDimensionlessSpinTwoZeroInj}{\ensuremath{0.3}}
\newcommand{\RunLowTauBadIncOneBHBOneEclipticLatitudeZeroInj}{\ensuremath{-0.3}}
\newcommand{\RunLowTauBadIncOneBHBOneEclipticLongitudeZeroInj}{\ensuremath{1.8}}
\newcommand{\RunLowTauBadIncOneBHBOneInclinationZeroInj}{\ensuremath{1.2}}
\newcommand{\RunLowTauBadIncOneBHBOneInitialOrbitalPhaseZeroInj}{\ensuremath{5.6}}
\newcommand{\RunLowTauBadIncOneBHBOneLuminosityDistanceZeroInj}{\ensuremath{122.570}}
\newcommand{\RunLowTauBadIncOneBHBOneMassRatioZeroInj}{\ensuremath{0.7}}
\newcommand{\RunLowTauBadIncOneBHBOneMergerTimeOrInitialOrbitalFrequencyZeroInj}{\ensuremath{9.4}}
\newcommand{\RunLowTauBadIncOneBHBOneMergerTimeZeroInj}{\ensuremath{3.6}}
\newcommand{\RunLowTauBadIncOneBHBOnePolarizationZeroInj}{\ensuremath{0.6}}
\newcommand{\RunLowTauBadIncOneBHBOneRedshiftZeroInj}{\ensuremath{0.03}}
\newcommand{\RunLowTauBadIncOneBHBOneRedshiftedChirpMassZeroInj}{\ensuremath{25.93}}
\newcommand{\RunLowTauBadIncOneBHBOneRedshiftedMassOneZeroInj}{\ensuremath{34.89}}
\newcommand{\RunLowTauBadIncOneBHBOneRedshiftedMassTwoZeroInj}{\ensuremath{25.55}}
\newcommand{\RunLowTauBadIncOneBHBOneReducedMassDifferenceZeroInj}{\ensuremath{0.2}}
\newcommand{\RunLowTauBadIncOneBHBOnecosInclinationZeroInj}{\ensuremath{0.3}}
\newcommand{\RunLowTauBadIncOneBHBOnesinEclipticLatitudeZeroInj}{\ensuremath{-0.3}}
\newcommand{\RunLowTauBadIncOneBHBOneDimensionlessSpinOneZeroPost}{\ensuremath{0.4^{+0.5}_{-0.5}}}
\newcommand{\RunLowTauBadIncOneBHBOneDimensionlessSpinTwoZeroPost}{\ensuremath{0.09^{+0.80}_{-0.87}}}
\newcommand{\RunLowTauBadIncOneBHBOneEclipticLatitudeZeroPost}{\ensuremath{-0.28^{+0.01}_{-0.01}}}
\newcommand{\RunLowTauBadIncOneBHBOneEclipticLongitudeZeroPost}{\ensuremath{1.753^{+0.004}_{-0.004}}}
\newcommand{\RunLowTauBadIncOneBHBOneInclinationZeroPost}{\ensuremath{1.1^{+0.3}_{-0.5}}}
\newcommand{\RunLowTauBadIncOneBHBOneInitialOrbitalPhaseZeroPost}{\ensuremath{3.5^{+2.1}_{-2.5}}}
\newcommand{\RunLowTauBadIncOneBHBOneLuminosityDistanceZeroPost}{\ensuremath{164^{+71}_{-48}}}
\newcommand{\RunLowTauBadIncOneBHBOneMassRatioZeroPost}{\ensuremath{0.7^{+0.3}_{-0.3}}}
\newcommand{\RunLowTauBadIncOneBHBOneMergerTimeOrInitialOrbitalFrequencyZeroPost}{\ensuremath{9.390248^{+0.000001}_{-0.000002}}}
\newcommand{\RunLowTauBadIncOneBHBOneMergerTimeZeroPost}{\ensuremath{3.62192^{+2\!\times\!10^{-5}}_{-0.00010}}}
\newcommand{\RunLowTauBadIncOneBHBOnePolarizationZeroPost}{\ensuremath{0.3^{+1.9}_{-2.8}}}
\newcommand{\RunLowTauBadIncOneBHBOneRedshiftZeroPost}{\ensuremath{0.04^{+0.02}_{-0.01}}}
\newcommand{\RunLowTauBadIncOneBHBOneRedshiftedChirpMassZeroPost}{\ensuremath{25.92937^{+0.00042}_{-0.00009}}}
\newcommand{\RunLowTauBadIncOneBHBOneRedshiftedMassOneZeroPost}{\ensuremath{35^{+12}_{-5}}}
\newcommand{\RunLowTauBadIncOneBHBOneRedshiftedMassTwoZeroPost}{\ensuremath{25^{+4}_{-5}}}
\newcommand{\RunLowTauBadIncOneBHBOneReducedMassDifferenceZeroPost}{\ensuremath{0.2^{+0.2}_{-0.2}}}
\newcommand{\RunLowTauBadIncOneBHBOnecosInclinationZeroPost}{\ensuremath{0.5^{+0.4}_{-0.2}}}
\newcommand{\RunLowTauBadIncOneBHBOnesinEclipticLatitudeZeroPost}{\ensuremath{-0.280^{+0.010}_{-0.010}}}
\newcommand{\RunLowTauBadIncOneBHBOneDimensionlessSpinOneZeroWidth}{\ensuremath{3.6}}
\newcommand{\RunLowTauBadIncOneBHBOneDimensionlessSpinTwoZeroWidth}{\ensuremath{6.4}}
\newcommand{\RunLowTauBadIncOneBHBOneEclipticLatitudeZeroWidth}{\ensuremath{-0.07}}
\newcommand{\RunLowTauBadIncOneBHBOneEclipticLongitudeZeroWidth}{\ensuremath{0.004}}
\newcommand{\RunLowTauBadIncOneBHBOneInclinationZeroWidth}{\ensuremath{0.6}}
\newcommand{\RunLowTauBadIncOneBHBOneInitialOrbitalPhaseZeroWidth}{\ensuremath{0.8}}
\newcommand{\RunLowTauBadIncOneBHBOneLuminosityDistanceZeroWidth}{\ensuremath{1.0}}
\newcommand{\RunLowTauBadIncOneBHBOneMassRatioZeroWidth}{\ensuremath{0.8}}
\newcommand{\RunLowTauBadIncOneBHBOneMergerTimeOrInitialOrbitalFrequencyZeroWidth}{\ensuremath{3\!\times\!10^{-7}}}
\newcommand{\RunLowTauBadIncOneBHBOneMergerTimeZeroWidth}{\ensuremath{3\!\times\!10^{-5}}}
\newcommand{\RunLowTauBadIncOneBHBOnePolarizationZeroWidth}{\ensuremath{7.4}}
\newcommand{\RunLowTauBadIncOneBHBOneRedshiftZeroWidth}{\ensuremath{0.9}}
\newcommand{\RunLowTauBadIncOneBHBOneRedshiftedChirpMassZeroWidth}{\ensuremath{2\!\times\!10^{-5}}}
\newcommand{\RunLowTauBadIncOneBHBOneRedshiftedMassOneZeroWidth}{\ensuremath{0.5}}
\newcommand{\RunLowTauBadIncOneBHBOneRedshiftedMassTwoZeroWidth}{\ensuremath{0.4}}
\newcommand{\RunLowTauBadIncOneBHBOneReducedMassDifferenceZeroWidth}{\ensuremath{2.6}}
\newcommand{\RunLowTauBadIncOneBHBOnecosInclinationZeroWidth}{\ensuremath{1.7}}
\newcommand{\RunLowTauBadIncOneBHBOnesinEclipticLatitudeZeroWidth}{\ensuremath{-0.07}}
\newcommand{\RunLowTauBadIncTwoSNR}{\ensuremath{10.92}}
\newcommand{\RunLowTauBadIncTwoBHBOneDimensionlessSpinOneZeroInj}{\ensuremath{0.4}}
\newcommand{\RunLowTauBadIncTwoBHBOneDimensionlessSpinTwoZeroInj}{\ensuremath{0.1}}
\newcommand{\RunLowTauBadIncTwoBHBOneEclipticLatitudeZeroInj}{\ensuremath{-0.4}}
\newcommand{\RunLowTauBadIncTwoBHBOneEclipticLongitudeZeroInj}{\ensuremath{4.4}}
\newcommand{\RunLowTauBadIncTwoBHBOneInclinationZeroInj}{\ensuremath{1.7}}
\newcommand{\RunLowTauBadIncTwoBHBOneInitialOrbitalPhaseZeroInj}{\ensuremath{4.5}}
\newcommand{\RunLowTauBadIncTwoBHBOneLuminosityDistanceZeroInj}{\ensuremath{92.95}}
\newcommand{\RunLowTauBadIncTwoBHBOneMassRatioZeroInj}{\ensuremath{0.7}}
\newcommand{\RunLowTauBadIncTwoBHBOneMergerTimeOrInitialOrbitalFrequencyZeroInj}{\ensuremath{8.2}}
\newcommand{\RunLowTauBadIncTwoBHBOneMergerTimeZeroInj}{\ensuremath{4.7}}
\newcommand{\RunLowTauBadIncTwoBHBOnePolarizationZeroInj}{\ensuremath{3.9}}
\newcommand{\RunLowTauBadIncTwoBHBOneRedshiftZeroInj}{\ensuremath{0.02}}
\newcommand{\RunLowTauBadIncTwoBHBOneRedshiftedChirpMassZeroInj}{\ensuremath{27.32}}
\newcommand{\RunLowTauBadIncTwoBHBOneRedshiftedMassOneZeroInj}{\ensuremath{36.36}}
\newcommand{\RunLowTauBadIncTwoBHBOneRedshiftedMassTwoZeroInj}{\ensuremath{27.21}}
\newcommand{\RunLowTauBadIncTwoBHBOneReducedMassDifferenceZeroInj}{\ensuremath{0.1}}
\newcommand{\RunLowTauBadIncTwoBHBOnecosInclinationZeroInj}{\ensuremath{-0.08}}
\newcommand{\RunLowTauBadIncTwoBHBOnesinEclipticLatitudeZeroInj}{\ensuremath{-0.4}}
\newcommand{\RunLowTauBadIncTwoBHBOneDimensionlessSpinOneZeroPost}{\ensuremath{0.1^{+0.7}_{-0.5}}}
\newcommand{\RunLowTauBadIncTwoBHBOneDimensionlessSpinTwoZeroPost}{\ensuremath{0.08^{+0.83}_{-0.92}}}
\newcommand{\RunLowTauBadIncTwoBHBOneEclipticLatitudeZeroPost}{\ensuremath{-0.392^{+0.009}_{-0.009}}}
\newcommand{\RunLowTauBadIncTwoBHBOneEclipticLongitudeZeroPost}{\ensuremath{4.426^{+0.005}_{-0.005}}}
\newcommand{\RunLowTauBadIncTwoBHBOneInclinationZeroPost}{\ensuremath{1.66^{+0.10}_{-0.09}}}
\newcommand{\RunLowTauBadIncTwoBHBOneInitialOrbitalPhaseZeroPost}{\ensuremath{2.9^{+1.7}_{-1.7}}}
\newcommand{\RunLowTauBadIncTwoBHBOneLuminosityDistanceZeroPost}{\ensuremath{99^{+21}_{-14}}}
\newcommand{\RunLowTauBadIncTwoBHBOneMassRatioZeroPost}{\ensuremath{0.5^{+0.4}_{-0.4}}}
\newcommand{\RunLowTauBadIncTwoBHBOneMergerTimeOrInitialOrbitalFrequencyZeroPost}{\ensuremath{8.238801^{+0.000002}_{-0.000002}}}
\newcommand{\RunLowTauBadIncTwoBHBOneMergerTimeZeroPost}{\ensuremath{4.70482^{+0.00010}_{-0.00052}}}
\newcommand{\RunLowTauBadIncTwoBHBOnePolarizationZeroPost}{\ensuremath{0.7^{+1.6}_{-1.6}}}
\newcommand{\RunLowTauBadIncTwoBHBOneRedshiftZeroPost}{\ensuremath{0.022^{+0.005}_{-0.003}}}
\newcommand{\RunLowTauBadIncTwoBHBOneRedshiftedChirpMassZeroPost}{\ensuremath{27.3231^{+0.0018}_{-3\!\times\!10^{-4}}}}
\newcommand{\RunLowTauBadIncTwoBHBOneRedshiftedMassOneZeroPost}{\ensuremath{44^{+39}_{-11}}}
\newcommand{\RunLowTauBadIncTwoBHBOneRedshiftedMassTwoZeroPost}{\ensuremath{23^{+7}_{-9}}}
\newcommand{\RunLowTauBadIncTwoBHBOneReducedMassDifferenceZeroPost}{\ensuremath{0.3^{+0.4}_{-0.3}}}
\newcommand{\RunLowTauBadIncTwoBHBOnecosInclinationZeroPost}{\ensuremath{-0.09^{+0.09}_{-0.10}}}
\newcommand{\RunLowTauBadIncTwoBHBOnesinEclipticLatitudeZeroPost}{\ensuremath{-0.382^{+0.008}_{-0.008}}}
\newcommand{\RunLowTauBadIncTwoBHBOneDimensionlessSpinOneZeroWidth}{\ensuremath{3.1}}
\newcommand{\RunLowTauBadIncTwoBHBOneDimensionlessSpinTwoZeroWidth}{\ensuremath{15.27}}
\newcommand{\RunLowTauBadIncTwoBHBOneEclipticLatitudeZeroWidth}{\ensuremath{-0.04}}
\newcommand{\RunLowTauBadIncTwoBHBOneEclipticLongitudeZeroWidth}{\ensuremath{2\!\times\!10^{-3}}}
\newcommand{\RunLowTauBadIncTwoBHBOneInclinationZeroWidth}{\ensuremath{0.1}}
\newcommand{\RunLowTauBadIncTwoBHBOneInitialOrbitalPhaseZeroWidth}{\ensuremath{0.8}}
\newcommand{\RunLowTauBadIncTwoBHBOneLuminosityDistanceZeroWidth}{\ensuremath{0.4}}
\newcommand{\RunLowTauBadIncTwoBHBOneMassRatioZeroWidth}{\ensuremath{1.0}}
\newcommand{\RunLowTauBadIncTwoBHBOneMergerTimeOrInitialOrbitalFrequencyZeroWidth}{\ensuremath{5\!\times\!10^{-7}}}
\newcommand{\RunLowTauBadIncTwoBHBOneMergerTimeZeroWidth}{\ensuremath{1\!\times\!10^{-4}}}
\newcommand{\RunLowTauBadIncTwoBHBOnePolarizationZeroWidth}{\ensuremath{0.8}}
\newcommand{\RunLowTauBadIncTwoBHBOneRedshiftZeroWidth}{\ensuremath{0.4}}
\newcommand{\RunLowTauBadIncTwoBHBOneRedshiftedChirpMassZeroWidth}{\ensuremath{8\!\times\!10^{-5}}}
\newcommand{\RunLowTauBadIncTwoBHBOneRedshiftedMassOneZeroWidth}{\ensuremath{1.4}}
\newcommand{\RunLowTauBadIncTwoBHBOneRedshiftedMassTwoZeroWidth}{\ensuremath{0.6}}
\newcommand{\RunLowTauBadIncTwoBHBOneReducedMassDifferenceZeroWidth}{\ensuremath{4.7}}
\newcommand{\RunLowTauBadIncTwoBHBOnecosInclinationZeroWidth}{\ensuremath{-2.3}}
\newcommand{\RunLowTauBadIncTwoBHBOnesinEclipticLatitudeZeroWidth}{\ensuremath{-0.04}}
\newcommand{\RunLowTauGoodIncOneSNR}{\ensuremath{11.12}}
\newcommand{\RunLowTauGoodIncOneBHBOneDimensionlessSpinOneZeroInj}{\ensuremath{0.6}}
\newcommand{\RunLowTauGoodIncOneBHBOneDimensionlessSpinTwoZeroInj}{\ensuremath{0.6}}
\newcommand{\RunLowTauGoodIncOneBHBOneEclipticLatitudeZeroInj}{\ensuremath{0.4}}
\newcommand{\RunLowTauGoodIncOneBHBOneEclipticLongitudeZeroInj}{\ensuremath{4.7}}
\newcommand{\RunLowTauGoodIncOneBHBOneInclinationZeroInj}{\ensuremath{2.9}}
\newcommand{\RunLowTauGoodIncOneBHBOneInitialOrbitalPhaseZeroInj}{\ensuremath{4.7}}
\newcommand{\RunLowTauGoodIncOneBHBOneLuminosityDistanceZeroInj}{\ensuremath{510.906}}
\newcommand{\RunLowTauGoodIncOneBHBOneMassRatioZeroInj}{\ensuremath{0.8}}
\newcommand{\RunLowTauGoodIncOneBHBOneMergerTimeOrInitialOrbitalFrequencyZeroInj}{\ensuremath{6.5}}
\newcommand{\RunLowTauGoodIncOneBHBOneMergerTimeZeroInj}{\ensuremath{4.2}}
\newcommand{\RunLowTauGoodIncOneBHBOnePolarizationZeroInj}{\ensuremath{6.0}}
\newcommand{\RunLowTauGoodIncOneBHBOneRedshiftZeroInj}{\ensuremath{0.1}}
\newcommand{\RunLowTauGoodIncOneBHBOneRedshiftedChirpMassZeroInj}{\ensuremath{43.21}}
\newcommand{\RunLowTauGoodIncOneBHBOneRedshiftedMassOneZeroInj}{\ensuremath{55.23}}
\newcommand{\RunLowTauGoodIncOneBHBOneRedshiftedMassTwoZeroInj}{\ensuremath{44.71}}
\newcommand{\RunLowTauGoodIncOneBHBOneReducedMassDifferenceZeroInj}{\ensuremath{0.1}}
\newcommand{\RunLowTauGoodIncOneBHBOnecosInclinationZeroInj}{\ensuremath{-1.0}}
\newcommand{\RunLowTauGoodIncOneBHBOnesinEclipticLatitudeZeroInj}{\ensuremath{0.4}}
\newcommand{\RunLowTauGoodIncOneBHBOneDimensionlessSpinOneZeroPost}{\ensuremath{0.6^{+0.4}_{-0.3}}}
\newcommand{\RunLowTauGoodIncOneBHBOneDimensionlessSpinTwoZeroPost}{\ensuremath{0.3^{+0.6}_{-1.0}}}
\newcommand{\RunLowTauGoodIncOneBHBOneEclipticLatitudeZeroPost}{\ensuremath{0.38^{+0.01}_{-0.01}}}
\newcommand{\RunLowTauGoodIncOneBHBOneEclipticLongitudeZeroPost}{\ensuremath{4.694^{+0.005}_{-0.005}}}
\newcommand{\RunLowTauGoodIncOneBHBOneInclinationZeroPost}{\ensuremath{2.5^{+0.5}_{-0.4}}}
\newcommand{\RunLowTauGoodIncOneBHBOneInitialOrbitalPhaseZeroPost}{\ensuremath{3.3^{+2.0}_{-2.0}}}
\newcommand{\RunLowTauGoodIncOneBHBOneLuminosityDistanceZeroPost}{\ensuremath{444^{+141}_{-138}}}
\newcommand{\RunLowTauGoodIncOneBHBOneMassRatioZeroPost}{\ensuremath{0.5^{+0.4}_{-0.3}}}
\newcommand{\RunLowTauGoodIncOneBHBOneMergerTimeOrInitialOrbitalFrequencyZeroPost}{\ensuremath{6.484078^{+0.000001}_{-0.000002}}}
\newcommand{\RunLowTauGoodIncOneBHBOneMergerTimeZeroPost}{\ensuremath{4.15061^{+7\!\times\!10^{-5}}_{-0.00023}}}
\newcommand{\RunLowTauGoodIncOneBHBOnePolarizationZeroPost}{\ensuremath{-0.1^{+2.0}_{-2.0}}}
\newcommand{\RunLowTauGoodIncOneBHBOneRedshiftZeroPost}{\ensuremath{0.09^{+0.03}_{-0.03}}}
\newcommand{\RunLowTauGoodIncOneBHBOneRedshiftedChirpMassZeroPost}{\ensuremath{43.2129^{+0.0015}_{-4\!\times\!10^{-4}}}}
\newcommand{\RunLowTauGoodIncOneBHBOneRedshiftedMassOneZeroPost}{\ensuremath{68^{+29}_{-17}}}
\newcommand{\RunLowTauGoodIncOneBHBOneRedshiftedMassTwoZeroPost}{\ensuremath{37^{+11}_{-9}}}
\newcommand{\RunLowTauGoodIncOneBHBOneReducedMassDifferenceZeroPost}{\ensuremath{0.3^{+0.3}_{-0.3}}}
\newcommand{\RunLowTauGoodIncOneBHBOnecosInclinationZeroPost}{\ensuremath{-0.8^{+0.3}_{-0.2}}}
\newcommand{\RunLowTauGoodIncOneBHBOnesinEclipticLatitudeZeroPost}{\ensuremath{0.371^{+0.009}_{-0.010}}}
\newcommand{\RunLowTauGoodIncOneBHBOneDimensionlessSpinOneZeroWidth}{\ensuremath{1.1}}
\newcommand{\RunLowTauGoodIncOneBHBOneDimensionlessSpinTwoZeroWidth}{\ensuremath{2.9}}
\newcommand{\RunLowTauGoodIncOneBHBOneEclipticLatitudeZeroWidth}{\ensuremath{0.05}}
\newcommand{\RunLowTauGoodIncOneBHBOneEclipticLongitudeZeroWidth}{\ensuremath{2\!\times\!10^{-3}}}
\newcommand{\RunLowTauGoodIncOneBHBOneInclinationZeroWidth}{\ensuremath{0.3}}
\newcommand{\RunLowTauGoodIncOneBHBOneInitialOrbitalPhaseZeroWidth}{\ensuremath{0.9}}
\newcommand{\RunLowTauGoodIncOneBHBOneLuminosityDistanceZeroWidth}{\ensuremath{0.5}}
\newcommand{\RunLowTauGoodIncOneBHBOneMassRatioZeroWidth}{\ensuremath{0.8}}
\newcommand{\RunLowTauGoodIncOneBHBOneMergerTimeOrInitialOrbitalFrequencyZeroWidth}{\ensuremath{5\!\times\!10^{-7}}}
\newcommand{\RunLowTauGoodIncOneBHBOneMergerTimeZeroWidth}{\ensuremath{7\!\times\!10^{-5}}}
\newcommand{\RunLowTauGoodIncOneBHBOnePolarizationZeroWidth}{\ensuremath{0.7}}
\newcommand{\RunLowTauGoodIncOneBHBOneRedshiftZeroWidth}{\ensuremath{0.5}}
\newcommand{\RunLowTauGoodIncOneBHBOneRedshiftedChirpMassZeroWidth}{\ensuremath{4\!\times\!10^{-5}}}
\newcommand{\RunLowTauGoodIncOneBHBOneRedshiftedMassOneZeroWidth}{\ensuremath{0.8}}
\newcommand{\RunLowTauGoodIncOneBHBOneRedshiftedMassTwoZeroWidth}{\ensuremath{0.5}}
\newcommand{\RunLowTauGoodIncOneBHBOneReducedMassDifferenceZeroWidth}{\ensuremath{5.0}}
\newcommand{\RunLowTauGoodIncOneBHBOnecosInclinationZeroWidth}{\ensuremath{-0.5}}
\newcommand{\RunLowTauGoodIncOneBHBOnesinEclipticLatitudeZeroWidth}{\ensuremath{0.05}}
\newcommand{\RunLowTauGoodIncTwoSNR}{\ensuremath{10.55}}
\newcommand{\RunLowTauGoodIncTwoBHBOneDimensionlessSpinOneZeroInj}{\ensuremath{0.02}}
\newcommand{\RunLowTauGoodIncTwoBHBOneDimensionlessSpinTwoZeroInj}{\ensuremath{0.2}}
\newcommand{\RunLowTauGoodIncTwoBHBOneEclipticLatitudeZeroInj}{\ensuremath{-0.4}}
\newcommand{\RunLowTauGoodIncTwoBHBOneEclipticLongitudeZeroInj}{\ensuremath{4.5}}
\newcommand{\RunLowTauGoodIncTwoBHBOneInclinationZeroInj}{\ensuremath{0.4}}
\newcommand{\RunLowTauGoodIncTwoBHBOneInitialOrbitalPhaseZeroInj}{\ensuremath{3.0}}
\newcommand{\RunLowTauGoodIncTwoBHBOneLuminosityDistanceZeroInj}{\ensuremath{254.146}}
\newcommand{\RunLowTauGoodIncTwoBHBOneMassRatioZeroInj}{\ensuremath{0.9}}
\newcommand{\RunLowTauGoodIncTwoBHBOneMergerTimeOrInitialOrbitalFrequencyZeroInj}{\ensuremath{9.7}}
\newcommand{\RunLowTauGoodIncTwoBHBOneMergerTimeZeroInj}{\ensuremath{2.6}}
\newcommand{\RunLowTauGoodIncTwoBHBOnePolarizationZeroInj}{\ensuremath{2.9}}
\newcommand{\RunLowTauGoodIncTwoBHBOneRedshiftZeroInj}{\ensuremath{0.06}}
\newcommand{\RunLowTauGoodIncTwoBHBOneRedshiftedChirpMassZeroInj}{\ensuremath{30.32}}
\newcommand{\RunLowTauGoodIncTwoBHBOneRedshiftedMassOneZeroInj}{\ensuremath{36.05}}
\newcommand{\RunLowTauGoodIncTwoBHBOneRedshiftedMassTwoZeroInj}{\ensuremath{33.66}}
\newcommand{\RunLowTauGoodIncTwoBHBOneReducedMassDifferenceZeroInj}{\ensuremath{0.03}}
\newcommand{\RunLowTauGoodIncTwoBHBOnecosInclinationZeroInj}{\ensuremath{0.9}}
\newcommand{\RunLowTauGoodIncTwoBHBOnesinEclipticLatitudeZeroInj}{\ensuremath{-0.4}}
\newcommand{\RunLowTauGoodIncTwoBHBOneDimensionlessSpinOneZeroPost}{\ensuremath{0.1^{+0.7}_{-0.6}}}
\newcommand{\RunLowTauGoodIncTwoBHBOneDimensionlessSpinTwoZeroPost}{\ensuremath{0.05^{+0.85}_{-0.90}}}
\newcommand{\RunLowTauGoodIncTwoBHBOneEclipticLatitudeZeroPost}{\ensuremath{-0.367^{+0.008}_{-0.008}}}
\newcommand{\RunLowTauGoodIncTwoBHBOneEclipticLongitudeZeroPost}{\ensuremath{4.513^{+0.004}_{-0.004}}}
\newcommand{\RunLowTauGoodIncTwoBHBOneInclinationZeroPost}{\ensuremath{0.7^{+0.4}_{-0.5}}}
\newcommand{\RunLowTauGoodIncTwoBHBOneInitialOrbitalPhaseZeroPost}{\ensuremath{3.7^{+2.0}_{-2.0}}}
\newcommand{\RunLowTauGoodIncTwoBHBOneLuminosityDistanceZeroPost}{\ensuremath{233^{+80}_{-74}}}
\newcommand{\RunLowTauGoodIncTwoBHBOneMassRatioZeroPost}{\ensuremath{0.7^{+0.2}_{-0.3}}}
\newcommand{\RunLowTauGoodIncTwoBHBOneMergerTimeOrInitialOrbitalFrequencyZeroPost}{\ensuremath{9.664925^{+0.000002}_{-0.000002}}}
\newcommand{\RunLowTauGoodIncTwoBHBOneMergerTimeZeroPost}{\ensuremath{2.58385^{+0.00001}_{-0.00006}}}
\newcommand{\RunLowTauGoodIncTwoBHBOnePolarizationZeroPost}{\ensuremath{0.6^{+2.0}_{-2.0}}}
\newcommand{\RunLowTauGoodIncTwoBHBOneRedshiftZeroPost}{\ensuremath{0.05^{+0.02}_{-0.02}}}
\newcommand{\RunLowTauGoodIncTwoBHBOneRedshiftedChirpMassZeroPost}{\ensuremath{30.32219^{+0.00045}_{-0.00009}}}
\newcommand{\RunLowTauGoodIncTwoBHBOneRedshiftedMassOneZeroPost}{\ensuremath{40^{+12}_{-5}}}
\newcommand{\RunLowTauGoodIncTwoBHBOneRedshiftedMassTwoZeroPost}{\ensuremath{30^{+4}_{-6}}}
\newcommand{\RunLowTauGoodIncTwoBHBOneReducedMassDifferenceZeroPost}{\ensuremath{0.1^{+0.2}_{-0.1}}}
\newcommand{\RunLowTauGoodIncTwoBHBOnecosInclinationZeroPost}{\ensuremath{0.8^{+0.2}_{-0.3}}}
\newcommand{\RunLowTauGoodIncTwoBHBOnesinEclipticLatitudeZeroPost}{\ensuremath{-0.358^{+0.007}_{-0.007}}}
\newcommand{\RunLowTauGoodIncTwoBHBOneDimensionlessSpinOneZeroWidth}{\ensuremath{65.59}}
\newcommand{\RunLowTauGoodIncTwoBHBOneDimensionlessSpinTwoZeroWidth}{\ensuremath{8.3}}
\newcommand{\RunLowTauGoodIncTwoBHBOneEclipticLatitudeZeroWidth}{\ensuremath{-0.04}}
\newcommand{\RunLowTauGoodIncTwoBHBOneEclipticLongitudeZeroWidth}{\ensuremath{2\!\times\!10^{-3}}}
\newcommand{\RunLowTauGoodIncTwoBHBOneInclinationZeroWidth}{\ensuremath{2.1}}
\newcommand{\RunLowTauGoodIncTwoBHBOneInitialOrbitalPhaseZeroWidth}{\ensuremath{1.4}}
\newcommand{\RunLowTauGoodIncTwoBHBOneLuminosityDistanceZeroWidth}{\ensuremath{0.6}}
\newcommand{\RunLowTauGoodIncTwoBHBOneMassRatioZeroWidth}{\ensuremath{0.6}}
\newcommand{\RunLowTauGoodIncTwoBHBOneMergerTimeOrInitialOrbitalFrequencyZeroWidth}{\ensuremath{4\!\times\!10^{-7}}}
\newcommand{\RunLowTauGoodIncTwoBHBOneMergerTimeZeroWidth}{\ensuremath{3\!\times\!10^{-5}}}
\newcommand{\RunLowTauGoodIncTwoBHBOnePolarizationZeroWidth}{\ensuremath{1.4}}
\newcommand{\RunLowTauGoodIncTwoBHBOneRedshiftZeroWidth}{\ensuremath{0.6}}
\newcommand{\RunLowTauGoodIncTwoBHBOneRedshiftedChirpMassZeroWidth}{\ensuremath{2\!\times\!10^{-5}}}
\newcommand{\RunLowTauGoodIncTwoBHBOneRedshiftedMassOneZeroWidth}{\ensuremath{0.5}}
\newcommand{\RunLowTauGoodIncTwoBHBOneRedshiftedMassTwoZeroWidth}{\ensuremath{0.3}}
\newcommand{\RunLowTauGoodIncTwoBHBOneReducedMassDifferenceZeroWidth}{\ensuremath{10.64}}
\newcommand{\RunLowTauGoodIncTwoBHBOnecosInclinationZeroWidth}{\ensuremath{0.5}}
\newcommand{\RunLowTauGoodIncTwoBHBOnesinEclipticLatitudeZeroWidth}{\ensuremath{-0.04}}
\newcommand{\RunMidTauBadIncOneSNR}{\ensuremath{10.59}}
\newcommand{\RunMidTauBadIncOneBHBOneDimensionlessSpinOneZeroInj}{\ensuremath{0.1}}
\newcommand{\RunMidTauBadIncOneBHBOneDimensionlessSpinTwoZeroInj}{\ensuremath{0.5}}
\newcommand{\RunMidTauBadIncOneBHBOneEclipticLatitudeZeroInj}{\ensuremath{-0.04}}
\newcommand{\RunMidTauBadIncOneBHBOneEclipticLongitudeZeroInj}{\ensuremath{0.9}}
\newcommand{\RunMidTauBadIncOneBHBOneInclinationZeroInj}{\ensuremath{1.4}}
\newcommand{\RunMidTauBadIncOneBHBOneInitialOrbitalPhaseZeroInj}{\ensuremath{4.4}}
\newcommand{\RunMidTauBadIncOneBHBOneLuminosityDistanceZeroInj}{\ensuremath{82.63}}
\newcommand{\RunMidTauBadIncOneBHBOneMassRatioZeroInj}{\ensuremath{0.8}}
\newcommand{\RunMidTauBadIncOneBHBOneMergerTimeOrInitialOrbitalFrequencyZeroInj}{\ensuremath{6.6}}
\newcommand{\RunMidTauBadIncOneBHBOneMergerTimeZeroInj}{\ensuremath{9.3}}
\newcommand{\RunMidTauBadIncOneBHBOnePolarizationZeroInj}{\ensuremath{0.006}}
\newcommand{\RunMidTauBadIncOneBHBOneRedshiftZeroInj}{\ensuremath{0.02}}
\newcommand{\RunMidTauBadIncOneBHBOneRedshiftedChirpMassZeroInj}{\ensuremath{26.07}}
\newcommand{\RunMidTauBadIncOneBHBOneRedshiftedMassOneZeroInj}{\ensuremath{33.97}}
\newcommand{\RunMidTauBadIncOneBHBOneRedshiftedMassTwoZeroInj}{\ensuremath{26.48}}
\newcommand{\RunMidTauBadIncOneBHBOneReducedMassDifferenceZeroInj}{\ensuremath{0.1}}
\newcommand{\RunMidTauBadIncOneBHBOnecosInclinationZeroInj}{\ensuremath{0.2}}
\newcommand{\RunMidTauBadIncOneBHBOnesinEclipticLatitudeZeroInj}{\ensuremath{-0.04}}
\newcommand{\RunMidTauBadIncOneBHBOneDimensionlessSpinOneZeroPost}{\ensuremath{0.002^{+0.86}_{-0.83}}}
\newcommand{\RunMidTauBadIncOneBHBOneDimensionlessSpinTwoZeroPost}{\ensuremath{0.05^{+0.85}_{-0.93}}}
\newcommand{\RunMidTauBadIncOneBHBOneEclipticLatitudeZeroPost}{\ensuremath{-0.01^{+0.08}_{-0.07}}}
\newcommand{\RunMidTauBadIncOneBHBOneEclipticLongitudeZeroPost}{\ensuremath{0.897^{+0.007}_{-0.007}}}
\newcommand{\RunMidTauBadIncOneBHBOneInclinationZeroPost}{\ensuremath{1.36^{+0.10}_{-0.15}}}
\newcommand{\RunMidTauBadIncOneBHBOneInitialOrbitalPhaseZeroPost}{\ensuremath{2.8^{+1.7}_{-1.7}}}
\newcommand{\RunMidTauBadIncOneBHBOneLuminosityDistanceZeroPost}{\ensuremath{90^{+28}_{-16}}}
\newcommand{\RunMidTauBadIncOneBHBOneMassRatioZeroPost}{\ensuremath{0.5^{+0.5}_{-0.3}}}
\newcommand{\RunMidTauBadIncOneBHBOneMergerTimeOrInitialOrbitalFrequencyZeroPost}{\ensuremath{6.577194^{+0.000002}_{-0.000002}}}
\newcommand{\RunMidTauBadIncOneBHBOneMergerTimeZeroPost}{\ensuremath{9.2772^{+0.0008}_{-0.0015}}}
\newcommand{\RunMidTauBadIncOneBHBOnePolarizationZeroPost}{\ensuremath{0.005^{+1.6}_{-1.6}}}
\newcommand{\RunMidTauBadIncOneBHBOneRedshiftZeroPost}{\ensuremath{0.020^{+0.006}_{-0.003}}}
\newcommand{\RunMidTauBadIncOneBHBOneRedshiftedChirpMassZeroPost}{\ensuremath{26.069^{+0.002}_{-1\!\times\!10^{-3}}}}
\newcommand{\RunMidTauBadIncOneBHBOneRedshiftedMassOneZeroPost}{\ensuremath{45^{+49}_{-14}}}
\newcommand{\RunMidTauBadIncOneBHBOneRedshiftedMassTwoZeroPost}{\ensuremath{21^{+8}_{-9}}}
\newcommand{\RunMidTauBadIncOneBHBOneReducedMassDifferenceZeroPost}{\ensuremath{0.4^{+0.4}_{-0.3}}}
\newcommand{\RunMidTauBadIncOneBHBOnecosInclinationZeroPost}{\ensuremath{0.21^{+0.14}_{-0.10}}}
\newcommand{\RunMidTauBadIncOneBHBOnesinEclipticLatitudeZeroPost}{\ensuremath{-0.01^{+0.08}_{-0.07}}}
\newcommand{\RunMidTauBadIncOneBHBOneDimensionlessSpinOneZeroWidth}{\ensuremath{13.02}}
\newcommand{\RunMidTauBadIncOneBHBOneDimensionlessSpinTwoZeroWidth}{\ensuremath{3.5}}
\newcommand{\RunMidTauBadIncOneBHBOneEclipticLatitudeZeroWidth}{\ensuremath{-4.3}}
\newcommand{\RunMidTauBadIncOneBHBOneEclipticLongitudeZeroWidth}{\ensuremath{0.01}}
\newcommand{\RunMidTauBadIncOneBHBOneInclinationZeroWidth}{\ensuremath{0.2}}
\newcommand{\RunMidTauBadIncOneBHBOneInitialOrbitalPhaseZeroWidth}{\ensuremath{0.8}}
\newcommand{\RunMidTauBadIncOneBHBOneLuminosityDistanceZeroWidth}{\ensuremath{0.5}}
\newcommand{\RunMidTauBadIncOneBHBOneMassRatioZeroWidth}{\ensuremath{1.0}}
\newcommand{\RunMidTauBadIncOneBHBOneMergerTimeOrInitialOrbitalFrequencyZeroWidth}{\ensuremath{6\!\times\!10^{-7}}}
\newcommand{\RunMidTauBadIncOneBHBOneMergerTimeZeroWidth}{\ensuremath{2\!\times\!10^{-4}}}
\newcommand{\RunMidTauBadIncOneBHBOnePolarizationZeroWidth}{\ensuremath{544.700}}
\newcommand{\RunMidTauBadIncOneBHBOneRedshiftZeroWidth}{\ensuremath{0.5}}
\newcommand{\RunMidTauBadIncOneBHBOneRedshiftedChirpMassZeroWidth}{\ensuremath{1\!\times\!10^{-4}}}
\newcommand{\RunMidTauBadIncOneBHBOneRedshiftedMassOneZeroWidth}{\ensuremath{1.8}}
\newcommand{\RunMidTauBadIncOneBHBOneRedshiftedMassTwoZeroWidth}{\ensuremath{0.7}}
\newcommand{\RunMidTauBadIncOneBHBOneReducedMassDifferenceZeroWidth}{\ensuremath{6.0}}
\newcommand{\RunMidTauBadIncOneBHBOnecosInclinationZeroWidth}{\ensuremath{1.3}}
\newcommand{\RunMidTauBadIncOneBHBOnesinEclipticLatitudeZeroWidth}{\ensuremath{-4.3}}
\newcommand{\RunMidTauBadIncTwoSNR}{\ensuremath{11.17}}
\newcommand{\RunMidTauBadIncTwoBHBOneDimensionlessSpinOneZeroInj}{\ensuremath{0.1}}
\newcommand{\RunMidTauBadIncTwoBHBOneDimensionlessSpinTwoZeroInj}{\ensuremath{0.1}}
\newcommand{\RunMidTauBadIncTwoBHBOneEclipticLatitudeZeroInj}{\ensuremath{-0.09}}
\newcommand{\RunMidTauBadIncTwoBHBOneEclipticLongitudeZeroInj}{\ensuremath{2.4}}
\newcommand{\RunMidTauBadIncTwoBHBOneInclinationZeroInj}{\ensuremath{1.6}}
\newcommand{\RunMidTauBadIncTwoBHBOneInitialOrbitalPhaseZeroInj}{\ensuremath{4.6}}
\newcommand{\RunMidTauBadIncTwoBHBOneLuminosityDistanceZeroInj}{\ensuremath{64.86}}
\newcommand{\RunMidTauBadIncTwoBHBOneMassRatioZeroInj}{\ensuremath{1.0}}
\newcommand{\RunMidTauBadIncTwoBHBOneMergerTimeOrInitialOrbitalFrequencyZeroInj}{\ensuremath{7.2}}
\newcommand{\RunMidTauBadIncTwoBHBOneMergerTimeZeroInj}{\ensuremath{9.1}}
\newcommand{\RunMidTauBadIncTwoBHBOnePolarizationZeroInj}{\ensuremath{3.7}}
\newcommand{\RunMidTauBadIncTwoBHBOneRedshiftZeroInj}{\ensuremath{0.01}}
\newcommand{\RunMidTauBadIncTwoBHBOneRedshiftedChirpMassZeroInj}{\ensuremath{23.00}}
\newcommand{\RunMidTauBadIncTwoBHBOneRedshiftedMassOneZeroInj}{\ensuremath{26.44}}
\newcommand{\RunMidTauBadIncTwoBHBOneRedshiftedMassTwoZeroInj}{\ensuremath{26.40}}
\newcommand{\RunMidTauBadIncTwoBHBOneReducedMassDifferenceZeroInj}{\ensuremath{7\!\times\!10^{-4}}}
\newcommand{\RunMidTauBadIncTwoBHBOnecosInclinationZeroInj}{\ensuremath{-0.03}}
\newcommand{\RunMidTauBadIncTwoBHBOnesinEclipticLatitudeZeroInj}{\ensuremath{-0.09}}
\newcommand{\RunMidTauBadIncTwoBHBOneDimensionlessSpinOneZeroPost}{\ensuremath{-0.2^{+0.9}_{-0.7}}}
\newcommand{\RunMidTauBadIncTwoBHBOneDimensionlessSpinTwoZeroPost}{\ensuremath{0.01^{+0.87}_{-0.90}}}
\newcommand{\RunMidTauBadIncTwoBHBOneEclipticLatitudeZeroPost}{\ensuremath{-0.07^{+0.16}_{-0.05}}}
\newcommand{\RunMidTauBadIncTwoBHBOneEclipticLongitudeZeroPost}{\ensuremath{2.354^{+0.005}_{-0.005}}}
\newcommand{\RunMidTauBadIncTwoBHBOneInclinationZeroPost}{\ensuremath{1.60^{+0.09}_{-0.09}}}
\newcommand{\RunMidTauBadIncTwoBHBOneInitialOrbitalPhaseZeroPost}{\ensuremath{3.0^{+1.7}_{-1.7}}}
\newcommand{\RunMidTauBadIncTwoBHBOneLuminosityDistanceZeroPost}{\ensuremath{69^{+13}_{-9}}}
\newcommand{\RunMidTauBadIncTwoBHBOneMassRatioZeroPost}{\ensuremath{0.5^{+0.4}_{-0.4}}}
\newcommand{\RunMidTauBadIncTwoBHBOneMergerTimeOrInitialOrbitalFrequencyZeroPost}{\ensuremath{7.158828^{+0.000002}_{-0.000002}}}
\newcommand{\RunMidTauBadIncTwoBHBOneMergerTimeZeroPost}{\ensuremath{9.1195^{+0.0006}_{-0.0012}}}
\newcommand{\RunMidTauBadIncTwoBHBOnePolarizationZeroPost}{\ensuremath{0.6^{+1.6}_{-1.6}}}
\newcommand{\RunMidTauBadIncTwoBHBOneRedshiftZeroPost}{\ensuremath{0.015^{+0.003}_{-0.002}}}
\newcommand{\RunMidTauBadIncTwoBHBOneRedshiftedChirpMassZeroPost}{\ensuremath{22.9990^{+0.0019}_{-0.0010}}}
\newcommand{\RunMidTauBadIncTwoBHBOneRedshiftedMassOneZeroPost}{\ensuremath{38^{+42}_{-11}}}
\newcommand{\RunMidTauBadIncTwoBHBOneRedshiftedMassTwoZeroPost}{\ensuremath{19^{+7}_{-8}}}
\newcommand{\RunMidTauBadIncTwoBHBOneReducedMassDifferenceZeroPost}{\ensuremath{0.3^{+0.4}_{-0.3}}}
\newcommand{\RunMidTauBadIncTwoBHBOnecosInclinationZeroPost}{\ensuremath{-0.03^{+0.09}_{-0.09}}}
\newcommand{\RunMidTauBadIncTwoBHBOnesinEclipticLatitudeZeroPost}{\ensuremath{-0.07^{+0.16}_{-0.05}}}
\newcommand{\RunMidTauBadIncTwoBHBOneDimensionlessSpinOneZeroWidth}{\ensuremath{12.12}}
\newcommand{\RunMidTauBadIncTwoBHBOneDimensionlessSpinTwoZeroWidth}{\ensuremath{13.85}}
\newcommand{\RunMidTauBadIncTwoBHBOneEclipticLatitudeZeroWidth}{\ensuremath{-2.3}}
\newcommand{\RunMidTauBadIncTwoBHBOneEclipticLongitudeZeroWidth}{\ensuremath{0.005}}
\newcommand{\RunMidTauBadIncTwoBHBOneInclinationZeroWidth}{\ensuremath{0.1}}
\newcommand{\RunMidTauBadIncTwoBHBOneInitialOrbitalPhaseZeroWidth}{\ensuremath{0.7}}
\newcommand{\RunMidTauBadIncTwoBHBOneLuminosityDistanceZeroWidth}{\ensuremath{0.3}}
\newcommand{\RunMidTauBadIncTwoBHBOneMassRatioZeroWidth}{\ensuremath{0.8}}
\newcommand{\RunMidTauBadIncTwoBHBOneMergerTimeOrInitialOrbitalFrequencyZeroWidth}{\ensuremath{6\!\times\!10^{-7}}}
\newcommand{\RunMidTauBadIncTwoBHBOneMergerTimeZeroWidth}{\ensuremath{2\!\times\!10^{-4}}}
\newcommand{\RunMidTauBadIncTwoBHBOnePolarizationZeroWidth}{\ensuremath{0.9}}
\newcommand{\RunMidTauBadIncTwoBHBOneRedshiftZeroWidth}{\ensuremath{0.3}}
\newcommand{\RunMidTauBadIncTwoBHBOneRedshiftedChirpMassZeroWidth}{\ensuremath{1\!\times\!10^{-4}}}
\newcommand{\RunMidTauBadIncTwoBHBOneRedshiftedMassOneZeroWidth}{\ensuremath{2.0}}
\newcommand{\RunMidTauBadIncTwoBHBOneRedshiftedMassTwoZeroWidth}{\ensuremath{0.6}}
\newcommand{\RunMidTauBadIncTwoBHBOneReducedMassDifferenceZeroWidth}{\ensuremath{991.012}}
\newcommand{\RunMidTauBadIncTwoBHBOnecosInclinationZeroWidth}{\ensuremath{-6.8}}
\newcommand{\RunMidTauBadIncTwoBHBOnesinEclipticLatitudeZeroWidth}{\ensuremath{-2.3}}
\newcommand{\RunMidTauGoodIncOneSNR}{\ensuremath{8.53}}
\newcommand{\RunMidTauGoodIncOneBHBOneDimensionlessSpinOneZeroInj}{\ensuremath{0.2}}
\newcommand{\RunMidTauGoodIncOneBHBOneDimensionlessSpinTwoZeroInj}{\ensuremath{0.5}}
\newcommand{\RunMidTauGoodIncOneBHBOneEclipticLatitudeZeroInj}{\ensuremath{0.1}}
\newcommand{\RunMidTauGoodIncOneBHBOneEclipticLongitudeZeroInj}{\ensuremath{5.3}}
\newcommand{\RunMidTauGoodIncOneBHBOneInclinationZeroInj}{\ensuremath{2.7}}
\newcommand{\RunMidTauGoodIncOneBHBOneInitialOrbitalPhaseZeroInj}{\ensuremath{3.8}}
\newcommand{\RunMidTauGoodIncOneBHBOneLuminosityDistanceZeroInj}{\ensuremath{344.103}}
\newcommand{\RunMidTauGoodIncOneBHBOneMassRatioZeroInj}{\ensuremath{0.9}}
\newcommand{\RunMidTauGoodIncOneBHBOneMergerTimeOrInitialOrbitalFrequencyZeroInj}{\ensuremath{5.4}}
\newcommand{\RunMidTauGoodIncOneBHBOneMergerTimeZeroInj}{\ensuremath{9.7}}
\newcommand{\RunMidTauGoodIncOneBHBOnePolarizationZeroInj}{\ensuremath{1.1}}
\newcommand{\RunMidTauGoodIncOneBHBOneRedshiftZeroInj}{\ensuremath{0.07}}
\newcommand{\RunMidTauGoodIncOneBHBOneRedshiftedChirpMassZeroInj}{\ensuremath{34.60}}
\newcommand{\RunMidTauGoodIncOneBHBOneRedshiftedMassOneZeroInj}{\ensuremath{42.76}}
\newcommand{\RunMidTauGoodIncOneBHBOneRedshiftedMassTwoZeroInj}{\ensuremath{36.97}}
\newcommand{\RunMidTauGoodIncOneBHBOneReducedMassDifferenceZeroInj}{\ensuremath{0.07}}
\newcommand{\RunMidTauGoodIncOneBHBOnecosInclinationZeroInj}{\ensuremath{-0.9}}
\newcommand{\RunMidTauGoodIncOneBHBOnesinEclipticLatitudeZeroInj}{\ensuremath{0.1}}
\newcommand{\RunMidTauGoodIncOneBHBOneDimensionlessSpinOneZeroPost}{\ensuremath{-0.02^{+0.89}_{-0.85}}}
\newcommand{\RunMidTauGoodIncOneBHBOneDimensionlessSpinTwoZeroPost}{\ensuremath{0.04^{+0.86}_{-0.91}}}
\newcommand{\RunMidTauGoodIncOneBHBOneEclipticLatitudeZeroPost}{\ensuremath{0.08^{+0.07}_{-0.19}}}
\newcommand{\RunMidTauGoodIncOneBHBOneEclipticLongitudeZeroPost}{\ensuremath{5.260^{+0.010}_{-0.010}}}
\newcommand{\RunMidTauGoodIncOneBHBOneInclinationZeroPost}{\ensuremath{2.5^{+0.5}_{-0.4}}}
\newcommand{\RunMidTauGoodIncOneBHBOneInitialOrbitalPhaseZeroPost}{\ensuremath{3.7^{+2.0}_{-2.0}}}
\newcommand{\RunMidTauGoodIncOneBHBOneLuminosityDistanceZeroPost}{\ensuremath{327^{+133}_{-113}}}
\newcommand{\RunMidTauGoodIncOneBHBOneMassRatioZeroPost}{\ensuremath{0.4^{+0.5}_{-0.3}}}
\newcommand{\RunMidTauGoodIncOneBHBOneMergerTimeOrInitialOrbitalFrequencyZeroPost}{\ensuremath{5.415500^{+0.000002}_{-0.000003}}}
\newcommand{\RunMidTauGoodIncOneBHBOneMergerTimeZeroPost}{\ensuremath{9.7198^{+0.0010}_{-0.0022}}}
\newcommand{\RunMidTauGoodIncOneBHBOnePolarizationZeroPost}{\ensuremath{-0.5^{+2.0}_{-2.0}}}
\newcommand{\RunMidTauGoodIncOneBHBOneRedshiftZeroPost}{\ensuremath{0.07^{+0.03}_{-0.02}}}
\newcommand{\RunMidTauGoodIncOneBHBOneRedshiftedChirpMassZeroPost}{\ensuremath{34.596^{+0.005}_{-0.002}}}
\newcommand{\RunMidTauGoodIncOneBHBOneRedshiftedMassOneZeroPost}{\ensuremath{62^{+72}_{-20}}}
\newcommand{\RunMidTauGoodIncOneBHBOneRedshiftedMassTwoZeroPost}{\ensuremath{26^{+12}_{-12}}}
\newcommand{\RunMidTauGoodIncOneBHBOneReducedMassDifferenceZeroPost}{\ensuremath{0.4^{+0.4}_{-0.4}}}
\newcommand{\RunMidTauGoodIncOneBHBOnecosInclinationZeroPost}{\ensuremath{-0.8^{+0.3}_{-0.2}}}
\newcommand{\RunMidTauGoodIncOneBHBOnesinEclipticLatitudeZeroPost}{\ensuremath{0.08^{+0.07}_{-0.19}}}
\newcommand{\RunMidTauGoodIncOneBHBOneDimensionlessSpinOneZeroWidth}{\ensuremath{9.1}}
\newcommand{\RunMidTauGoodIncOneBHBOneDimensionlessSpinTwoZeroWidth}{\ensuremath{3.8}}
\newcommand{\RunMidTauGoodIncOneBHBOneEclipticLatitudeZeroWidth}{\ensuremath{2.4}}
\newcommand{\RunMidTauGoodIncOneBHBOneEclipticLongitudeZeroWidth}{\ensuremath{0.004}}
\newcommand{\RunMidTauGoodIncOneBHBOneInclinationZeroWidth}{\ensuremath{0.3}}
\newcommand{\RunMidTauGoodIncOneBHBOneInitialOrbitalPhaseZeroWidth}{\ensuremath{1.1}}
\newcommand{\RunMidTauGoodIncOneBHBOneLuminosityDistanceZeroWidth}{\ensuremath{0.7}}
\newcommand{\RunMidTauGoodIncOneBHBOneMassRatioZeroWidth}{\ensuremath{0.9}}
\newcommand{\RunMidTauGoodIncOneBHBOneMergerTimeOrInitialOrbitalFrequencyZeroWidth}{\ensuremath{9\!\times\!10^{-7}}}
\newcommand{\RunMidTauGoodIncOneBHBOneMergerTimeZeroWidth}{\ensuremath{3\!\times\!10^{-4}}}
\newcommand{\RunMidTauGoodIncOneBHBOnePolarizationZeroWidth}{\ensuremath{3.6}}
\newcommand{\RunMidTauGoodIncOneBHBOneRedshiftZeroWidth}{\ensuremath{0.7}}
\newcommand{\RunMidTauGoodIncOneBHBOneRedshiftedChirpMassZeroWidth}{\ensuremath{2\!\times\!10^{-4}}}
\newcommand{\RunMidTauGoodIncOneBHBOneRedshiftedMassOneZeroWidth}{\ensuremath{2.1}}
\newcommand{\RunMidTauGoodIncOneBHBOneRedshiftedMassTwoZeroWidth}{\ensuremath{0.6}}
\newcommand{\RunMidTauGoodIncOneBHBOneReducedMassDifferenceZeroWidth}{\ensuremath{10.47}}
\newcommand{\RunMidTauGoodIncOneBHBOnecosInclinationZeroWidth}{\ensuremath{-0.6}}
\newcommand{\RunMidTauGoodIncOneBHBOnesinEclipticLatitudeZeroWidth}{\ensuremath{2.4}}
\newcommand{\RunMidTauGoodIncTwoSNR}{\ensuremath{9.28}}
\newcommand{\RunMidTauGoodIncTwoBHBOneDimensionlessSpinOneZeroInj}{\ensuremath{0.1}}
\newcommand{\RunMidTauGoodIncTwoBHBOneDimensionlessSpinTwoZeroInj}{\ensuremath{0.4}}
\newcommand{\RunMidTauGoodIncTwoBHBOneEclipticLatitudeZeroInj}{\ensuremath{0.2}}
\newcommand{\RunMidTauGoodIncTwoBHBOneEclipticLongitudeZeroInj}{\ensuremath{0.4}}
\newcommand{\RunMidTauGoodIncTwoBHBOneInclinationZeroInj}{\ensuremath{2.7}}
\newcommand{\RunMidTauGoodIncTwoBHBOneInitialOrbitalPhaseZeroInj}{\ensuremath{2.9}}
\newcommand{\RunMidTauGoodIncTwoBHBOneLuminosityDistanceZeroInj}{\ensuremath{301.756}}
\newcommand{\RunMidTauGoodIncTwoBHBOneMassRatioZeroInj}{\ensuremath{0.6}}
\newcommand{\RunMidTauGoodIncTwoBHBOneMergerTimeOrInitialOrbitalFrequencyZeroInj}{\ensuremath{5.8}}
\newcommand{\RunMidTauGoodIncTwoBHBOneMergerTimeZeroInj}{\ensuremath{9.0}}
\newcommand{\RunMidTauGoodIncTwoBHBOnePolarizationZeroInj}{\ensuremath{0.6}}
\newcommand{\RunMidTauGoodIncTwoBHBOneRedshiftZeroInj}{\ensuremath{0.06}}
\newcommand{\RunMidTauGoodIncTwoBHBOneRedshiftedChirpMassZeroInj}{\ensuremath{32.31}}
\newcommand{\RunMidTauGoodIncTwoBHBOneRedshiftedMassOneZeroInj}{\ensuremath{49.86}}
\newcommand{\RunMidTauGoodIncTwoBHBOneRedshiftedMassTwoZeroInj}{\ensuremath{28.08}}
\newcommand{\RunMidTauGoodIncTwoBHBOneReducedMassDifferenceZeroInj}{\ensuremath{0.3}}
\newcommand{\RunMidTauGoodIncTwoBHBOnecosInclinationZeroInj}{\ensuremath{-0.9}}
\newcommand{\RunMidTauGoodIncTwoBHBOnesinEclipticLatitudeZeroInj}{\ensuremath{0.2}}
\newcommand{\RunMidTauGoodIncTwoBHBOneDimensionlessSpinOneZeroPost}{\ensuremath{-0.04^{+0.89}_{-0.81}}}
\newcommand{\RunMidTauGoodIncTwoBHBOneDimensionlessSpinTwoZeroPost}{\ensuremath{0.04^{+0.86}_{-0.92}}}
\newcommand{\RunMidTauGoodIncTwoBHBOneEclipticLatitudeZeroPost}{\ensuremath{0.22^{+0.02}_{-0.03}}}
\newcommand{\RunMidTauGoodIncTwoBHBOneEclipticLongitudeZeroPost}{\ensuremath{0.431^{+0.008}_{-0.008}}}
\newcommand{\RunMidTauGoodIncTwoBHBOneInclinationZeroPost}{\ensuremath{2.5^{+0.5}_{-0.4}}}
\newcommand{\RunMidTauGoodIncTwoBHBOneInitialOrbitalPhaseZeroPost}{\ensuremath{3.5^{+2.0}_{-2.0}}}
\newcommand{\RunMidTauGoodIncTwoBHBOneLuminosityDistanceZeroPost}{\ensuremath{281^{+104}_{-95}}}
\newcommand{\RunMidTauGoodIncTwoBHBOneMassRatioZeroPost}{\ensuremath{0.5^{+0.5}_{-0.3}}}
\newcommand{\RunMidTauGoodIncTwoBHBOneMergerTimeOrInitialOrbitalFrequencyZeroPost}{\ensuremath{5.826694^{+0.000002}_{-0.000003}}}
\newcommand{\RunMidTauGoodIncTwoBHBOneMergerTimeZeroPost}{\ensuremath{8.9610^{+0.0008}_{-0.0017}}}
\newcommand{\RunMidTauGoodIncTwoBHBOnePolarizationZeroPost}{\ensuremath{-0.4^{+2.0}_{-2.0}}}
\newcommand{\RunMidTauGoodIncTwoBHBOneRedshiftZeroPost}{\ensuremath{0.06^{+0.02}_{-0.02}}}
\newcommand{\RunMidTauGoodIncTwoBHBOneRedshiftedChirpMassZeroPost}{\ensuremath{32.311^{+0.004}_{-0.002}}}
\newcommand{\RunMidTauGoodIncTwoBHBOneRedshiftedMassOneZeroPost}{\ensuremath{56^{+65}_{-17}}}
\newcommand{\RunMidTauGoodIncTwoBHBOneRedshiftedMassTwoZeroPost}{\ensuremath{25^{+10}_{-12}}}
\newcommand{\RunMidTauGoodIncTwoBHBOneReducedMassDifferenceZeroPost}{\ensuremath{0.4^{+0.4}_{-0.3}}}
\newcommand{\RunMidTauGoodIncTwoBHBOnecosInclinationZeroPost}{\ensuremath{-0.8^{+0.3}_{-0.2}}}
\newcommand{\RunMidTauGoodIncTwoBHBOnesinEclipticLatitudeZeroPost}{\ensuremath{0.22^{+0.02}_{-0.03}}}
\newcommand{\RunMidTauGoodIncTwoBHBOneDimensionlessSpinOneZeroWidth}{\ensuremath{13.07}}
\newcommand{\RunMidTauGoodIncTwoBHBOneDimensionlessSpinTwoZeroWidth}{\ensuremath{4.8}}
\newcommand{\RunMidTauGoodIncTwoBHBOneEclipticLatitudeZeroWidth}{\ensuremath{0.2}}
\newcommand{\RunMidTauGoodIncTwoBHBOneEclipticLongitudeZeroWidth}{\ensuremath{0.04}}
\newcommand{\RunMidTauGoodIncTwoBHBOneInclinationZeroWidth}{\ensuremath{0.3}}
\newcommand{\RunMidTauGoodIncTwoBHBOneInitialOrbitalPhaseZeroWidth}{\ensuremath{1.4}}
\newcommand{\RunMidTauGoodIncTwoBHBOneLuminosityDistanceZeroWidth}{\ensuremath{0.7}}
\newcommand{\RunMidTauGoodIncTwoBHBOneMassRatioZeroWidth}{\ensuremath{1.4}}
\newcommand{\RunMidTauGoodIncTwoBHBOneMergerTimeOrInitialOrbitalFrequencyZeroWidth}{\ensuremath{8\!\times\!10^{-7}}}
\newcommand{\RunMidTauGoodIncTwoBHBOneMergerTimeZeroWidth}{\ensuremath{3\!\times\!10^{-4}}}
\newcommand{\RunMidTauGoodIncTwoBHBOnePolarizationZeroWidth}{\ensuremath{7.1}}
\newcommand{\RunMidTauGoodIncTwoBHBOneRedshiftZeroWidth}{\ensuremath{0.6}}
\newcommand{\RunMidTauGoodIncTwoBHBOneRedshiftedChirpMassZeroWidth}{\ensuremath{2\!\times\!10^{-4}}}
\newcommand{\RunMidTauGoodIncTwoBHBOneRedshiftedMassOneZeroWidth}{\ensuremath{1.7}}
\newcommand{\RunMidTauGoodIncTwoBHBOneRedshiftedMassTwoZeroWidth}{\ensuremath{0.8}}
\newcommand{\RunMidTauGoodIncTwoBHBOneReducedMassDifferenceZeroWidth}{\ensuremath{2.7}}
\newcommand{\RunMidTauGoodIncTwoBHBOnecosInclinationZeroWidth}{\ensuremath{-0.6}}
\newcommand{\RunMidTauGoodIncTwoBHBOnesinEclipticLatitudeZeroWidth}{\ensuremath{0.2}}
\newcommand{\RunIMBHPurpleSNR}{\ensuremath{12.11}}
\newcommand{\RunIMBHPurpleBHBOneDimensionlessSpinOneZeroAbsWidth}{\ensuremath{1.8}}
\newcommand{\RunIMBHPurpleBHBOneDimensionlessSpinTwoZeroAbsWidth}{\ensuremath{1.8}}
\newcommand{\RunIMBHPurpleBHBOneEclipticLatitudeZeroAbsWidth}{\ensuremath{0.3}}
\newcommand{\RunIMBHPurpleBHBOneEclipticLongitudeZeroAbsWidth}{\ensuremath{0.1}}
\newcommand{\RunIMBHPurpleBHBOneInclinationZeroAbsWidth}{\ensuremath{0.8}}
\newcommand{\RunIMBHPurpleBHBOneInitialOrbitalPhaseZeroAbsWidth}{\ensuremath{4.1}}
\newcommand{\RunIMBHPurpleBHBOneLuminosityDistanceZeroAbsWidth}{\ensuremath{493.070}}
\newcommand{\RunIMBHPurpleBHBOneMassRatioZeroAbsWidth}{\ensuremath{0.8}}
\newcommand{\RunIMBHPurpleBHBOneMergerTimeOrInitialOrbitalFrequencyZeroAbsWidth}{\ensuremath{0.000003}}
\newcommand{\RunIMBHPurpleBHBOneMergerTimeZeroAbsWidth}{\ensuremath{0.01}}
\newcommand{\RunIMBHPurpleBHBOnePolarizationZeroAbsWidth}{\ensuremath{4.1}}
\newcommand{\RunIMBHPurpleBHBOneRedshiftZeroAbsWidth}{\ensuremath{0.09}}
\newcommand{\RunIMBHPurpleBHBOneRedshiftedChirpMassZeroAbsWidth}{\ensuremath{0.8}}
\newcommand{\RunIMBHPurpleBHBOneRedshiftedMassOneZeroAbsWidth}{\ensuremath{4273.1031}}
\newcommand{\RunIMBHPurpleBHBOneRedshiftedMassTwoZeroAbsWidth}{\ensuremath{879.841}}
\newcommand{\RunIMBHPurpleBHBOneReducedMassDifferenceZeroAbsWidth}{\ensuremath{0.8}}
\newcommand{\RunIMBHPurpleBHBOnecosInclinationZeroAbsWidth}{\ensuremath{0.5}}
\newcommand{\RunIMBHPurpleBHBOnesinEclipticLatitudeZeroAbsWidth}{\ensuremath{0.3}}
\newcommand{\RunIMBHPurpleBHBOneDimensionlessSpinOneZeroInj}{\ensuremath{0.0}}
\newcommand{\RunIMBHPurpleBHBOneDimensionlessSpinTwoZeroInj}{\ensuremath{0.0}}
\newcommand{\RunIMBHPurpleBHBOneEclipticLatitudeZeroInj}{\ensuremath{0.0}}
\newcommand{\RunIMBHPurpleBHBOneEclipticLongitudeZeroInj}{\ensuremath{3.1}}
\newcommand{\RunIMBHPurpleBHBOneInclinationZeroInj}{\ensuremath{1.6}}
\newcommand{\RunIMBHPurpleBHBOneInitialOrbitalPhaseZeroInj}{\ensuremath{1}}
\newcommand{\RunIMBHPurpleBHBOneLuminosityDistanceZeroInj}{\ensuremath{1012.2935}}
\newcommand{\RunIMBHPurpleBHBOneMassRatioZeroInj}{\ensuremath{1.0}}
\newcommand{\RunIMBHPurpleBHBOneMergerTimeOrInitialOrbitalFrequencyZeroInj}{\ensuremath{0.6}}
\newcommand{\RunIMBHPurpleBHBOneMergerTimeZeroInj}{\ensuremath{5.1}}
\newcommand{\RunIMBHPurpleBHBOnePolarizationZeroInj}{\ensuremath{0.0}}
\newcommand{\RunIMBHPurpleBHBOneRedshiftZeroInj}{\ensuremath{0.2}}
\newcommand{\RunIMBHPurpleBHBOneRedshiftedChirpMassZeroInj}{\ensuremath{1200}}
\newcommand{\RunIMBHPurpleBHBOneRedshiftedMassOneZeroInj}{\ensuremath{1378.5759}}
\newcommand{\RunIMBHPurpleBHBOneRedshiftedMassTwoZeroInj}{\ensuremath{1378.3002}}
\newcommand{\RunIMBHPurpleBHBOneReducedMassDifferenceZeroInj}{\ensuremath{1\!\times\!10^{-4}}}
\newcommand{\RunIMBHPurpleBHBOnecosInclinationZeroInj}{\ensuremath{1}}
\newcommand{\RunIMBHPurpleBHBOnesinEclipticLatitudeZeroInj}{\ensuremath{0.0}}
\newcommand{\RunIMBHPurpleBHBOneDimensionlessSpinOneZeroPost}{\ensuremath{-0.05^{+0.92}_{-0.84}}}
\newcommand{\RunIMBHPurpleBHBOneDimensionlessSpinTwoZeroPost}{\ensuremath{0.0001^{+0.89}_{-0.89}}}
\newcommand{\RunIMBHPurpleBHBOneEclipticLatitudeZeroPost}{\ensuremath{-0.001^{+0.14}_{-0.14}}}
\newcommand{\RunIMBHPurpleBHBOneEclipticLongitudeZeroPost}{\ensuremath{3.14^{+0.06}_{-0.06}}}
\newcommand{\RunIMBHPurpleBHBOneInclinationZeroPost}{\ensuremath{0.7^{+0.4}_{-0.5}}}
\newcommand{\RunIMBHPurpleBHBOneInitialOrbitalPhaseZeroPost}{\ensuremath{2.9^{+2.0}_{-2.0}}}
\newcommand{\RunIMBHPurpleBHBOneLuminosityDistanceZeroPost}{\ensuremath{842^{+246}_{-246}}}
\newcommand{\RunIMBHPurpleBHBOneMassRatioZeroPost}{\ensuremath{0.4^{+0.4}_{-0.4}}}
\newcommand{\RunIMBHPurpleBHBOneMergerTimeOrInitialOrbitalFrequencyZeroPost}{\ensuremath{0.583997^{+0.000001}_{-0.000001}}}
\newcommand{\RunIMBHPurpleBHBOneMergerTimeZeroPost}{\ensuremath{9.999^{+0.004}_{-0.008}}}
\newcommand{\RunIMBHPurpleBHBOnePolarizationZeroPost}{\ensuremath{-0.3^{+2.0}_{-2.0}}}
\newcommand{\RunIMBHPurpleBHBOneRedshiftZeroPost}{\ensuremath{0.17^{+0.04}_{-0.05}}}
\newcommand{\RunIMBHPurpleBHBOneRedshiftedChirpMassZeroPost}{\ensuremath{1200.0723^{+0.5611}_{-0.2757}}}
\newcommand{\RunIMBHPurpleBHBOneRedshiftedMassOneZeroPost}{\ensuremath{2250^{+3500}_{-800}}}
\newcommand{\RunIMBHPurpleBHBOneRedshiftedMassTwoZeroPost}{\ensuremath{880^{+430}_{-450}}}
\newcommand{\RunIMBHPurpleBHBOneReducedMassDifferenceZeroPost}{\ensuremath{0.4^{+0.4}_{-0.4}}}
\newcommand{\RunIMBHPurpleBHBOnecosInclinationZeroPost}{\ensuremath{0.8^{+0.2}_{-0.3}}}
\newcommand{\RunIMBHPurpleBHBOnesinEclipticLatitudeZeroPost}{\ensuremath{-0.001^{+0.138}_{-0.142}}}
\newcommand{\RunIMBHPurpleBHBOneEclipticLongitudeZeroWidth}{\ensuremath{0.04}}
\newcommand{\RunIMBHPurpleBHBOneInclinationZeroWidth}{\ensuremath{0.5}}
\newcommand{\RunIMBHPurpleBHBOneInitialOrbitalPhaseZeroWidth}{\ensuremath{4.1}}
\newcommand{\RunIMBHPurpleBHBOneLuminosityDistanceZeroWidth}{\ensuremath{0.5}}
\newcommand{\RunIMBHPurpleBHBOneMassRatioZeroWidth}{\ensuremath{0.8}}
\newcommand{\RunIMBHPurpleBHBOneMergerTimeOrInitialOrbitalFrequencyZeroWidth}{\ensuremath{5\!\times\!10^{-6}}}
\newcommand{\RunIMBHPurpleBHBOneMergerTimeZeroWidth}{\ensuremath{2\!\times\!10^{-3}}}
\newcommand{\RunIMBHPurpleBHBOneRedshiftZeroWidth}{\ensuremath{0.5}}
\newcommand{\RunIMBHPurpleBHBOneRedshiftedChirpMassZeroWidth}{\ensuremath{7\!\times\!10^{-4}}}
\newcommand{\RunIMBHPurpleBHBOneRedshiftedMassOneZeroWidth}{\ensuremath{3.1}}
\newcommand{\RunIMBHPurpleBHBOneRedshiftedMassTwoZeroWidth}{\ensuremath{0.6}}
\newcommand{\RunIMBHPurpleBHBOneReducedMassDifferenceZeroWidth}{\ensuremath{8110.7581}}
\newcommand{\RunIMBHPurpleBHBOnecosInclinationZeroWidth}{\ensuremath{0.5}}
\newcommand{\RunIMBHTealSNR}{\ensuremath{11.03}}
\newcommand{\RunIMBHTealBHBOneDimensionlessSpinOneZeroAbsWidth}{\ensuremath{1.8}}
\newcommand{\RunIMBHTealBHBOneDimensionlessSpinTwoZeroAbsWidth}{\ensuremath{1.8}}
\newcommand{\RunIMBHTealBHBOneEclipticLatitudeZeroAbsWidth}{\ensuremath{0.3}}
\newcommand{\RunIMBHTealBHBOneEclipticLongitudeZeroAbsWidth}{\ensuremath{0.2}}
\newcommand{\RunIMBHTealBHBOneInclinationZeroAbsWidth}{\ensuremath{0.8}}
\newcommand{\RunIMBHTealBHBOneInitialOrbitalPhaseZeroAbsWidth}{\ensuremath{4.1}}
\newcommand{\RunIMBHTealBHBOneLuminosityDistanceZeroAbsWidth}{\ensuremath{239.617}}
\newcommand{\RunIMBHTealBHBOneMassRatioZeroAbsWidth}{\ensuremath{0.8}}
\newcommand{\RunIMBHTealBHBOneMergerTimeOrInitialOrbitalFrequencyZeroAbsWidth}{\ensuremath{0.000002}}
\newcommand{\RunIMBHTealBHBOneMergerTimeZeroAbsWidth}{\ensuremath{0.02}}
\newcommand{\RunIMBHTealBHBOnePolarizationZeroAbsWidth}{\ensuremath{4.1}}
\newcommand{\RunIMBHTealBHBOneRedshiftZeroAbsWidth}{\ensuremath{0.05}}
\newcommand{\RunIMBHTealBHBOneRedshiftedChirpMassZeroAbsWidth}{\ensuremath{0.7}}
\newcommand{\RunIMBHTealBHBOneRedshiftedMassOneZeroAbsWidth}{\ensuremath{4283.3750}}
\newcommand{\RunIMBHTealBHBOneRedshiftedMassTwoZeroAbsWidth}{\ensuremath{821.101}}
\newcommand{\RunIMBHTealBHBOneReducedMassDifferenceZeroAbsWidth}{\ensuremath{0.8}}
\newcommand{\RunIMBHTealBHBOnecosInclinationZeroAbsWidth}{\ensuremath{0.5}}
\newcommand{\RunIMBHTealBHBOnesinEclipticLatitudeZeroAbsWidth}{\ensuremath{0.3}}
\newcommand{\RunIMBHTealBHBOneDimensionlessSpinOneZeroInj}{\ensuremath{0.0}}
\newcommand{\RunIMBHTealBHBOneDimensionlessSpinTwoZeroInj}{\ensuremath{0.0}}
\newcommand{\RunIMBHTealBHBOneEclipticLatitudeZeroInj}{\ensuremath{0.0}}
\newcommand{\RunIMBHTealBHBOneEclipticLongitudeZeroInj}{\ensuremath{3.1}}
\newcommand{\RunIMBHTealBHBOneInclinationZeroInj}{\ensuremath{1.6}}
\newcommand{\RunIMBHTealBHBOneInitialOrbitalPhaseZeroInj}{\ensuremath{1}}
\newcommand{\RunIMBHTealBHBOneLuminosityDistanceZeroInj}{\ensuremath{475.822}}
\newcommand{\RunIMBHTealBHBOneMassRatioZeroInj}{\ensuremath{1.0}}
\newcommand{\RunIMBHTealBHBOneMergerTimeOrInitialOrbitalFrequencyZeroInj}{\ensuremath{0.5}}
\newcommand{\RunIMBHTealBHBOneMergerTimeZeroInj}{\ensuremath{17.30}}
\newcommand{\RunIMBHTealBHBOnePolarizationZeroInj}{\ensuremath{0.0}}
\newcommand{\RunIMBHTealBHBOneRedshiftZeroInj}{\ensuremath{0.10}}
\newcommand{\RunIMBHTealBHBOneRedshiftedChirpMassZeroInj}{\ensuremath{1100.0}}
\newcommand{\RunIMBHTealBHBOneRedshiftedMassOneZeroInj}{\ensuremath{1263.6946}}
\newcommand{\RunIMBHTealBHBOneRedshiftedMassTwoZeroInj}{\ensuremath{1263.4418}}
\newcommand{\RunIMBHTealBHBOneReducedMassDifferenceZeroInj}{\ensuremath{1\!\times\!10^{-4}}}
\newcommand{\RunIMBHTealBHBOnecosInclinationZeroInj}{\ensuremath{1}}
\newcommand{\RunIMBHTealBHBOnesinEclipticLatitudeZeroInj}{\ensuremath{0.0}}
\newcommand{\RunIMBHTealBHBOneDimensionlessSpinOneZeroPost}{\ensuremath{-0.02^{+0.91}_{-0.89}}}
\newcommand{\RunIMBHTealBHBOneDimensionlessSpinTwoZeroPost}{\ensuremath{0.004^{+0.89}_{-0.90}}}
\newcommand{\RunIMBHTealBHBOneEclipticLatitudeZeroPost}{\ensuremath{-0.00003^{+0.15277}_{-0.15471}}}
\newcommand{\RunIMBHTealBHBOneEclipticLongitudeZeroPost}{\ensuremath{3.14^{+0.08}_{-0.08}}}
\newcommand{\RunIMBHTealBHBOneInclinationZeroPost}{\ensuremath{0.7^{+0.4}_{-0.5}}}
\newcommand{\RunIMBHTealBHBOneInitialOrbitalPhaseZeroPost}{\ensuremath{2.9^{+2.0}_{-2.0}}}
\newcommand{\RunIMBHTealBHBOneLuminosityDistanceZeroPost}{\ensuremath{396^{+122}_{-117}}}
\newcommand{\RunIMBHTealBHBOneMassRatioZeroPost}{\ensuremath{0.5^{+0.4}_{-0.4}}}
\newcommand{\RunIMBHTealBHBOneMergerTimeOrInitialOrbitalFrequencyZeroPost}{\ensuremath{0.475491^{+0.000001}_{-0.000001}}}
\newcommand{\RunIMBHTealBHBOneMergerTimeZeroPost}{\ensuremath{19.998^{+0.006}_{-0.016}}}
\newcommand{\RunIMBHTealBHBOnePolarizationZeroPost}{\ensuremath{-0.3^{+2.1}_{-2.0}}}
\newcommand{\RunIMBHTealBHBOneRedshiftZeroPost}{\ensuremath{0.08^{+0.02}_{-0.02}}}
\newcommand{\RunIMBHTealBHBOneRedshiftedChirpMassZeroPost}{\ensuremath{1100.0675^{+0.5290}_{-0.1950}}}
\newcommand{\RunIMBHTealBHBOneRedshiftedMassOneZeroPost}{\ensuremath{2128^{+3500}_{-800}}}
\newcommand{\RunIMBHTealBHBOneRedshiftedMassTwoZeroPost}{\ensuremath{787^{+413}_{-407}}}
\newcommand{\RunIMBHTealBHBOneReducedMassDifferenceZeroPost}{\ensuremath{0.5^{+0.4}_{-0.4}}}
\newcommand{\RunIMBHTealBHBOnecosInclinationZeroPost}{\ensuremath{0.8^{+0.2}_{-0.3}}}
\newcommand{\RunIMBHTealBHBOnesinEclipticLatitudeZeroPost}{\ensuremath{-0.00003^{+0.15218}_{-0.15409}}}
\newcommand{\RunIMBHTealBHBOneEclipticLongitudeZeroWidth}{\ensuremath{0.05}}
\newcommand{\RunIMBHTealBHBOneInclinationZeroWidth}{\ensuremath{0.5}}
\newcommand{\RunIMBHTealBHBOneInitialOrbitalPhaseZeroWidth}{\ensuremath{4.1}}
\newcommand{\RunIMBHTealBHBOneLuminosityDistanceZeroWidth}{\ensuremath{0.5}}
\newcommand{\RunIMBHTealBHBOneMassRatioZeroWidth}{\ensuremath{0.8}}
\newcommand{\RunIMBHTealBHBOneMergerTimeOrInitialOrbitalFrequencyZeroWidth}{\ensuremath{5\!\times\!10^{-6}}}
\newcommand{\RunIMBHTealBHBOneMergerTimeZeroWidth}{\ensuremath{1\!\times\!10^{-3}}}
\newcommand{\RunIMBHTealBHBOneRedshiftZeroWidth}{\ensuremath{0.5}}
\newcommand{\RunIMBHTealBHBOneRedshiftedChirpMassZeroWidth}{\ensuremath{7\!\times\!10^{-4}}}
\newcommand{\RunIMBHTealBHBOneRedshiftedMassOneZeroWidth}{\ensuremath{3.4}}
\newcommand{\RunIMBHTealBHBOneRedshiftedMassTwoZeroWidth}{\ensuremath{0.6}}
\newcommand{\RunIMBHTealBHBOneReducedMassDifferenceZeroWidth}{\ensuremath{8220.7046}}
\newcommand{\RunIMBHTealBHBOnecosInclinationZeroWidth}{\ensuremath{0.5}}

\begin{sidewaystable}
\caption{Injection parameters for the selected sources as described in Section~\ref{subsec:param-estim}.
Where relevant, parameter values are specified in detector frame. For simplicity, we omit introducing new symbols for each of those. 
\label{tab:injections}
}
    \centering
    \small
    \addtolength{\tabcolsep}{-0.4em}
    \renewcommand{\arraystretch}{1.2} 
    \begin{tabular}{|r|cccc|cc|cc|cccc|ccc|c|}
        \hline
        Run 
        & $\mathcal{M}_{c} [M_\odot]$ 
        & $m_1 [M_\odot]$
        & $m_2 [M_\odot]$
        & $q$ 
        & $\chi_1$ 
        & $\chi_2$ 
        & $f^{0}_{\rm orb} [{\rm mHz}]$ 
        & $\tau_c[{\rm yr}]$
        & $z$
        & $d_L [{\rm Mpc]}$
        & $\sin b$
        & $l$
        & $\cos{\iota}$
        & $\psi$
        & $\phi_{\rm orb}$
        & SNR
        \\
        \hline
        High $\tau_c$
        & \RunHiTauGoodIncOneBHBOneRedshiftedChirpMassZeroInj
        & \RunHiTauGoodIncOneBHBOneRedshiftedMassOneZeroInj
        & \RunHiTauGoodIncOneBHBOneRedshiftedMassTwoZeroInj
        & \RunHiTauGoodIncOneBHBOneMassRatioZeroInj
        & \RunHiTauGoodIncOneBHBOneDimensionlessSpinOneZeroInj 
        & \RunHiTauGoodIncOneBHBOneDimensionlessSpinTwoZeroInj
        & \RunHiTauGoodIncOneBHBOneMergerTimeOrInitialOrbitalFrequencyZeroInj 
        & \RunHiTauGoodIncOneBHBOneMergerTimeZeroInj
        & \RunHiTauGoodIncOneBHBOneRedshiftZeroInj
        & \RunHiTauGoodIncOneBHBOneLuminosityDistanceZeroInj
        & \RunHiTauGoodIncOneBHBOnesinEclipticLatitudeZeroInj
        & \RunHiTauGoodIncOneBHBOneEclipticLongitudeZeroInj
        & \RunHiTauGoodIncOneBHBOnecosInclinationZeroInj
        & \RunHiTauGoodIncOneBHBOnePolarizationZeroInj
        & \RunHiTauGoodIncOneBHBOneInitialOrbitalPhaseZeroInj
        & 
        \RunHiTauGoodIncOneSNR
        \\
        & \RunHiTauGoodIncTwoBHBOneRedshiftedChirpMassZeroInj
        & \RunHiTauGoodIncTwoBHBOneRedshiftedMassOneZeroInj
        & \RunHiTauGoodIncTwoBHBOneRedshiftedMassTwoZeroInj
        & \RunHiTauGoodIncTwoBHBOneMassRatioZeroInj
        & \RunHiTauGoodIncTwoBHBOneDimensionlessSpinOneZeroInj 
        & \RunHiTauGoodIncTwoBHBOneDimensionlessSpinTwoZeroInj
        & \RunHiTauGoodIncTwoBHBOneMergerTimeOrInitialOrbitalFrequencyZeroInj 
        & \RunHiTauGoodIncTwoBHBOneMergerTimeZeroInj
        & \RunHiTauGoodIncTwoBHBOneRedshiftZeroInj
        & \RunHiTauGoodIncTwoBHBOneLuminosityDistanceZeroInj
        & \RunHiTauGoodIncTwoBHBOnesinEclipticLatitudeZeroInj
        & \RunHiTauGoodIncTwoBHBOneEclipticLongitudeZeroInj
        & \RunHiTauGoodIncTwoBHBOnecosInclinationZeroInj
        & \RunHiTauGoodIncTwoBHBOnePolarizationZeroInj
        & \RunHiTauGoodIncTwoBHBOneInitialOrbitalPhaseZeroInj
        & 
        \RunHiTauGoodIncTwoSNR
        \\
        & \RunHiTauBadIncOneBHBOneRedshiftedChirpMassZeroInj
        & \RunHiTauBadIncOneBHBOneRedshiftedMassOneZeroInj
        & \RunHiTauBadIncOneBHBOneRedshiftedMassTwoZeroInj
        & \RunHiTauBadIncOneBHBOneMassRatioZeroInj
        & \RunHiTauBadIncOneBHBOneDimensionlessSpinOneZeroInj 
        & \RunHiTauBadIncOneBHBOneDimensionlessSpinTwoZeroInj
        & \RunHiTauBadIncOneBHBOneMergerTimeOrInitialOrbitalFrequencyZeroInj 
        & \RunHiTauBadIncOneBHBOneMergerTimeZeroInj
        & \RunHiTauBadIncOneBHBOneRedshiftZeroInj
        & \RunHiTauBadIncOneBHBOneLuminosityDistanceZeroInj
        & \RunHiTauBadIncOneBHBOnesinEclipticLatitudeZeroInj
        & \RunHiTauBadIncOneBHBOneEclipticLongitudeZeroInj
        & \RunHiTauBadIncOneBHBOnecosInclinationZeroInj
        & \RunHiTauBadIncOneBHBOnePolarizationZeroInj
        & \RunHiTauBadIncOneBHBOneInitialOrbitalPhaseZeroInj
        & 
        \RunHiTauBadIncOneSNR
        \\
        & \RunHiTauBadIncTwoBHBOneRedshiftedChirpMassZeroInj
        & \RunHiTauBadIncTwoBHBOneRedshiftedMassOneZeroInj
        & \RunHiTauBadIncTwoBHBOneRedshiftedMassTwoZeroInj
        & \RunHiTauBadIncTwoBHBOneMassRatioZeroInj
        & \RunHiTauBadIncTwoBHBOneDimensionlessSpinOneZeroInj 
        & \RunHiTauBadIncTwoBHBOneDimensionlessSpinTwoZeroInj
        & \RunHiTauBadIncTwoBHBOneMergerTimeOrInitialOrbitalFrequencyZeroInj 
        & \RunHiTauBadIncTwoBHBOneMergerTimeZeroInj
        & \RunHiTauBadIncTwoBHBOneRedshiftZeroInj
        & \RunHiTauBadIncTwoBHBOneLuminosityDistanceZeroInj
        & \RunHiTauBadIncTwoBHBOnesinEclipticLatitudeZeroInj
        & \RunHiTauBadIncTwoBHBOneEclipticLongitudeZeroInj
        & \RunHiTauBadIncTwoBHBOnecosInclinationZeroInj
        & \RunHiTauBadIncTwoBHBOnePolarizationZeroInj
        & \RunHiTauBadIncTwoBHBOneInitialOrbitalPhaseZeroInj
        & \RunHiTauBadIncTwoSNR
        \\
        \hline
        Mid $\tau_c$
        & \RunMidTauGoodIncOneBHBOneRedshiftedChirpMassZeroInj
        & \RunMidTauGoodIncOneBHBOneRedshiftedMassOneZeroInj
        & \RunMidTauGoodIncOneBHBOneRedshiftedMassTwoZeroInj
        & \RunMidTauGoodIncOneBHBOneMassRatioZeroInj
        & \RunMidTauGoodIncOneBHBOneDimensionlessSpinOneZeroInj 
        & \RunMidTauGoodIncOneBHBOneDimensionlessSpinTwoZeroInj
        & \RunMidTauGoodIncOneBHBOneMergerTimeOrInitialOrbitalFrequencyZeroInj 
        & \RunMidTauGoodIncOneBHBOneMergerTimeZeroInj
        & \RunMidTauGoodIncOneBHBOneRedshiftZeroInj
        & \RunMidTauGoodIncOneBHBOneLuminosityDistanceZeroInj
        & \RunMidTauGoodIncOneBHBOnesinEclipticLatitudeZeroInj
        & \RunMidTauGoodIncOneBHBOneEclipticLongitudeZeroInj
        & \RunMidTauGoodIncOneBHBOnecosInclinationZeroInj
        & \RunMidTauGoodIncOneBHBOnePolarizationZeroInj
        & \RunMidTauGoodIncOneBHBOneInitialOrbitalPhaseZeroInj
        & 
        \RunMidTauGoodIncOneSNR
        \\
        & \RunMidTauGoodIncTwoBHBOneRedshiftedChirpMassZeroInj
        & \RunMidTauGoodIncTwoBHBOneRedshiftedMassOneZeroInj
        & \RunMidTauGoodIncTwoBHBOneRedshiftedMassTwoZeroInj
        & \RunMidTauGoodIncTwoBHBOneMassRatioZeroInj
        & \RunMidTauGoodIncTwoBHBOneDimensionlessSpinOneZeroInj 
        & \RunMidTauGoodIncTwoBHBOneDimensionlessSpinTwoZeroInj
        & \RunMidTauGoodIncTwoBHBOneMergerTimeOrInitialOrbitalFrequencyZeroInj 
        & \RunMidTauGoodIncTwoBHBOneMergerTimeZeroInj
        & \RunMidTauGoodIncTwoBHBOneRedshiftZeroInj
        & \RunMidTauGoodIncTwoBHBOneLuminosityDistanceZeroInj
        & \RunMidTauGoodIncTwoBHBOnesinEclipticLatitudeZeroInj
        & \RunMidTauGoodIncTwoBHBOneEclipticLongitudeZeroInj
        & \RunMidTauGoodIncTwoBHBOnecosInclinationZeroInj
        & \RunMidTauGoodIncTwoBHBOnePolarizationZeroInj
        & \RunMidTauGoodIncTwoBHBOneInitialOrbitalPhaseZeroInj
        & 
        \RunMidTauGoodIncTwoSNR
        \\
        & \RunMidTauBadIncOneBHBOneRedshiftedChirpMassZeroInj
        & \RunMidTauBadIncOneBHBOneRedshiftedMassOneZeroInj
        & \RunMidTauBadIncOneBHBOneRedshiftedMassTwoZeroInj
        & \RunMidTauBadIncOneBHBOneMassRatioZeroInj
        & \RunMidTauBadIncOneBHBOneDimensionlessSpinOneZeroInj 
        & \RunMidTauBadIncOneBHBOneDimensionlessSpinTwoZeroInj
        & \RunMidTauBadIncOneBHBOneMergerTimeOrInitialOrbitalFrequencyZeroInj 
        & \RunMidTauBadIncOneBHBOneMergerTimeZeroInj
        & \RunMidTauBadIncOneBHBOneRedshiftZeroInj
        & \RunMidTauBadIncOneBHBOneLuminosityDistanceZeroInj
        & \RunMidTauBadIncOneBHBOnesinEclipticLatitudeZeroInj
        & \RunMidTauBadIncOneBHBOneEclipticLongitudeZeroInj
        & \RunMidTauBadIncOneBHBOnecosInclinationZeroInj
        & \RunMidTauBadIncOneBHBOnePolarizationZeroInj
        & \RunMidTauBadIncOneBHBOneInitialOrbitalPhaseZeroInj
        & 
        \RunMidTauBadIncOneSNR
        \\
        & \RunMidTauBadIncTwoBHBOneRedshiftedChirpMassZeroInj
        & \RunMidTauBadIncTwoBHBOneRedshiftedMassOneZeroInj
        & \RunMidTauBadIncTwoBHBOneRedshiftedMassTwoZeroInj
        & \RunMidTauBadIncTwoBHBOneMassRatioZeroInj
        & \RunMidTauBadIncTwoBHBOneDimensionlessSpinOneZeroInj 
        & \RunMidTauBadIncTwoBHBOneDimensionlessSpinTwoZeroInj
        & \RunMidTauBadIncTwoBHBOneMergerTimeOrInitialOrbitalFrequencyZeroInj 
        & \RunMidTauBadIncTwoBHBOneMergerTimeZeroInj 
        & \RunMidTauBadIncTwoBHBOneRedshiftZeroInj 
        & \RunMidTauBadIncTwoBHBOneLuminosityDistanceZeroInj
        & \RunMidTauBadIncTwoBHBOnesinEclipticLatitudeZeroInj
        & \RunMidTauBadIncTwoBHBOneEclipticLongitudeZeroInj
        & \RunMidTauBadIncTwoBHBOnecosInclinationZeroInj
        & \RunMidTauBadIncTwoBHBOnePolarizationZeroInj
        & \RunMidTauBadIncTwoBHBOneInitialOrbitalPhaseZeroInj
        & 
        \RunMidTauBadIncTwoSNR
        \\
        \hline
        Low $\tau_c$
        & \RunLowTauGoodIncOneBHBOneRedshiftedChirpMassZeroInj
        & \RunLowTauGoodIncOneBHBOneRedshiftedMassOneZeroInj
        & \RunLowTauGoodIncOneBHBOneRedshiftedMassTwoZeroInj
        & \RunLowTauGoodIncOneBHBOneMassRatioZeroInj
        & \RunLowTauGoodIncOneBHBOneDimensionlessSpinOneZeroInj 
        & \RunLowTauGoodIncOneBHBOneDimensionlessSpinTwoZeroInj
        & \RunLowTauGoodIncOneBHBOneMergerTimeOrInitialOrbitalFrequencyZeroInj 
        & \RunLowTauGoodIncOneBHBOneMergerTimeZeroInj 
        & \RunLowTauGoodIncOneBHBOneRedshiftZeroInj 
        & \RunLowTauGoodIncOneBHBOneLuminosityDistanceZeroInj
        & \RunLowTauGoodIncOneBHBOnesinEclipticLatitudeZeroInj
        & \RunLowTauGoodIncOneBHBOneEclipticLongitudeZeroInj
        & \RunLowTauGoodIncOneBHBOnecosInclinationZeroInj
        & \RunLowTauGoodIncOneBHBOnePolarizationZeroInj
        & \RunLowTauGoodIncOneBHBOneInitialOrbitalPhaseZeroInj
        & 
        \RunLowTauGoodIncOneSNR
        \\
        & \RunLowTauGoodIncTwoBHBOneRedshiftedChirpMassZeroInj
        & \RunLowTauGoodIncTwoBHBOneRedshiftedMassOneZeroInj
        & \RunLowTauGoodIncTwoBHBOneRedshiftedMassTwoZeroInj
        & \RunLowTauGoodIncTwoBHBOneMassRatioZeroInj
        & \RunLowTauGoodIncTwoBHBOneDimensionlessSpinOneZeroInj 
        & \RunLowTauGoodIncTwoBHBOneDimensionlessSpinTwoZeroInj
        & \RunLowTauGoodIncTwoBHBOneMergerTimeOrInitialOrbitalFrequencyZeroInj 
        & \RunLowTauGoodIncTwoBHBOneMergerTimeZeroInj 
        & \RunLowTauGoodIncTwoBHBOneRedshiftZeroInj 
        & \RunLowTauGoodIncTwoBHBOneLuminosityDistanceZeroInj
        & \RunLowTauGoodIncTwoBHBOnesinEclipticLatitudeZeroInj
        & \RunLowTauGoodIncTwoBHBOneEclipticLongitudeZeroInj
        & \RunLowTauGoodIncTwoBHBOnecosInclinationZeroInj
        & \RunLowTauGoodIncTwoBHBOnePolarizationZeroInj
        & \RunLowTauGoodIncTwoBHBOneInitialOrbitalPhaseZeroInj
        & 
        \RunLowTauGoodIncTwoSNR
        \\
        & \RunLowTauBadIncOneBHBOneRedshiftedChirpMassZeroInj
        & \RunLowTauBadIncOneBHBOneRedshiftedMassOneZeroInj
        & \RunLowTauBadIncOneBHBOneRedshiftedMassTwoZeroInj
        & \RunLowTauBadIncOneBHBOneMassRatioZeroInj
        & \RunLowTauBadIncOneBHBOneDimensionlessSpinOneZeroInj 
        & \RunLowTauBadIncOneBHBOneDimensionlessSpinTwoZeroInj
        & \RunLowTauBadIncOneBHBOneMergerTimeOrInitialOrbitalFrequencyZeroInj
        & \RunLowTauBadIncOneBHBOneMergerTimeZeroInj
        & \RunLowTauBadIncOneBHBOneRedshiftZeroInj
        & \RunLowTauBadIncOneBHBOneLuminosityDistanceZeroInj
        & \RunLowTauBadIncOneBHBOnesinEclipticLatitudeZeroInj
        & \RunLowTauBadIncOneBHBOneEclipticLongitudeZeroInj
        & \RunLowTauBadIncOneBHBOnecosInclinationZeroInj
        & \RunLowTauBadIncOneBHBOnePolarizationZeroInj
        & \RunLowTauBadIncOneBHBOneInitialOrbitalPhaseZeroInj
        &
        \RunLowTauBadIncOneSNR
        \\
        & \RunLowTauBadIncTwoBHBOneRedshiftedChirpMassZeroInj
        & \RunLowTauBadIncTwoBHBOneRedshiftedMassOneZeroInj
        & \RunLowTauBadIncTwoBHBOneRedshiftedMassTwoZeroInj
        & \RunLowTauBadIncTwoBHBOneMassRatioZeroInj
        & \RunLowTauBadIncTwoBHBOneDimensionlessSpinOneZeroInj 
        & \RunLowTauBadIncTwoBHBOneDimensionlessSpinTwoZeroInj
        & \RunLowTauBadIncTwoBHBOneMergerTimeOrInitialOrbitalFrequencyZeroInj 
        & \RunLowTauBadIncTwoBHBOneMergerTimeZeroInj
        & \RunLowTauBadIncTwoBHBOneRedshiftZeroInj 
        & \RunLowTauBadIncTwoBHBOneLuminosityDistanceZeroInj
        & \RunLowTauBadIncTwoBHBOnesinEclipticLatitudeZeroInj
        & \RunLowTauBadIncTwoBHBOneEclipticLongitudeZeroInj
        & \RunLowTauBadIncTwoBHBOnecosInclinationZeroInj
        & \RunLowTauBadIncTwoBHBOnePolarizationZeroInj
        & \RunLowTauBadIncTwoBHBOneInitialOrbitalPhaseZeroInj
        &
        \RunLowTauBadIncTwoSNR
        \\
        \hline
        \textcolor{teal}{\large \ding{73}}
        & \RunIMBHTealBHBOneRedshiftedChirpMassZeroInj
        & \RunIMBHTealBHBOneRedshiftedMassOneZeroInj
        & \RunIMBHTealBHBOneRedshiftedMassTwoZeroInj
        & \RunIMBHTealBHBOneMassRatioZeroInj
        & \RunIMBHTealBHBOneDimensionlessSpinOneZeroInj 
        & \RunIMBHTealBHBOneDimensionlessSpinTwoZeroInj
        & \RunIMBHTealBHBOneMergerTimeOrInitialOrbitalFrequencyZeroInj
        & \RunIMBHTealBHBOneMergerTimeZeroInj
        & \RunIMBHTealBHBOneRedshiftZeroInj
        & \RunIMBHTealBHBOneLuminosityDistanceZeroInj
        & \RunIMBHTealBHBOnesinEclipticLatitudeZeroInj
        & \RunIMBHTealBHBOneEclipticLongitudeZeroInj
        & \RunIMBHTealBHBOnecosInclinationZeroInj
        & \RunIMBHTealBHBOnePolarizationZeroInj
        & \RunIMBHTealBHBOneInitialOrbitalPhaseZeroInj
        & \RunIMBHTealSNR\\
        \textcolor{magenta}{\large \ding{73}}       
        & \RunIMBHPurpleBHBOneRedshiftedChirpMassZeroInj
        & \RunIMBHPurpleBHBOneRedshiftedMassOneZeroInj
        & \RunIMBHPurpleBHBOneRedshiftedMassTwoZeroInj
        & \RunIMBHPurpleBHBOneMassRatioZeroInj
        & \RunIMBHPurpleBHBOneDimensionlessSpinOneZeroInj 
        & \RunIMBHPurpleBHBOneDimensionlessSpinTwoZeroInj
        & \RunIMBHPurpleBHBOneMergerTimeOrInitialOrbitalFrequencyZeroInj 
        & \RunIMBHPurpleBHBOneMergerTimeZeroInj 
        & \RunIMBHPurpleBHBOneRedshiftZeroInj 
        & \RunIMBHPurpleBHBOneLuminosityDistanceZeroInj
        & \RunIMBHPurpleBHBOnesinEclipticLatitudeZeroInj
        & \RunIMBHPurpleBHBOneEclipticLongitudeZeroInj
        & \RunIMBHPurpleBHBOnecosInclinationZeroInj
        & \RunIMBHPurpleBHBOnePolarizationZeroInj
        & \RunIMBHPurpleBHBOneInitialOrbitalPhaseZeroInj
        & \RunIMBHPurpleSNR\\
        \textcolor{black}{\large \ding{73}}
        & \RunIMBHBlackBHBOneRedshiftedChirpMassZeroInj
        & \RunIMBHBlackBHBOneRedshiftedMassOneZeroInj
        & \RunIMBHBlackBHBOneRedshiftedMassTwoZeroInj
        & \RunIMBHBlackBHBOneMassRatioZeroInj
        & \RunIMBHBlackBHBOneDimensionlessSpinOneZeroInj 
        & \RunIMBHBlackBHBOneDimensionlessSpinTwoZeroInj
        & \RunIMBHBlackBHBOneMergerTimeOrInitialOrbitalFrequencyZeroInj
        & \RunIMBHBlackBHBOneMergerTimeZeroInj
        & \RunIMBHBlackBHBOneRedshiftZeroInj
        & \RunIMBHBlackBHBOneLuminosityDistanceZeroInj
        & \RunIMBHBlackBHBOnesinEclipticLatitudeZeroInj
        & \RunIMBHBlackBHBOneEclipticLongitudeZeroInj
        & \RunIMBHBlackBHBOnecosInclinationZeroInj
        & \RunIMBHBlackBHBOnePolarizationZeroInj
        & \RunIMBHBlackBHBOneInitialOrbitalPhaseZeroInj
        & \RunIMBHBlackSNR\\
        \hline
    \end{tabular}
\end{sidewaystable}

\begin{table}
\caption{Recovered intrinsic parameters for selected sources as discussed in Section~\ref{subsec:param-estim}. Injection parameters are listed in~\cref{tab:injections}. Quoted fractional errors are computed as the ratio between the posterior $90\%$ confidence intervals and corresponding median. 
Point estimates $m^{+\Delta u}_{-\Delta l}$ denote posterior median, upper and lower widths corresponding to $90\%$ confidence intervals.
\label{tab:pe-intrinsic}
}
\centering
    \addtolength{\tabcolsep}{-0.4em}
    \renewcommand{\arraystretch}{1.3} 
    \begin{tabular}{|c|crrr|rr|cc|}
        \hline
        Run 
        & $\Delta \mathcal{M}_c /\mathcal{M}_c $ 
        & $m_1 [M_\odot]$
        & $m_2 [M_\odot]$
        & $q$\,\,\,\,\,\,\,\,\,
        & $\chi_1$\,\,\,\,\,\, 
        & $\chi_2$ \,\,\,\,\,\,
        & $\Delta f^0_{\rm orb}/f^0_{\rm orb} $ 
        & $\Delta \tau_c / \tau_c$
        \\
        \hline
        High $\tau_c$
        & \RunHiTauGoodIncOneBHBOneRedshiftedChirpMassZeroWidth
        & \RunHiTauGoodIncOneBHBOneRedshiftedMassOneZeroPost
        & \RunHiTauGoodIncOneBHBOneRedshiftedMassTwoZeroPost
        & \RunHiTauGoodIncOneBHBOneMassRatioZeroPost
        & \RunHiTauGoodIncOneBHBOneDimensionlessSpinOneZeroPost 
        & \RunHiTauGoodIncOneBHBOneDimensionlessSpinTwoZeroPost
        & \RunHiTauGoodIncOneBHBOneMergerTimeOrInitialOrbitalFrequencyZeroWidth 
        & \RunHiTauGoodIncOneBHBOneMergerTimeZeroWidth
        \\
        & \RunHiTauGoodIncTwoBHBOneRedshiftedChirpMassZeroWidth
        & \RunHiTauGoodIncTwoBHBOneRedshiftedMassOneZeroPost
        & \RunHiTauGoodIncTwoBHBOneRedshiftedMassTwoZeroPost
        & \RunHiTauGoodIncTwoBHBOneMassRatioZeroPost
        & \RunHiTauGoodIncTwoBHBOneDimensionlessSpinOneZeroPost 
        & \RunHiTauGoodIncTwoBHBOneDimensionlessSpinTwoZeroPost
        & \RunHiTauGoodIncTwoBHBOneMergerTimeOrInitialOrbitalFrequencyZeroWidth 
        & \RunHiTauGoodIncTwoBHBOneMergerTimeZeroWidth
        \\
        & \RunHiTauBadIncOneBHBOneRedshiftedChirpMassZeroWidth
        & \RunHiTauBadIncOneBHBOneRedshiftedMassOneZeroPost
        & \RunHiTauBadIncOneBHBOneRedshiftedMassTwoZeroPost
        & \RunHiTauBadIncOneBHBOneMassRatioZeroPost
        & \RunHiTauBadIncOneBHBOneDimensionlessSpinOneZeroPost 
        & \RunHiTauBadIncOneBHBOneDimensionlessSpinTwoZeroPost
        & \RunHiTauBadIncOneBHBOneMergerTimeOrInitialOrbitalFrequencyZeroWidth 
        & \RunHiTauBadIncOneBHBOneMergerTimeZeroWidth
        \\
        & \RunHiTauBadIncTwoBHBOneRedshiftedChirpMassZeroWidth
        & \RunHiTauBadIncTwoBHBOneRedshiftedMassOneZeroPost
        & \RunHiTauBadIncTwoBHBOneRedshiftedMassTwoZeroPost
        & \RunHiTauBadIncTwoBHBOneMassRatioZeroPost
        & \RunHiTauBadIncTwoBHBOneDimensionlessSpinOneZeroPost 
        & \RunHiTauBadIncTwoBHBOneDimensionlessSpinTwoZeroPost
        & \RunHiTauBadIncTwoBHBOneMergerTimeOrInitialOrbitalFrequencyZeroWidth 
        & \RunHiTauBadIncTwoBHBOneMergerTimeZeroWidth
        \\
        \hline
        Mid $\tau_c$
        & \RunMidTauGoodIncOneBHBOneRedshiftedChirpMassZeroWidth
        & \RunMidTauGoodIncOneBHBOneRedshiftedMassOneZeroPost
        & \RunMidTauGoodIncOneBHBOneRedshiftedMassTwoZeroPost
        & \RunMidTauGoodIncOneBHBOneMassRatioZeroPost
        & \RunMidTauGoodIncOneBHBOneDimensionlessSpinOneZeroPost 
        & \RunMidTauGoodIncOneBHBOneDimensionlessSpinTwoZeroPost
        & \RunMidTauGoodIncOneBHBOneMergerTimeOrInitialOrbitalFrequencyZeroWidth 
        & \RunMidTauGoodIncOneBHBOneMergerTimeZeroWidth
        \\
        & \RunMidTauGoodIncTwoBHBOneRedshiftedChirpMassZeroWidth
        & \RunMidTauGoodIncTwoBHBOneRedshiftedMassOneZeroPost
        & \RunMidTauGoodIncTwoBHBOneRedshiftedMassTwoZeroPost
        & \RunMidTauGoodIncTwoBHBOneMassRatioZeroPost
        & \RunMidTauGoodIncTwoBHBOneDimensionlessSpinOneZeroPost 
        & \RunMidTauGoodIncTwoBHBOneDimensionlessSpinTwoZeroPost
        & \RunMidTauGoodIncTwoBHBOneMergerTimeOrInitialOrbitalFrequencyZeroWidth 
        & \RunMidTauGoodIncTwoBHBOneMergerTimeZeroWidth
        \\
        & \RunMidTauBadIncOneBHBOneRedshiftedChirpMassZeroWidth
        & \RunMidTauBadIncOneBHBOneRedshiftedMassOneZeroPost
        & \RunMidTauBadIncOneBHBOneRedshiftedMassTwoZeroPost
        & \RunMidTauBadIncOneBHBOneMassRatioZeroPost
        & \RunMidTauBadIncOneBHBOneDimensionlessSpinOneZeroPost 
        & \RunMidTauBadIncOneBHBOneDimensionlessSpinTwoZeroPost
        & \RunMidTauBadIncOneBHBOneMergerTimeOrInitialOrbitalFrequencyZeroWidth 
        & \RunMidTauBadIncOneBHBOneMergerTimeZeroWidth
        \\
        & \RunMidTauBadIncTwoBHBOneRedshiftedChirpMassZeroWidth
        & \RunMidTauBadIncTwoBHBOneRedshiftedMassOneZeroPost
        & \RunMidTauBadIncTwoBHBOneRedshiftedMassTwoZeroPost
        & \RunMidTauBadIncTwoBHBOneMassRatioZeroPost
        & \RunMidTauBadIncTwoBHBOneDimensionlessSpinOneZeroPost 
        & \RunMidTauBadIncTwoBHBOneDimensionlessSpinTwoZeroPost
        & \RunMidTauBadIncTwoBHBOneMergerTimeOrInitialOrbitalFrequencyZeroWidth 
        & \RunMidTauBadIncTwoBHBOneMergerTimeZeroWidth
        \\
        \hline
        Low $\tau_c$
        & \RunLowTauGoodIncOneBHBOneRedshiftedChirpMassZeroWidth
        & \RunLowTauGoodIncOneBHBOneRedshiftedMassOneZeroPost
        & \RunLowTauGoodIncOneBHBOneRedshiftedMassTwoZeroPost
        & \RunLowTauGoodIncOneBHBOneMassRatioZeroPost
        & \RunLowTauGoodIncOneBHBOneDimensionlessSpinOneZeroPost 
        & \RunLowTauGoodIncOneBHBOneDimensionlessSpinTwoZeroPost
        & \RunLowTauGoodIncOneBHBOneMergerTimeOrInitialOrbitalFrequencyZeroWidth 
        & \RunLowTauGoodIncOneBHBOneMergerTimeZeroWidth
        \\
        & \RunLowTauGoodIncTwoBHBOneRedshiftedChirpMassZeroWidth
        & \RunLowTauGoodIncTwoBHBOneRedshiftedMassOneZeroPost
        & \RunLowTauGoodIncTwoBHBOneRedshiftedMassTwoZeroPost
        & \RunLowTauGoodIncTwoBHBOneMassRatioZeroPost
        & \RunLowTauGoodIncTwoBHBOneDimensionlessSpinOneZeroPost 
        & \RunLowTauGoodIncTwoBHBOneDimensionlessSpinTwoZeroPost
        & \RunLowTauGoodIncTwoBHBOneMergerTimeOrInitialOrbitalFrequencyZeroWidth 
        & \RunLowTauGoodIncTwoBHBOneMergerTimeZeroWidth
        \\
        & \RunLowTauBadIncOneBHBOneRedshiftedChirpMassZeroWidth
        & \RunLowTauBadIncOneBHBOneRedshiftedMassOneZeroPost
        & \RunLowTauBadIncOneBHBOneRedshiftedMassTwoZeroPost
        & \RunLowTauBadIncOneBHBOneMassRatioZeroPost
        & \RunLowTauBadIncOneBHBOneDimensionlessSpinOneZeroPost 
        & \RunLowTauBadIncOneBHBOneDimensionlessSpinTwoZeroPost
        & \RunLowTauBadIncOneBHBOneMergerTimeOrInitialOrbitalFrequencyZeroWidth
        & \RunLowTauBadIncOneBHBOneMergerTimeZeroWidth
        \\
        & \RunLowTauBadIncTwoBHBOneRedshiftedChirpMassZeroWidth
        & \RunLowTauBadIncTwoBHBOneRedshiftedMassOneZeroPost
        & \RunLowTauBadIncTwoBHBOneRedshiftedMassTwoZeroPost
        & \RunLowTauBadIncTwoBHBOneMassRatioZeroPost
        & \RunLowTauBadIncTwoBHBOneDimensionlessSpinOneZeroPost
        & \RunLowTauBadIncTwoBHBOneDimensionlessSpinTwoZeroPost
        & \RunLowTauBadIncTwoBHBOneMergerTimeOrInitialOrbitalFrequencyZeroWidth
        & \RunLowTauBadIncTwoBHBOneMergerTimeZeroWidth
        \\
        \hline
        \textcolor{teal}{\large \ding{73}}
        & \RunIMBHTealBHBOneRedshiftedChirpMassZeroWidth
        & \RunIMBHTealBHBOneRedshiftedMassOneZeroPost
        & \RunIMBHTealBHBOneRedshiftedMassTwoZeroPost
        & \RunIMBHTealBHBOneMassRatioZeroPost
        & \RunIMBHTealBHBOneDimensionlessSpinOneZeroPost
        & \RunIMBHTealBHBOneDimensionlessSpinTwoZeroPost
        & \RunIMBHTealBHBOneMergerTimeOrInitialOrbitalFrequencyZeroWidth
        & \RunIMBHTealBHBOneMergerTimeZeroWidth 
        \\
        \textcolor{magenta}{\large \ding{73}}
        & \RunIMBHPurpleBHBOneRedshiftedChirpMassZeroWidth
        & \RunIMBHPurpleBHBOneRedshiftedMassOneZeroPost
        & \RunIMBHPurpleBHBOneRedshiftedMassTwoZeroPost
        & \RunIMBHPurpleBHBOneMassRatioZeroPost
        & \RunIMBHPurpleBHBOneDimensionlessSpinOneZeroPost
        & \RunIMBHPurpleBHBOneDimensionlessSpinTwoZeroPost
        & \RunIMBHPurpleBHBOneMergerTimeOrInitialOrbitalFrequencyZeroWidth
        & \RunIMBHPurpleBHBOneMergerTimeZeroWidth 
        \\
        \textcolor{black}{\large \ding{73}}
        & \RunIMBHBlackBHBOneRedshiftedChirpMassZeroWidth
        & \RunIMBHBlackBHBOneRedshiftedMassOneZeroPost
        & \RunIMBHBlackBHBOneRedshiftedMassTwoZeroPost
        & \RunIMBHBlackBHBOneMassRatioZeroPost
        & \RunIMBHBlackBHBOneDimensionlessSpinOneZeroPost
        & \RunIMBHBlackBHBOneDimensionlessSpinTwoZeroPost
        & \RunIMBHBlackBHBOneMergerTimeOrInitialOrbitalFrequencyZeroWidth
        & \RunIMBHBlackBHBOneMergerTimeZeroWidth 
        \\
        \hline
    \end{tabular}
\end{table}

\begin{table}
\caption{Recovered extrinsic parameters for selected sources as discussed in Section~\ref{subsec:param-estim}. Quoted fractional errors are computed as the ratio between the posterior $90\%$ confidence intervals and corresponding median. 
Point estimates $m^{+\Delta u}_{-\Delta l}$ denote posterior median, upper and lower widths corresponding to a $90\%$ confidence interval. \label{tab:pe-extrinsic}}
    \centering
    \addtolength{\tabcolsep}{-0.4em}
    \renewcommand{\arraystretch}{1.4} 
    \begin{tabular}{|c|cccc|rrr|}
        \hline
        Run 
        & $z$
        & $d_L [{\rm Mpc]}$
        & $\sin b$
        & $l$
        & $\cos{\iota}$
        & $\psi$
        & $\phi_{\rm orb}$
        \\
        \hline
        High $\tau_c$
        & \RunHiTauGoodIncOneBHBOneRedshiftZeroPost
        & \RunHiTauGoodIncOneBHBOneLuminosityDistanceZeroPost
        & \RunHiTauGoodIncOneBHBOnesinEclipticLatitudeZeroPost
        & \RunHiTauGoodIncOneBHBOneEclipticLongitudeZeroPost
        & \RunHiTauGoodIncOneBHBOnecosInclinationZeroPost
        & \RunHiTauGoodIncOneBHBOnePolarizationZeroPost
        & \RunHiTauGoodIncOneBHBOneInitialOrbitalPhaseZeroPost
        \\
        & \RunHiTauGoodIncTwoBHBOneRedshiftZeroPost
        & \RunHiTauGoodIncTwoBHBOneLuminosityDistanceZeroPost
        & \RunHiTauGoodIncTwoBHBOnesinEclipticLatitudeZeroPost
        & \RunHiTauGoodIncTwoBHBOneEclipticLongitudeZeroPost
        & \RunHiTauGoodIncTwoBHBOnecosInclinationZeroPost
        & \RunHiTauGoodIncTwoBHBOnePolarizationZeroPost
        & \RunHiTauGoodIncTwoBHBOneInitialOrbitalPhaseZeroPost
        \\
        & \RunHiTauBadIncOneBHBOneRedshiftZeroPost
        & \RunHiTauBadIncOneBHBOneLuminosityDistanceZeroPost
        & \RunHiTauBadIncOneBHBOnesinEclipticLatitudeZeroPost
        & \RunHiTauBadIncOneBHBOneEclipticLongitudeZeroPost
        & \RunHiTauBadIncOneBHBOnecosInclinationZeroPost
        & \RunHiTauBadIncOneBHBOnePolarizationZeroPost
        & \RunHiTauBadIncOneBHBOneInitialOrbitalPhaseZeroPost
        \\
        & \RunHiTauBadIncTwoBHBOneRedshiftZeroPost
        & \RunHiTauBadIncTwoBHBOneLuminosityDistanceZeroPost
        & \RunHiTauBadIncTwoBHBOnesinEclipticLatitudeZeroPost
        & \RunHiTauBadIncTwoBHBOneEclipticLongitudeZeroPost
        & \RunHiTauBadIncTwoBHBOnecosInclinationZeroPost
        & \RunHiTauBadIncTwoBHBOnePolarizationZeroPost
        & \RunHiTauBadIncTwoBHBOneInitialOrbitalPhaseZeroPost
        \\
        \hline
        Mid $\tau_c$
        & \RunMidTauGoodIncOneBHBOneRedshiftZeroPost
        & \RunMidTauGoodIncOneBHBOneLuminosityDistanceZeroPost
        & \RunMidTauGoodIncOneBHBOnesinEclipticLatitudeZeroPost
        & \RunMidTauGoodIncOneBHBOneEclipticLongitudeZeroPost
        & \RunMidTauGoodIncOneBHBOnecosInclinationZeroPost
        & \RunMidTauGoodIncOneBHBOnePolarizationZeroPost
        & \RunMidTauGoodIncOneBHBOneInitialOrbitalPhaseZeroPost
        \\
        & \RunMidTauGoodIncTwoBHBOneRedshiftZeroPost
        & \RunMidTauGoodIncTwoBHBOneLuminosityDistanceZeroPost
        & \RunMidTauGoodIncTwoBHBOnesinEclipticLatitudeZeroPost
        & \RunMidTauGoodIncTwoBHBOneEclipticLongitudeZeroPost
        & \RunMidTauGoodIncTwoBHBOnecosInclinationZeroPost
        & \RunMidTauGoodIncTwoBHBOnePolarizationZeroPost
        & \RunMidTauGoodIncTwoBHBOneInitialOrbitalPhaseZeroPost
        \\
        & \RunMidTauBadIncOneBHBOneRedshiftZeroPost
        & \RunMidTauBadIncOneBHBOneLuminosityDistanceZeroPost
        & \RunMidTauBadIncOneBHBOnesinEclipticLatitudeZeroPost
        & \RunMidTauBadIncOneBHBOneEclipticLongitudeZeroPost
        & \RunMidTauBadIncOneBHBOnecosInclinationZeroPost
        & \RunMidTauBadIncOneBHBOnePolarizationZeroPost
        & \RunMidTauBadIncOneBHBOneInitialOrbitalPhaseZeroPost
        \\
        & \RunMidTauBadIncTwoBHBOneRedshiftZeroPost 
        & \RunMidTauBadIncTwoBHBOneLuminosityDistanceZeroPost
        & \RunMidTauBadIncTwoBHBOnesinEclipticLatitudeZeroPost
        & \RunMidTauBadIncTwoBHBOneEclipticLongitudeZeroPost
        & \RunMidTauBadIncTwoBHBOnecosInclinationZeroPost
        & \RunMidTauBadIncTwoBHBOnePolarizationZeroPost
        & \RunMidTauBadIncTwoBHBOneInitialOrbitalPhaseZeroPost
        \\
        \hline
        Low $\tau_c$
        & \RunLowTauGoodIncOneBHBOneRedshiftZeroPost 
        & \RunLowTauGoodIncOneBHBOneLuminosityDistanceZeroPost
        & \RunLowTauGoodIncOneBHBOnesinEclipticLatitudeZeroPost
        & \RunLowTauGoodIncOneBHBOneEclipticLongitudeZeroPost
        & \RunLowTauGoodIncOneBHBOnecosInclinationZeroPost
        & \RunLowTauGoodIncOneBHBOnePolarizationZeroPost
        & \RunLowTauGoodIncOneBHBOneInitialOrbitalPhaseZeroPost
        \\
        & \RunLowTauGoodIncTwoBHBOneRedshiftZeroPost 
        & \RunLowTauGoodIncTwoBHBOneLuminosityDistanceZeroPost
        & \RunLowTauGoodIncTwoBHBOnesinEclipticLatitudeZeroPost
        & \RunLowTauGoodIncTwoBHBOneEclipticLongitudeZeroPost
        & \RunLowTauGoodIncTwoBHBOnecosInclinationZeroPost
        & \RunLowTauGoodIncTwoBHBOnePolarizationZeroPost
        & \RunLowTauGoodIncTwoBHBOneInitialOrbitalPhaseZeroPost
        \\
        & \RunLowTauBadIncOneBHBOneRedshiftZeroPost
        & \RunLowTauBadIncOneBHBOneLuminosityDistanceZeroPost
        & \RunLowTauBadIncOneBHBOnesinEclipticLatitudeZeroPost
        & \RunLowTauBadIncOneBHBOneEclipticLongitudeZeroPost
        & \RunLowTauBadIncOneBHBOnecosInclinationZeroPost
        & \RunLowTauBadIncOneBHBOnePolarizationZeroPost
        & \RunLowTauBadIncOneBHBOneInitialOrbitalPhaseZeroPost
        \\
        & \RunLowTauBadIncTwoBHBOneRedshiftZeroPost 
        & \RunLowTauBadIncTwoBHBOneLuminosityDistanceZeroPost
        & \RunLowTauBadIncTwoBHBOnesinEclipticLatitudeZeroPost
        & \RunLowTauBadIncTwoBHBOneEclipticLongitudeZeroPost
        & \RunLowTauBadIncTwoBHBOnecosInclinationZeroPost
        & \RunLowTauBadIncTwoBHBOnePolarizationZeroPost
        & \RunLowTauBadIncTwoBHBOneInitialOrbitalPhaseZeroPost
        \\
        \hline
        \textcolor{teal}{\large \ding{73}}
        & \RunIMBHTealBHBOneRedshiftZeroPost
        & \RunIMBHTealBHBOneLuminosityDistanceZeroPost
        & \RunIMBHTealBHBOnesinEclipticLatitudeZeroPost
        & \RunIMBHTealBHBOneEclipticLongitudeZeroPost
        & \RunIMBHTealBHBOnecosInclinationZeroPost
        & \RunIMBHTealBHBOnePolarizationZeroPost
        & \RunIMBHTealBHBOneInitialOrbitalPhaseZeroPost
        \\
        \textcolor{magenta}{\large \ding{73}}
        & \RunIMBHPurpleBHBOneRedshiftZeroPost
        & \RunIMBHPurpleBHBOneLuminosityDistanceZeroPost
        & \RunIMBHPurpleBHBOnesinEclipticLatitudeZeroPost
        & \RunIMBHPurpleBHBOneEclipticLongitudeZeroPost
        & \RunIMBHPurpleBHBOnecosInclinationZeroPost
        & \RunIMBHPurpleBHBOnePolarizationZeroPost
        & \RunIMBHPurpleBHBOneInitialOrbitalPhaseZeroPost
       \\
        \textcolor{black}{\large \ding{73}}
        & \RunIMBHBlackBHBOneRedshiftZeroPost
        & \RunIMBHBlackBHBOneLuminosityDistanceZeroPost
        & \RunIMBHBlackBHBOnesinEclipticLatitudeZeroPost
        & \RunIMBHBlackBHBOneEclipticLongitudeZeroPost
        & \RunIMBHBlackBHBOnecosInclinationZeroPost
        & \RunIMBHBlackBHBOnePolarizationZeroPost
        & \RunIMBHBlackBHBOneInitialOrbitalPhaseZeroPost
        \\
        \hline
    \end{tabular}
\end{table}

\section{Conclusions}
\label{sec:conclusions}

In this paper we have analysed the potential of LISA to characterise the population of \srcnames{} observed by the LVK detector network. 
We have constructed synthetic catalogues of circular \srcnames{}, based on the \PP population inferred from the GWTC-3 catalogue, as in~\cite{Babak:2023lro}. 
The distribution of source parameters follows the posteriors on the chirp mass, mass ratio and spin models detailed in~\cite{KAGRA:2021duu}, while the extrinsic source parameters are drawn from uniform and isotropic distributions. 
The sources we find detectable by LISA cluster at low redshift, and therefore the dependence of their merger rate with redshift can be approximated with a single power law, for which we have also sampled over the parameter posteriors derived in~\cite{KAGRA:2021duu}.

While in~\cite{Babak:2023lro} these synthetic catalogues were used to forecast the SGWB from unresolved \srcnames{} in the LISA band, in this paper we have re-generated the low-redshift tail of these catalogues to use the reciprocal information that can be extracted from them, i.e.~the number and characteristics of the sources resolvable by LISA. 
As a selection criterion for detectability, we have adopted an SNR threshold $\rho_0$ set at 8 for direct LISA detection, or at 6 and 4 for archival searches, assuming 4\,yr of LISA mission duration operating at nominal sensitivity.
We have found that the number of expected resolvable sources with $\rho_0=8$ has a median value of about 5.
With $10\%$ probability, this number will be higher than 10; 
however, with $2\%$ probability, no sources will be detected by LISA.
Decreasing the SNR detection threshold increases the number of sources: the LISA data stream is expected to contain at least 20 sources with $\rho_0= 4$ with probability above $90\%$,
showing that archival searches will provide a relatively high number of additional \srcname{} detections.
For $\rho_0= 6$, at least 5 detectable sources are expected at $90\%$ probability, and the probability of detecting zero sources is extremely small, of the order of $0.03\%$.

We have also analysed the subset of sources for which multiband observations can occur within the LISA mission or shortly after. 
It turns out that the number of sources merging within $15\,\mathrm{yr}$ from the time LISA starts operating, is around a third of the total population detectable by LISA, independently of the assumed SNR threshold $\rho_0$.
In particular, the number of multiband detectable sources by LISA alone with $\rho_0=8$ is compatible with 0 at $21\%$ probability and has a median of {$2$}. 
For $\rho_0=6$, such a number is still compatible with zero though at $10\%$ probability only, and has a median value of 4. It instead reaches 5 with a probability above $90\%$ for $\rho_0= 4$. 
Such sources with SNR just above $\rho_0=4$ or 6 cannot be resolved by LISA alone, but thanks to their multiband property, they are expected to be recovered through archival searches~\cite{Ewing:2020brd}. 

The number of sources itself is an important observable, though subject to large Poisson uncertainties for the highest threshold SNRs: 
a number of resolved signals highly inconsistent with the above predictions could hint at missing features in the sBHB population model (e.g.\ higher-than-expected likelihood of larger masses, local effects in the merger rate), especially when combined with the corresponding prediction for the sBHB confusion noise, which LISA should be able to measure with high precision~\cite{Babak:2023lro} (see, however, \cite{Staelens:2023xjn}).
Indeed, we have found sizable correlations between the number of detectable events in LISA and some of the population parameters, most notably a strong, negative correlation with $\kappa$, the power-law index of the merger rate dependence on redshift (a lower $\kappa$ produces an excess of events at low redshift). Milder correlations appear with the power-law index of the primary mass distribution $\alpha$, and with the mixture parameter $\lambda_{\rm peak}$ determining the Gaussian bump superimposed on the power law, where each of them increases the number of higher-mass sources.
Remarkably, the amplitude of the \srcnames{} confusion noise presents similar correlations but with opposite sign, which makes a combined measurement of the number of resolvable \srcnames{} and their confusion noise a powerful tool to unveil the characteristics of the \srcnames{} population.

We have moreover shown that the subset of \srcnames{} detectable by LISA and compatible with \cat have chirp mass $10\,\msun\lesssim \mathcal{M}_c\lesssim 100\, \msun$, residual time-to-coalescence $4 \,\mathrm{yr}\lesssim \tau_c\lesssim 100\,\mathrm{yr}$, and redshift $z\lesssim 0.1$. 
Hence, LISA can explore the tail of the \srcname{} population at very low redshift and relatively high chirp mass, towards 100\,$\msun$.
Even though the number of such sources which are detectable by LISA is rather small (not surprisingly since they pertain to the tail of the distribution), their detection is still relevant and complementary to the observations available before LISA flies: LVK at design sensitivity would need an unrealistic number of observing runs to equivalently probe this region of the parameter space. 
This also suggests that LISA would be best equipped to characterise a population with higher chirp mass than the \PP high-mass cutoff, if it existed.
To prove this, while remaining agnostic about the population, we have constructed for the first time the LISA waterfall plot for chirp mass smaller than $4 \times 10^3 \msun$. We found that in this range of masses it strongly depends on the residual time-to-coalescence and the source inclination. 
For sources remaining in band at least for the entire mission duration, as mass increases from $10\,\msun$, SNR increases with mass, increasing in turn the reach in redshift;
after a peak at $\mathcal{M}_c\sim 100\,\msun$, the tendency inverts as the binary signal is displaced to frequencies where the Galactic foreground dominates the noise budget, reducing the source's SNR;
at $\mathcal{M}_c\gtrsim 10^3\,\msun$, the sensitivity improves again as the binary's signal mostly occur in the high-frequency part of the LISA noise curve, where the Galactic foreground is subdominant.
For sources remaining in band for less than the mission duration, sensitivity grows along with chirp mass for the whole mass range, because their SNR is always dominated by the signal at high frequency.
However, as already discussed, sources with such a low $\tau_c$ are extremely rare: not only they are intrinsically less numerous, but also have smaller accumulated SNR, despite their chirping that makes detection easier.

The comparison of our LISA and LVK waterfall plots in the chirp mass region $\mathcal O (1)$ -- $\mathcal{O}(10^3) \,\msun$ 
has shown that the complementarity of the observatories mainly resides in the region $\mathcal{M}_c\gtrsim 10^3\,\msun$, where the LVK sensitivity drops.
However, by performing parameter estimation on a subset of sources selected from the catalogues (i.e.~drawn from the \PP population), we have shown that LISA has great potential in characterising \srcnames{} with sufficiently high SNR (between 8 and 12), granting a much better precision in the parameter inference than LVK. 
Therefore, even if LISA measurements will not be directly informative on the population due to the very low number of resolvable sources, they can help characterising a few, low-redshift candidates with great precision. 

We have reached such conclusions by performing parameter estimation 
on \srcnames{} with $\tau_c\lesssim 15$ yr and SNR above the threshold $\rho_0=8$. 
This implies the sources to be
close by ($z\lesssim 0.1$) and 
emitting in the high-frequency part of the LISA sensitivity ($f_{\rm orb}^0\geq 4.4$ mHz). 
The selected sources turn out to be
characterised with exquisite precision.
As far as the chirp mass, initial orbital frequency and residual time-to-coalescence are concerned, their relative reconstruction errors are all below the per-mille level by at least one order of magnitude. For sources with $\tau_c<10\,{\rm yr}$, the residual time-to-coalescence can be reconstructed with an uncertainty of a few hours,  
showing that multiband detection can be feasible
at a sensitivity similar to current ground-based interferometers, such that not many sources are expected to be detectable over that time interval.

Additionally we explore the sensitivity of LISA to a population of sources with intermediate values of the chirp mass $\mathcal{M}_c= \mathcal{O}(10^3)\,\msun$, by analysing three representative sources with $\mathcal{M}_c= 10^3\,\msun$ and $z=0.1,\,0.2$ and $0.8$, to which LVK is not sensitive.
The outcome is that LISA can reconstruct the chirp mass and residual time-to-coalescence of these sources with relative error 
below the per-mille level, and down to $\mathcal{O}(10^{-6})$ for the source with $z=0.8$, for which redshift can be determined down to a precision of $7\%$.
Therefore, if such a population exists, LISA will be able to fully characterise it and distinguish it from the \srcnames{} probed by LVK.

In this paper we have restricted the analysis to the LISA nominal mission duration of 4\,yr, and assumed LVK network to have design sensitivity. 
Nevertheless, it is certainly worth extending our analysis to longer mission times and the co-existence of Einstein Telescope and Cosmic Explorer, 
which are expected to be much more sensitive than LVK to the high-mass tail 
of the \srcname{} population. It would also be of great interest to revise the sBHB-related science that LISA can achieve, using our catalogues. Several tests for general relativity and cosmography with LISA have been proposed~\cite{LISACosmologyWorkingGroup:2022jok, LISA:2022yao}, but their expected outcome strongly depends on the properties and number of the sBHBs that LISA can resolve. 
Our priority in the short-term future is to account for 
the next LVK sBHB population inference analysis, which may provide surprises. 
These are not expected to modify the qualitative conclusions of the present paper, 
but with small adjustments to the methods and codes described here, we can 
readily obtain reliable, quantitative updates. 

The full set of synthetic populations generated for and used in this work, together with a Python notebook demonstrating parts of the present analysis, is available at \url{https://zenodo.org/records/13974091} \cite{torrado_2024_13974091}.
The data and code to reproduce figures throughout the paper is available at~\cite{sobbh-data-release}.

\acknowledgments

The authors would like to thank S.~Babak, I.Dvorkin, A.~Klein, L.~Lehoucq, P.Marcoccia, and A.~Ricciardone for useful comments and stimulating discussions.
RB would like to thank the \textsc{Balrog} developers for valuable feedback throughout.
RB acknowledges support through the Italian Space Agency grant \emph{Phase A activity for LISA mission, Agreement n. 2017-29-H.0}
and by the ICSC National Research Center funded by NextGenerationEU. 
JT acknowledges financial support from the Supporting TAlent in ReSearch@University of Padova (STARS@UNIPD) for the project ``Constraining Cosmology and Astrophysics with Gravitational Waves, Cosmic Microwave Background and Large-Scale Structure cross-correlations''.
CC acknowledges support from the Swiss National Science Foundation (SNSF Project Funding grant \href{https://data.snf.ch/grants/grant/212125}{212125})
NK acknowledges support from the European Union’s Horizon 2020
research and innovation programme under the Marie Skłodowska-Curie grant agreement No 101065596.
GN is partly supported by the grant Project
No. 302640 of the Research Council of Norway.
MP acknowledges the hospitality of
Imperial College London, which provided office space during some parts of this project.
AS acknowledges financial support provided under the European Union’s H2020 ERC Consolidator Grant ``Binary Massive Black Hole Astrophysics'' (B Massive, Grant Agreement: 818691).
Computational resources were provided by the University of Birmingham BlueBEAR High Performance Computing facility, CINECA with allocations through INFN, Bicocca, ISCRA project HP10BEQ9JB, Google Cloud research credit program, award no.GCP19980904, and the CloudVeneto initiative of the Università di Padova and the INFN -- Sezione di Padova.

\bibliographystyle{JHEP}
\bibliography{references}
\end{document}